\documentclass[aps, prd, showpacs, floatfix, superscriptaddress, twocolumn, nofootinbib, reprint, preprintnumbers]{revtex4-2}

\usepackage[section]{placeins}

\usepackage{enumitem}
\setlist[enumerate]{leftmargin=14pt, labelsep=0.20em, itemsep=0.em}
\usepackage{subcaption}
\usepackage{siunitx}
\usepackage{lipsum, multirow, microtype, amsmath, amssymb, newfloat, bm, color}
\usepackage{graphicx} 
\usepackage{dcolumn}  

\usepackage[dvipsnames]{xcolor}

\usepackage{orcidlink}
\newtheorem{theorem}{Theorem}

\usepackage{hyperref}
\usepackage{hypcap}
\hypersetup{
  colorlinks=true,
  citecolor=red,
  linkcolor=magenta,
  urlcolor=NavyBlue,
  allbordercolors={0 0 0},
  pdfborderstyle={/S/U/W 1}
}

\newcommand{\norm}[1]{\ensuremath{\lVert#1\rVert}}

\def \Hz	    {\mathrm{Hz}}

\def \pycbc     {{\textsc{PyCBC}}}

\def \param     {\vec \Lambda}

\def \ngauss    {N_{\text{g}}}
\def \nunif     {N_{\text{u}}}
\def \relbin    {{\texttt{RelBin}}}
\def \TaylorF   {{\texttt{TaylorF2}}}

\newcommand{\IITGn}{Indian Institute of Technology Gandhinagar, Palaj Gandhinagar, Gujarat 382055, India.\vspace*{4pt}}
\newcommand{\IISERT}{Indian Institute of Science Education and Research Tirupati, Andhra Pradesh 517501, India.\vspace*{3pt}}
\newcommand{\NIKHEF}{Nikhef, Science Park 105, 1098 XG Amsterdam, The Netherlands.}
\newcommand{\UU}{Institute for Gravitational and Subatomic Physics (GRASP), \mbox{Utrecht University}, Princetonplein 1, 3584 CC Utrecht, The Netherlands.}
\newcommand{\IWF}{Space Research Institute, Austrian Academy of Sciences, Schmiedlstrasse 6, 8042 Graz, Austria}

\begin{document}


\title{
{Prompt sky localization of compact binary sources using meshfree approximation}
}
\author{\sc{Lalit Pathak}\orcidlink{0000-0002-9523-7945}} 
\email{lalit.pathak@iitgn.ac.in} \affiliation{\IITGn}
\author{\sc{Sanket Munishwar}\orcidlink{0009-0007-6954-7556}}
\email{sanket.m@students.iisertirupati.ac.in} \affiliation{\IISERT}
\author{\sc{Amit Reza}\orcidlink{0000-0001-7934-0259}} 
\email{areza@nikhef.nl} \affiliation{\NIKHEF} \affiliation{\UU} \affiliation{\IWF}
\author{\sc{Anand S. Sengupta}\orcidlink{0000-0002-3212-0475}\vspace*{3pt}} 
\email{asengupta@iitgn.ac.in} \affiliation{\IITGn}
\begin{abstract}
The number of gravitational wave signals from the merger of compact binary systems detected in the network of advanced LIGO and Virgo detectors is expected to increase considerably in the upcoming science runs. Once a confident detection is made, it is crucial to reconstruct the source's properties rapidly, particularly the sky position and chirp mass, to follow up on these transient sources with telescopes operating at different electromagnetic bands for multi-messenger astronomy. In this context, we present a rapid parameter estimation (PE) method aided by mesh-free approximations to accurately reconstruct properties of compact binary sources from data gathered by a network of gravitational wave detectors. This approach builds upon our previous algorithm [L. Pathak \textit{et al.}, Fast likelihood evaluation using meshfree approximations for reconstructing compact binary sources, \href{https://journals.aps.org/prd/abstract/10.1103/PhysRevD.108.064055}{Phys. Rev. D \textbf{108}, 064055 (2023)}] to expedite the evaluation of the likelihood function and extend it to enable coherent network PE in a ten-dimensional parameter space, including sky position and polarization angle. Additionally, we propose an optimized interpolation node placement strategy during the start-up stage to enhance the accuracy of the marginalized posterior distributions.
With this updated method, we can estimate the properties of binary neutron star (BNS) sources in approximately 2.4~(2.7) min for the \TaylorF~(\texttt{IMRPhenomD}) signal model by utilizing 64 CPU cores on a shared memory architecture. 
Furthermore, our approach can be integrated into existing parameter estimation pipelines, providing a valuable tool for the broader scientific community. We also highlight some areas for improvements to this algorithm in the future, which includes overcoming the limitations due to narrow prior bounds.
\end{abstract}

\maketitle 

\section{Introduction}
\label{sec:intro}
Gravitational wave signals emitted by compact binary systems made of neutron stars and black holes are expected to take up to a few minutes to sweep through the sensitive band of the advanced LIGO and Virgo detectors, leading up to the epoch of their eventual cataclysmic merger. The merger of two neutron stars/Neutron star-black holes (NSBH)~\cite{Eichler:1989ve, 1986ApJ_308L_47G, kilonovae_metzger, Li_1998, PhysRevD.107.124007} can be followed by the emission of electromagnetic (EM) radiation at different wavelengths that fade over different time scales ranging from a few minutes to several months, as seen in the fortuitous detection of the binary neutron star system (BNS) GW170817~\cite{abbott2017gw170817} by the network of advanced LIGO-Virgo detectors~\cite{Harry:2010zz, Abbott_2020, Acernese_2014, 2015, PhysRevD.93.112004} in the second science run. This discovery resulted in the first multi-wavelength observation of post-merger electromagnetic (EM) emissions from the merger of two neutron stars. The successful localization and discovery, in this case, were primarily due to the source's fortunate proximity: GW170817 was relatively close (about 40 Mpc), falling well within the sky- and orientation-averaged ranges of the two LIGO detectors, resulting in the loudest signal ever detected by a GW network. The discovery of the EM counterpart of GW170817 improved understanding of the physics of short gamma-ray bursts (SGRBs)~\cite{lamb2018grb}, confirmation of the formation of heavy elements after the merger and other areas of new physics, and provided invaluable opportunities to explore the mysteries of the universe. Therefore, BNS systems are the prime targets for multi-messenger astronomy.

In upcoming observing runs, the LIGO-Virgo-Kagra detectors~\cite{Harry:2010zz, Acernese_2014, 2015, PhysRevD.93.112004} are expected to be sensitive above $10\,\Hz$, enabling a more extensive exploration of cosmic volume compared to their current versions. At design sensitivity~\cite{abbott2020prospects}, these detectors are expected to have a BNS range of $\sim 330$ Mpc (advanced LIGO), $\sim 150 - 260$ Mpc (advanced Virgo) and $\sim 155$ Mpc (KAGRA). 
Consequently, the detection rate is expected to increase to the point where signals from compact binary mergers can be observed daily. It is project that there could be $10^{+52}_{-10}$/$1^{+91}_{-1}$ BNS/NSBH events detected in the O4 run~\cite{abbott2020prospects}. To efficiently observe EM counterparts in the future, it will be crucial to prioritize events based on their chirp masses, as suggested by Margalit et al.~\cite{Margalit_2019}. Furthermore, future GW events could provide opportunities to study various physical properties of binary systems, such as those with eccentricity~\cite{Lenon_2020}, spin precession~\cite{PhysRevD.49.6274, Hannam_2022}, and more. As such, prompt localization and source reconstruction would be vital for studying electromagnetic (EM) counterparts of binary neutron star systems (BNS) or neutron star-black hole (NSBH) binary systems in the future. Moreover, a fast PE technique can facilitate the execution of extensive inference investigations~\cite{wolfe2023small}, like population inference, within a reasonable time frame, which could be impractical when employing conventional brute-force PE methods.

The standard technique for PE uses Bayesian inference, which involves calculating the likelihood of observing the data under the hypothesis that an astrophysical signal (with some model parameters) is embedded in additive Gaussian noise with zero mean and unit variance. This likelihood calculation involves two computationally expensive parts: first, generating the model (template) waveforms~\cite{PhysRevD.103.104056, PhysRevResearch.1.033015, ramosbuades2023seobnrv5phm} at sample points proposed by the sampling algorithm and, thereafter, determining the overlap between the data and the waveform. Generating these templates is the most computationally expensive part of the likelihood calculation, especially for low-mass systems like binary neutron stars (BNS). This computational burden is amplified by the improved sensitivity of detectors, resulting in a larger number of waveform cycles within the detectors' sensitive band for BNS systems. Additionally, incorporating physical effects into the waveform can further increase the computational cost of generating the waveform and sampling from the expanded parameter space. 
Wherever possible, factorizing~\cite{islam2022factorized} the log-likelihood function into pieces that exclusively depend on the extrinsic and intrinsic parameters, respectively, can lead to a significant reduction in the number of waveform generation and thereby speed up the PE analysis. For example, such factorizations can be made for non-precessing waveforms, as considered in this work. 

In the current operational framework of the LIGO-Virgo-KAGRA (LVK) collaboration, the primary tool for rapid sky localization is {\texttt{BAYESTAR}}~\cite{singer2016rapid} - a Bayesian, non-Markov Chain Monte Carlo (MCMC) sky localization algorithm. This method exhibits remarkable speed, furnishing posterior probability density distributions across sky coordinates within a few tens of seconds following the detection of a gravitational wave signal from the merger of a compact binary source.
However, through a comprehensive Bayesian PE analysis encompassing both intrinsic and extrinsic model parameters, Finstad et al.~\cite{Finstad_2020} have established that the accuracy of sky localization can be significantly augmented, achieving an enhancement of approximately ${14 \, \text{deg}^2}$ over those obtained from the {\texttt{BAYESTAR}} algorithm alone.
Beyond the evident advantages of sky-location precision, a full PE analysis also furnishes a highly accurate estimation of the chirp mass for compact binary systems. This additional information plays a pivotal role in making judicious decisions regarding electromagnetic (EM) follow-up observations - which would be crucial in future observing runs of the network of advanced detectors, further underscoring the merits of developing fast-PE algorithms. Deep-learning-based sky-localization tools (CBC-SkyNet~\cite{chatterjee2022premerger}) have been recently developed that can obtain ``pre-merger'' sky localization areas that are comparable in accuracy to {\texttt{BAYESTAR}}.

Numerous rapid PE algorithms have surfaced in the gravitational-wave literature, which revolves around two overarching concepts:
\begin{enumerate}
    \item[i.] Likelihood-Based Approaches: The first category involves approaches that rapidly evaluate the likelihood function. This set of methods encompasses various methods such as Reduced Order Models (ROMs)~\cite{canizares2013gravitational, canizares2015accelerated, smith2016fast, Morisaki_2020, morisaki2023rapid}, Heterodyning (or relative binning)~\cite{Venumadhav2018, cornish2021heterodyned, islam2022factorized}, the simple-pe~\cite{fairhurst2023fast} algorithm, and methods based on Gaussian process regression (GPR)~\cite{rasmussen2006gaussian} like RIFT~\cite{https://doi.org/10.48550/arxiv.1805.10457}. Other techniques, such as adaptive frequency resolution-based likelihood evaluation~\cite{Morisaki:2021ngj} and mass-spin reparametrization-based rapid PE~\cite{Lee:2022jpn}, have also been proposed. Improved algorithms allowing for more efficient reduced-order quadrature bases have been recently proposed~\cite{effroqs23}.
    
    \item[ii.] Likelihood "Free" Approaches: The second category consists of methods that bypass direct likelihood evaluation and instead learn the posterior distributions using advanced machine-learning (ML) techniques such as Deep Learning, Normalizing Flows, and Variational Inference~\cite{Chua_2020, Green_2020, green2020complete, gabbard2022bayesian}.
\end{enumerate}
Hybrid techniques that combine likelihood heterodyning with tools to enhance the convergence of gradient-based MCMC samplers have also been proposed~\cite{wong2023fast}. Other techniques involving score-based diffusion models~\cite{score_based_likelihood} to learn an empirical noise distribution directly from the detector data have also been proposed.

In this context, our mesh-free approach aligns with the first category of rapid PE methods since it is designed to swiftly assess the likelihood. 
Note that the RIFT~\cite{https://doi.org/10.48550/arxiv.1805.10457} method also interpolates the likelihood directly using GPR. In comparison, the meshfree method first represents the likelihood function on an orthonormal basis and then uses radial basis functions to interpolate the coefficients. So, inspite of the apparent similarity, the context in which the “interpolation” technique is applied is completely different in the two methods. The use of GPR (a supervised training method) to ``learn'' the likelihood function at different points of the parameter space is in the spirit of ML methods in contrast with the classical approach followed by the meshfree method.

In our previous work~\cite{pathak2022rapid}, we introduced the meshfree method as a means of rapidly inferring parameters for compact binary coalescence (CBC) sources for a single detector by considering only six parameters: the two-component masses, two aligned spins, luminosity distance, and coalescence time.
In this study, (a) we extend the method to include additional parameters such as sky location, inclination, and polarization, thereby enabling a coherent multi-detector PE analysis. (b) Furthermore, we have modified the Radial Basis Function (RBF) meshfree interpolation node placement scheme, which directly impacts the accuracy of the PE results. In contrast to our previous work, which utilized uniformly distributed nodes over the sample space, we now employ a combination of multivariate Gaussian and uniform distributions for efficient node placement. 
(c) Our results demonstrate that when run on $32\, (64)$ cores, the meshfree method can produce accurate marginalized posteriors for the GW170817 event and locate it in the sky within $\sim 3.4\, (2.4)$ minutes of detecting the event. The resulting posteriors obtained from the meshfree method are statistically indistinguishable from those obtained using the brute force method implemented in \pycbc~\cite{alex_nitz_2022_6912865}. Please note that the brute force method is not optimized and is used only as a yardstick for accuracy.

The rest of the paper is structured into the following sections: Section \ref{sec:param_est} introduces the PE basics, where we briefly discuss the Bayesian approach for inferring GW parameters from the data and define the coherent network likelihood. Section \ref{sec:meshfree} explains the start-up and online stages and discusses the likelihood evaluation procedure using meshfree interpolation. The next section \ref{subsec:gw17} presents the results of a detailed analysis of the meshfree method for the GW170817 BNS event. 
In Section \ref{subsec:simulated Data}, we examine the effectiveness of our method by testing it on simulated events covering a wide range of signal-to-noise ratios (SNR). 
Finally, we summarize the results in Section~\ref{sec:concl_outl} and discuss the limitations of the current implementation (such as restricting the sampler over narrow prior bounds). We suggest some ideas for overcoming this limitation in follow-up studies.

\section{Parameter estimation}
\label{sec:param_est}
\subsection{Bayesian inference}
\label{subsec:bayes_infe}
Given a stretch of GW strain data $\boldsymbol{d}$ from detectors containing a GW signal $\boldsymbol{h}(\vec \Lambda)$, embedded in additive Gaussian noise $\boldsymbol{n}$, we want to estimate the posterior distribution ${p(\vec \Lambda \mid \boldsymbol{d})}$ over the source parameters $\vec \Lambda$. The posterior distribution, in turn, is related to the likelihood function ${\mathcal{L}(\boldsymbol{d}\mid \vec \Lambda)}$ via the well-known Bayes' theorem:
\begin{equation} 
p(\vec \Lambda \mid \boldsymbol{d}) = \frac{\mathcal{L}(\boldsymbol{d} \mid \vec \Lambda) \ p(\vec \Lambda)}{p(\boldsymbol{d})}
\label{eq:Bayes}
\end{equation}
where, 
$p(\vec \Lambda)$ is the prior distribution over model parameters ${\param \equiv \{ \vec \lambda, \vec \theta, t_{c} \}}$. Here $\vec \lambda$ is a set of intrinsic parameters such as the component masses $m_{1,2}$ and dimensionless aligned spins $\chi_{1z,2z}$ while $\vec \theta$ represents the set of extrinsic parameters such as the source's sky location i.e. right ascension ($\alpha$) and declination ($\delta$), the inclination ($\iota$) of the orbital plane of the binary with respect to the line of sight, the polarization angle ($\psi$), the luminosity distance ($d_{L}$) of the source from the Earth and the geocentric epoch of coalescence $t_c$. 

In principle, given a numerical prescription for calculating the likelihood function ${\mathcal{L}(\boldsymbol{d}\mid \vec \Lambda)}$ under a waveform model and assume prior distributions over the model parameters, we can evaluate the left-hand side of Eq.~\eqref{eq:Bayes} at any given point in the sample space upto an overall normalization factor. However, with a large number of parameters, $\vec \Lambda$ (typically $\sim 15$ parameters), evaluating ${p(\vec \Lambda \mid \boldsymbol{d})}$ on a fine grid over the sample space becomes increasingly tedious and eventually intractable with finite computational resources. Therefore, a more intelligent strategy is employed where one uses stochastic sampling techniques to estimate the posterior distribution ${p(\vec \Lambda \mid \boldsymbol{d})}$. There are a number of schemes to sample the posterior distribution, such as Markov Chain Monte Carlo (MCMC)~\cite{Foreman_Mackey_2013} and its variants, Nested Sampling~\cite{skilling2006nested} algorithms, etc. In this paper, we use {\sc{dynesty}}~\cite{speagle2020dynesty, sergey_koposov_2023_7600689}, an extensively used Python implementation of the nested sampling algorithm for GW data analysis. In this work, we have only focused on speeding up the PE algorithm by quickly evaluating the likelihood function at any point proposed by the sampling algorithm. Admittedly, another aspect of designing a fast-PE algorithm would involve optimizing the sampling algorithm itself - which is not considered in this work. It is easy to see that such improvements to the sampling algorithm can positively impact the performance of many fast-PE methods (including ours). For example, a significant improvement in computation time has been demonstrated by efficiently populating the parameter space as proposed by the {\sc{VARAHA}} sampling technique~\cite{varaha}.

\subsection{The likelihood function}
\label{sec:likelihood}

Let  $\boldsymbol{d}^{(i)}$ be the strain data recorded at the $i^{\text{th}}$ detector containing an astrophysical GW signal $\tilde{h}^{(i)}(\vec\Lambda)$. We pause to remark that complex $\tilde{h}^{(i)}(\vec \Lambda)$ will denote the frequency-domain Fourier transform (FT) of the signal $h^{(i)}(\vec \Lambda)$. Assuming uncorrelated noise among the $N_{\text{d}}$ detectors in the network, the coherent log-likelihood~\cite{lalinference} is given by 
\begin{multline}
\label{eq:multigenlikelihood}
\ln \mathcal{L}(\vec\Lambda) =  \sum_{i=1}^{N_{\text{d}}} {\langle \boldsymbol{d}^{(i)} \mid \tilde{h}^{(i)}(\vec\Lambda)\rangle} \\ 
- \frac{1}{2} \sum_{i=1}^{N_{\text{d}}} \left [ \| \tilde{h}^{(i)}(\vec\Lambda)\|^2 + \| \boldsymbol{d}^{(i)} \|^2 \right ].
\end{multline} 

For a non-precessing GW signal model, one can use the relation ${\tilde{h}_{\times} \propto -j\ \tilde{h}_{+}}$ between the two polarization states to express the signal at the $i^{\text{th}}$ detector as:
\begin{equation}
\label{eq:detframwaveform}
\begin{split}
\tilde{h}^{(i)}(\vec\Lambda) \equiv \tilde{h}(\vec\Lambda, t^{(i)}) 
    &= \mathcal{A}^{(i)} \tilde{h}_{+}(\vec \lambda, t^{(i)}), \\
    &= \mathcal{A}^{(i)}\, \tilde{h}_{+}(\vec \lambda)\, e^{-j\,2\pi f t_c} \, e^{-j\,2\pi f \Delta t^{(i)}}.
\end{split}
\end{equation}
The complex amplitude of the signal $\mathcal{A}^{(i)}$ depends only on the extrinsic parameters $\vec\theta \in \vec \Lambda$ through the antenna pattern functions, luminosity distance, and the inclination angle:
\begin{multline}
\mathcal{A}^{(i)} = 
    \frac{1}{d_L} \left[ \frac{1+\cos^2 \iota}{2}  F^{(i)}_{+}(\alpha, \delta, \psi) \right.\\
            \left. - \: j \cos \iota \ F^{(i)}_{\times}(\alpha, \delta, \psi) \right],
\end{multline}
where ${F^{(i)}_{+}(\alpha, \delta, \psi)}$ and ${F^{(i)}_{\times}(\alpha, \delta, \psi)}$ are respectively the `plus' and `cross' antenna pattern functions of the $i^{\text{th}}$ detector. The antenna pattern functions of a detector describe the angular response of the detector to incoming GW signals. It arises from contracting the position and geometry dependent `detector tensor'~\cite{detectorTensor} with the metric perturbations (GW), thereby mapping the latter to a GW strain amplitude $h^{(i)}(\vec\Lambda)$ recorded at the detector.

As indicated by the Eq.~\eqref{eq:detframwaveform}, the signal acquires an additional phase difference during its projection from the geocentric frame to the detector frame, which corresponds to a time delay denoted by $\Delta t^{(i)}$. This temporal offset originates due to the relative positioning of the $i^{\text{th}}$ detector in relation to the Earth's center, and it can be expressed explicitly as follows:
\begin{equation}
\label{eq:time_delay}
    \Delta t^{(i)} \equiv t^{(i)} - t_{c} 
    = \frac{\vec x^{(i)} \cdot \hat{N}(\alpha, \delta) }{c},
\end{equation}
where $\vec x^{(i)}$ is a vector pointing from the Earth's center to the location of the $i^{\text{th}}$ detector, $t^{(i)}$ is the time at the $i^{\text{th}}$ detector, and $\hat{N}(\alpha, \delta)$ is the direction of the GW propagation~\cite{PhysRevD.92.023002}. 
In our analysis, we consider the log-likelihood function marginalized over the coalescence phase~\cite{thrane_2019} parameter, which can be written as follows:
\begin{multline}
\label{eq:multiphaselikelihood1}
    \left.\ln \mathcal{L}(\vec\Lambda \mid \boldsymbol{d}^{(i)})\right|_{\phi_c} 
    = \ln I_{0}\left[\left|\sum_{i=1}^{N_{d}}\langle \boldsymbol{d}^{(i)}\mid \tilde{h}^{(i)}(\vec \Lambda)\rangle\right|\right] \\
    - \frac{1}{2}\sum_{i=1}^{N_{d}}\left[ \| \tilde{h}^{(i)}(\vec\Lambda)\|^2 + \| \boldsymbol{d}^{(i)} \|^2 \right],
\end{multline}
where $I_0(\cdot)$ is the modified Bessel function of the first kind. By marginalizing over extrinsic parameters, we effectively reduce the dimensionality of the problem, resulting in accelerated likelihood calculations and enhanced sampling convergence. Substituting Eq.~\eqref{eq:detframwaveform} in the above equation, we get
\begin{multline}
\label{eq:multiphaselikelihood2}
    \left.\ln \mathcal{L}(\vec\Lambda \mid \boldsymbol{d}^{(i)})\right|_{\phi_c} 
    = \ln I_{0}\left[\left|\sum_{i=1}^{N_{d}}{{\mathcal{A}}^{(i)}}^{*}\, \langle \boldsymbol{d}^{(i)} \mid \tilde{h}_{+} (\vec\lambda, t^{(i)}) \rangle \right|\right] \\
    - \frac{1}{2}\sum_{i=1}^{N_{d}}\left[ \left|\mathcal{A}^{(i)}\right|^2 \sigma^2(\vec \lambda)^{(i)} + \| \boldsymbol{d}^{(i)} \|^2 \right];
\end{multline}
where, 
$\langle \boldsymbol{d}^{(i)} \mid \tilde{h}_{+}(\vec \lambda, t^{(i)}) \rangle$ is the complex overlap integral, while ${\sigma^2(\vec\lambda)^{(i)} \equiv \langle \tilde{h}_{+}(\vec \lambda, t^{(i)}) \mid \tilde{h}_{+}(\vec \lambda, t^{(i)}) \rangle}$ is the squared norm of the template $\: \tilde{h}_{+} (\vec\lambda)$. ${\sigma^2(\vec\lambda)^{(i)}}$ depends on the noise power spectral density (PSD) of the $i^{\text{th}}$ detector. 
The squared norm of the data vector, $\| \boldsymbol{d}^{(i)} \|^2$, remains constant during the PE analysis and, therefore, doesn't impact the overall `shape' of the likelihood. As such, it can be excluded from the subsequent analysis.

\section{Meshfree likelihood interpolation}
\label{sec:meshfree}
\begin{figure*}
\centering
\includegraphics[width=\textwidth, clip=True]{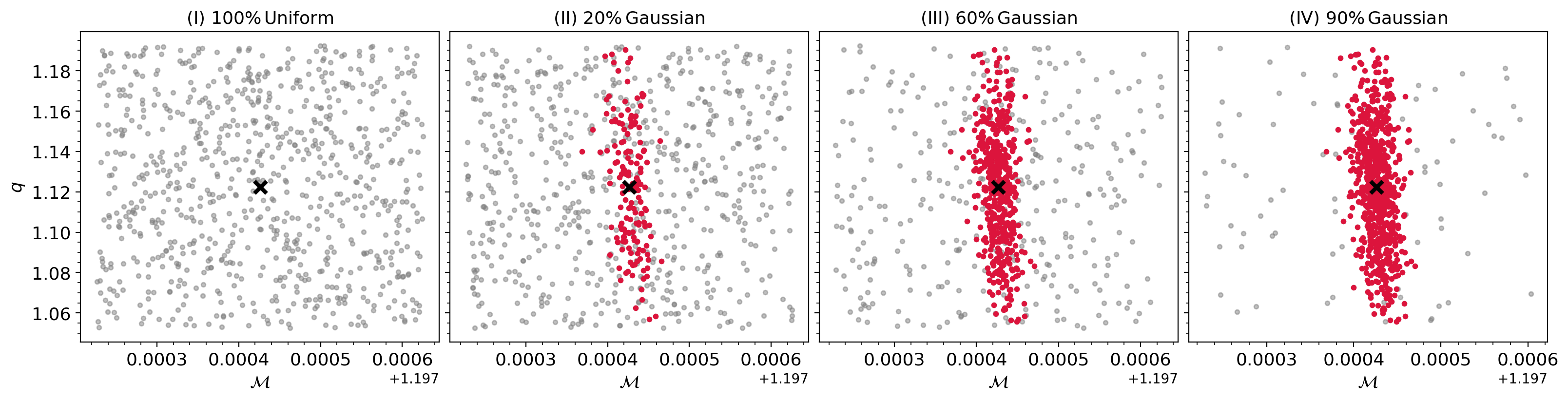}
\caption{
The figure displays randomly selected RBF interpolation nodes in the intrinsic parameter space with varying fractions from Gaussian and uniform distributions (2D slice in $\mathcal{M}$ and $q$ coordinates). Randomly placed nodes drawn from a Gaussian distribution are denoted in red, while those from a uniform distribution are shown in gray. The covariance matrix of the Gaussian distribution can be estimated semi-analytically given the signal model. Sub-figure-(I) represents a choice with $100 \%$ (all) nodes drawn from a uniform distribution, and the other sub-figures exhibit mixtures of different percentages of nodes from the uniform and Gaussian distributions. The center (marked as a ``black cross") of the sample space is determined through an optimization routine based on the highest network signal-to-noise ratio (SNR). This search is guided by the best-matched trigger obtained from the upstream search pipeline.
}
\label{fig:nodes_distr}
\end{figure*}

As discussed in our previous work~\cite{pathak2022rapid}, the meshfree method comprises two distinct stages:
\begin{enumerate}
    \item Start-up Stage: During this phase, we create radial basis function (RBF) interpolants for the pertinent quantities.
    \item Online Stage: In this stage, we swiftly compute the likelihood function at query points proposed by the sampling algorithm.
\end{enumerate}
Next, we will delve into the various stages that constitute meshfree interpolation.
%

\subsection{Start-up stage} 
\label{subsec:startup-stage}
During the start-up stage, we generate RBF interpolants for the relevant quantities, which enable rapid likelihood calculations in the online stage. The start-up stage can be further divided into the following parts: 
\begin{enumerate}
    \item[i.] Placement of RBF nodes over the intrinsic parameter space, $\lambda$. These discrete nodes are denoted as ${\vec \lambda^n: n = 1, 2, ..., N.}$ 
    
    Note that we set $\Delta t^{(i)} = 0$ for all the detectors while calculating ${\vec z^{(i)}(\vec \lambda^{n})}$. Additional time delays $\Delta t^{(i)}$ for the signal to reach the $i^\text{th}$ detector depending on the source's location in the sky   
    can be taken care of during the online stage while sampling the posterior distribution over the sky location parameters, as explained later. 

    \item[iii.] Singular value decomposition (SVD) of the complex time-series matrix ${ \mathcal{Z}^{(i)}}$ at the $i^\text{th}$ detector, where the $n^{th}$ row is defined as ${\vec z^{(i)}(\vec \lambda^{n})}$, and 

    \item[iv.] Generation of RBF interpolating functions for the SVD coefficients and the template norm square $\sigma^{2}$ for each detector. 
\end{enumerate}
We now describe some of these steps in greater detail.

\subsubsection{RBF nodes placement}
\label{subsec:nodesplacement}
\begin{figure}[htpb]
\includegraphics[width=\columnwidth, clip=True]{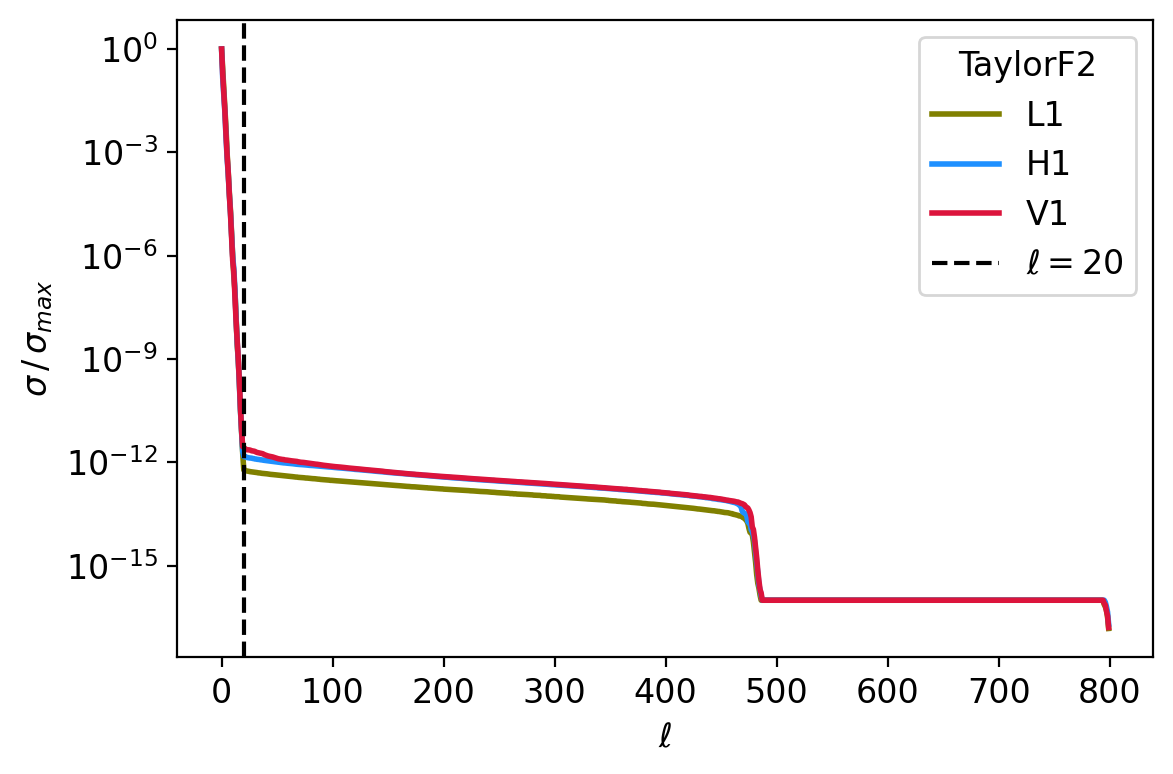}
\caption{
The figure shows the profile of the singular values normalized by the highest singular value. Notably, there is a sharp decline in the singular values (indicated by a black dashed line) due to the large correlation between the time-series $\vec z^{(i)}(\vec \lambda^n)$ at different RBF interpolation nodes, revealing that only the linear combination of the top-few singular vectors (around 20) is sufficient for effectively reconstructing the time series $\vec z^{(i)}(\vec \lambda^q)$ with minimal reconstruction error. The 3.5~PN \TaylorF~post-Newtonian signal model is assumed.
}
\label{fig:sing_values}
\end{figure}

The accurate construction of meshfree RBF interpolants relies significantly on the strategy of randomly distributing the RBF nodes, denoted as $\vec \lambda^{n}$, across the sample space. However, the task of finding an optimal set of nodes poses a considerable challenge. Typically, a hypercube in the sample space is selected around a reference point during the sampling of the posterior distribution. This reference point, commonly known as trigger or best-matched template $\vec \lambda^{\ast}$ and trigger time (geocentric) $t_{\text{trig}}$ are derived from an upstream search pipeline~\cite{GstLAL_2010, usman2016pycbc}.

In our previous work, we opted for a uniform distribution of nodes within the hypercube centered around the reference point (injection in the case of a simulated BNS event). This approach works well for events with low SNRs since the corresponding likelihood function exhibited a relatively flatter profile across the parameter space. However, for high SNR events, where the likelihood profile exhibits a sharp peak in certain regions of the parameter space, uniformly distributing nodes within the hypercube may not adequately cover the support of the posterior distribution, particularly in areas where we seek higher accuracy in interpolated likelihood estimates. As a result, we adopted a combination of nodes drawn from both a multivariate Gaussian distribution and a uniform distribution across the sample space within the hypercube. The multivariate Gaussian distribution is defined by a mean vector $\vec \mu$ and a covariance matrix $\boldsymbol{\Sigma}$, and its expression is as follows:
\begin{equation}
    p(\vec \lambda) \propto \exp\left[-(\vec \lambda - \vec \mu)^T \: \boldsymbol{\Sigma}^{-1} \: (\vec \lambda - \vec \mu)\right]
\end{equation}
where $\mu$ can be selected as the trigger point $\vec \lambda^{\ast}$, and $\boldsymbol{\Sigma}$ can be computed using the inverse of the Fisher matrix $\Gamma$ evaluated at $\vec \lambda^{\ast}$. While the distribution of nodes generated using the Fisher matrix may not perfectly match the actual posterior distribution across the sample space, it does exhibit significant overlap with the true distribution, especially in the vicinity of the likelihood peak.

Since the trigger originates from a search pipeline that uses a discrete set of templates over a grid (also known as a template bank)~\cite{temp_bank_soumen, PhysRevD.103.084047}, we can select a reference point, denoted as $\vec \lambda_{\text{ref}}$, in close proximity to the trigger point ($\vec \lambda^{\ast}$), which may have a higher network SNR. We employ an optimization strategy~\cite{2020SciPy-NMeth} aimed at maximizing the network SNR in order to identify this reference point. Subsequently, a portion of the interpolation nodes are generated from a multivariate Gaussian distribution, $\mathcal{N}(\vec \lambda_{\text{ref}}, \boldsymbol{\Sigma})$, where $\boldsymbol{\Sigma}$ is computed from the inverse of the Fisher matrix evaluated at $\vec \lambda_{\text{ref}}$, to adequately cover the region around the peak. For regions of the sample space where Gaussian nodes are sparse or possibly nonexistent, we spray a uniform distribution of samples across the hypercube. This approach ensures high accuracy in interpolated likelihood estimates throughout the entire sample space.

For a visual representation illustrating the various combinations of Gaussian and uniform nodes, please refer to Figure~\ref{fig:nodes_distr}. In particular, note that choosing a small fraction of random RBF nodes from the multivariate Gaussian distribution leads to more accurate posterior distributions over certain parameters (see Appendix~\ref{appendix:A}).

\subsubsection{SVD of ${\mathcal{Z}^{(i)}}$ and Interpolants generation}
\label{subsec:svdinterpolantsgeneration}
After obtaining the appropriate RBF nodes $\vec \lambda^{n}$, we can efficiently compute the time-series ${\vec z^{(i)}(\vec \lambda^{n}) \equiv z^{(i)}(\vec \lambda^{n}, t_c)}$ using Fast Fourier Transform (FFT) based circular correlations where $t_c$ are the uniformly spaced discrete-time shifts taken from a specified range around a reference coalescence time $t_{\text{trig}}$. Given that the predominant portion of the posterior support originates from the vicinity of the likelihood peak, we opt for ${t_{c} \in [t_{\text{trig}} \pm 0.15\; \text{s} ]}$ as the interval for sampling $t_{c}$.

Similarly, we compute the template norm square ${\sigma^2(\vec \lambda^{n})^{(i)}}$ at all the RBF nodes $\vec \lambda^{n}$.

Our objective is to identify the basis vectors capable of spanning the space of $\vec z^{(i)}(\vec \lambda^{n})$. This can be achieved by stacking (row-wise) the time-series vectors $\vec z^{(i)}(\vec \lambda^{n})$ for each $\vec \lambda^{n}$, where $n = 1, 2, \cdots, N$. Subsequently, performing the Singular Value Decomposition (SVD) on the resultant matrix ${\mathcal{Z}^{(i)}}$ yields our desired basis vectors ${\vec u_\mu}$. It can be shown that any vector $\vec z^{(i)}(\vec \lambda^{n})$ within the specified parameter range can be expressed as a linear combination of these basis vectors ${\vec u_\mu}$,
\begin{equation}
\label{eq:svdbasistimeseries}
\vec z^{(i)}(\vec \lambda^{n}) = \sum_{\mu = 1}^N\, C^{n (i)}_{\mu}\ \vec u^{(i)}_{\mu}
\end{equation}
where, ${C^{n (i)}_{\mu} \equiv C^{(i)}_{\mu}(\vec \lambda^{n})}, \ \mu=1, 2, \cdots N$ are the $N$ SVD coefficients. These coefficients $C^{n (i)}_{\mu}$ are characterized as smooth functions over $\vec \lambda^{n}$ within a sufficiently narrow boundary encompassing the posterior support.
Therefore, each of these $N$ coefficients can be treated as a scalar-valued, smooth function over the $d$-dimensional intrinsic parameter space and expressed as a linear combination of the radial basis functions~\cite{doi:10.1142/6437} (RBFs) $\phi$ centered at the interpolation nodes:
\begin{equation}
\label{eq:rbfcoeff}
C^{q (i)}_{\mu} = \sum_{n=1}^N\, a^{(i)}_{n}\, \phi(\|\vec \lambda^q - \vec \lambda^{n}\|_2) + \sum_{j = 1}^{M}\, b^{(i)}_{j}\, p_j(\vec \lambda^q)
\end{equation} 
where $\phi$ is the RBF kernel centered at ${\vec \lambda^{n} \in \mathbb{R}^d}$, and $\left \{  p_{j} \right \}$ denotes the monomials that span the space of polynomials with a predetermined degree $\nu$ in $d$-dimensions. 
The inclusion of these monomial terms has been shown to enhance the accuracy of RBF interpolation~\cite{2016JCoPh}. 

Similarly, we can also express $\sigma^2(\vec \lambda^{q})^{(i)}$ in terms of the RBF functions and monomials by treating it as smoothly varying scalar-valued function over the interpolation domain.

From Eq.~\eqref{eq:rbfcoeff}, we find that there are a total of $(N + M)$ coefficients to be solved for each interpolating function. 
The SVD coefficients $C_{\mu}^{q (i)}$ and $\sigma^2(\vec \lambda^{q})^{(i)}$ are known at each of the $N$ RBF nodes $\vec \lambda^n$. These provide $N$ interpolation conditions. To uniquely determine the coefficients $a_{n}$ and $b_j$, we impose $M$ additional conditions of the form ${\sum_{j=1}^M a^{(i)}_j p_j(\vec \lambda^q) = 0}$. This leads to the following system of $N + M$ equations:
\begin{equation}
	\begin{bmatrix}
		\boldsymbol{\Phi} & \boldsymbol{P} \\
		\boldsymbol{P}^{T} & \boldsymbol{O} 
	\end{bmatrix} \
		\begin{bmatrix}
		\boldsymbol{a^{(i)}} \\ 
		\boldsymbol{b^{(i)}}
	\end{bmatrix} \ 
	=  
	\begin{bmatrix}
		C^{n (i)}_{\mu} \\ 
		\boldsymbol{0} 
	\end{bmatrix}
\label{Eq:rbf-sle}
\end{equation}
where the matrices $\boldsymbol{\Phi}$ and $\boldsymbol{P}$ have components ${\Phi_{ij} = \phi(\norm{\vec\lambda^{i}-\vec\lambda^{j}}_2)}$ and ${P_{ij} = p_j(\vec\lambda^{i})}$ respectively; ${\boldsymbol{O}_{M\times M}}$ is a zero-matrix and ${\boldsymbol{0}_{M\times 1}}$ is a zero-vector.

The additional $M$ conditions arise due to the fact that $\boldsymbol{\Phi}$ needs to be a real, symmetric, and conditionally positive definite matrix of order $\nu + 1$, and the following theorem (reproduced here from \cite{doi:10.1142/6437} for completeness) guarantees the uniqueness of the solutions:
\begin{theorem}
\label{thm:rbf_solve}
\textit{Let $\boldsymbol{A}$ be a real symmetric $N \times N$ matrix that is conditionally positive definite of order one, and let $\boldsymbol{B} = [1,...,1]^T$ be an $N \times 1$ matrix (column vector). Then, the system of linear equations}
\begin{equation*}
	\begin{bmatrix}
		\boldsymbol{A} & \boldsymbol{B} \\
		\boldsymbol{B}^T & \boldsymbol{O} 
	\end{bmatrix} \
		\begin{bmatrix}
		\boldsymbol{c} \\ 
		\boldsymbol{d}
	\end{bmatrix} \ 
	=  
	\begin{bmatrix}
		\boldsymbol{y} \\ 
		\boldsymbol{0} 
	\end{bmatrix}
\label{Eq:theorem_sle}
\end{equation*}
\textit{is uniquely solvable.}
\end{theorem}
Although Theorem~\ref{thm:rbf_solve} is applicable for cases where $\nu = 0$, it can be generalized to polynomials of higher degrees (For a proof, refer to Theorem $7.2$ in~\cite{doi:10.1142/6437}). 

Now, the set of equations Eq.~\eqref{Eq:rbf-sle} can be uniquely solved for the unknown coefficients $\boldsymbol{a}^{(i)}$ and $\boldsymbol{b}^{(i)}$ to determine the meshfree interpolating functions  corresponding to the coefficients $C_{\mu}^{(i)}$. 
From the sharply decreasing spectrum of eigenvalues shown in Fig.~\ref{fig:sing_values}, it is clear that only the ``top-few'' basis vectors are sufficient to reconstruct $\vec z^{(i)}(\vec \lambda^q)$ at minimal reconstruction error, we generate only top-$\ell$ meshfree interpolants of $C_{\mu}^{q(i)}$ where $\mu = 1,....,\ell$. 
Note that $\ell$ is chosen based on the spectrum of eigenvalues of the matrix ${\mathcal{Z}^{(i)}}$. We have chosen $\ell=20$ where the normalized eigenvalues fall~(see Fig.~\ref{fig:sing_values}) to $\sim 10^{-12}$. However, this is a subjective choice.

In a similar way, we generate the meshfree interpolants for the scalar-valued function $\sigma^2(\vec \lambda^q)^{(i)}$ as well. 

Note that to optimize computational efficiency, we need to generate a minimum number of RBF nodes $\vec \lambda^{n}$ given the degree ($\nu$) of the polynomial and dimension ($d$) of the intrinsic parameter space, and it is given by $N_{\text{min}} = \binom{\nu + d}{\nu}$. Increasing the number of nodes increases not only the computational cost of generating the interpolants but also their evaluation cost during the online stage, which is evident from Eq.~\eqref{eq:rbfcoeff} as the number of coefficients $a_n^{(i)}$ increase linearly with the number of RBF nodes $N$. In our study, we empirically choose at least twice (or more) the minimum number of required nodes.   

At the end of the start-up stage, we thus have uniquely determined $(\ell +1)$ interpolating functions

\subsection{Online stage}
\label{subsec:online-stage}
Once we have the meshfree interpolants ready, we can evaluate the meshfree interpolated values of $C^{q (i)}_{\mu}$ and $\sigma^2(\vec \lambda^q)^{(i)}$ at any query point $\vec \lambda^q$ rapidly. We then combine the interpolated values of the coefficients $C^{q (i)}_{\mu}$ with the corresponding basis vectors $\vec u_{\mu}$ according to Eq.~\eqref{eq:svdbasistimeseries}, restricting the summation to the first $\ell$ terms.

In the startup stage~\ref{subsec:startup-stage}, we set ${\Delta t^{(i)} = 0}$ while calculating ${\vec z^{(i)}(\vec \lambda^n)}$ and didn't include the time-delay relative to the Earth's center, which depends on the sky location. Now, we incorporate it while reconstructing the interpolated time-series ${\vec z^{(i)}(\vec \lambda^q)}$. Moreover, we focus on constructing only ${\vec z^{(i)}(\vec \lambda^q)}$ at approximately 10 time-samples centered around query time $t^{q (i)}$ (Eq.~\eqref{eq:time_delay}). To ensure sub-sample accuracy, we fit these samples with a cubic spline, which incurs negligible computational cost. Subsequently, we calculate ${\vec z^{(i)}(\vec \lambda^q)}$ at the query $t^{q (i)}$ using the cubic-spline interpolant. Similarly, we evaluate the interpolated value of $\sigma^2(\vec \lambda^q)^{(i)}$. Finally, combining ${\vec z^{(i)}(\vec \lambda^q)}$ and ${\sigma^2(\vec \lambda^q)^{(i)}}$ with coefficients dependent on extrinsic parameters $\vec \theta$, as exemplified in Eq.~\eqref{eq:multiphaselikelihood2}, enables the computation of $\ln \mathcal{L}_{\text{RBF}}$.

\section{Numerical experiments}
\label{sec:numerical}

\subsection{GW170817 event}
\label{subsec:gw17}

\begin{figure*}[t]
\begin{subfigure}{0.49\linewidth}
    \includegraphics[width=\linewidth]{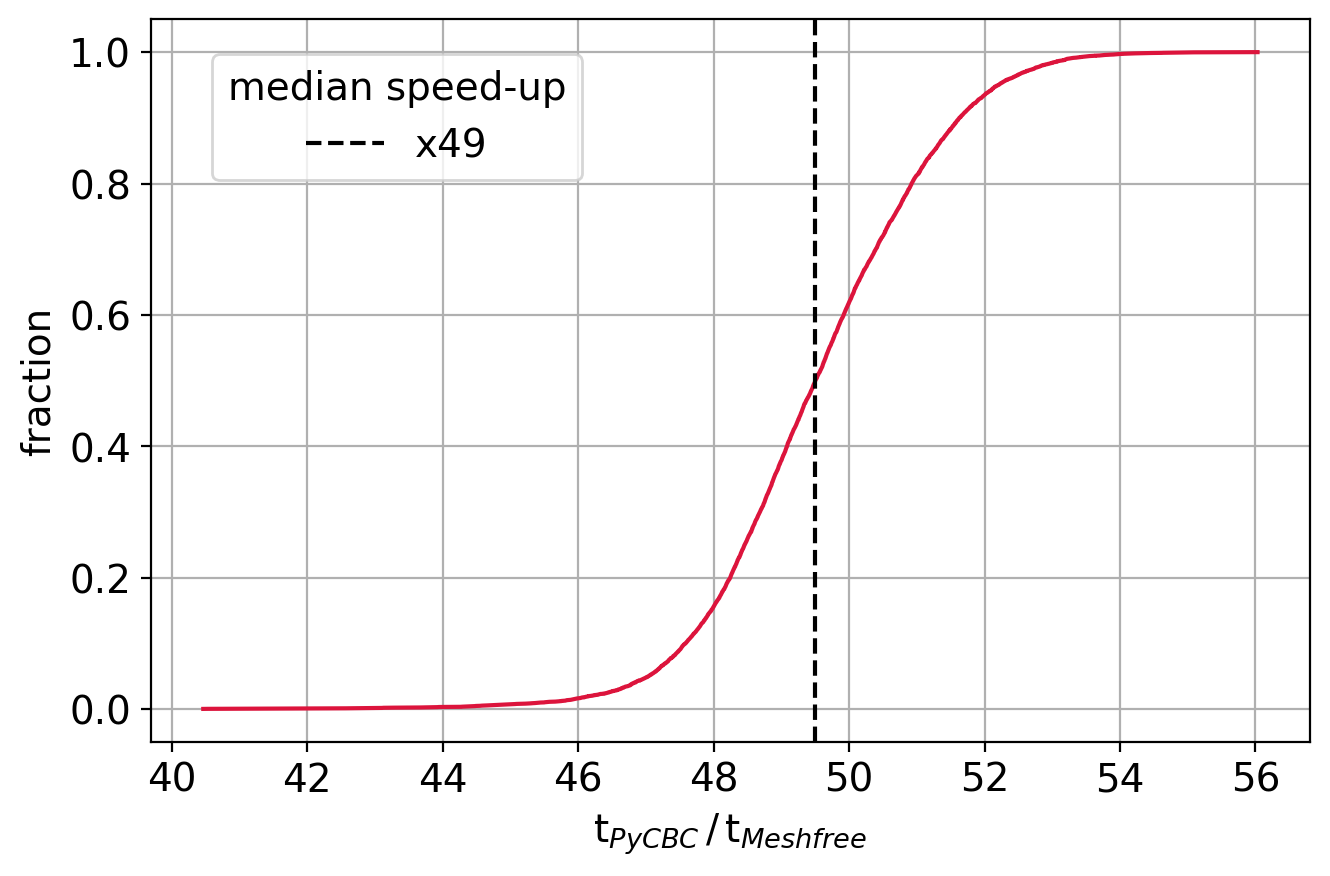}
    \caption{\TaylorF~signal model}
    \label{figspeedup:TF2}
\end{subfigure}\hfill
\begin{subfigure}{0.49\linewidth}
    \includegraphics[width=\linewidth]{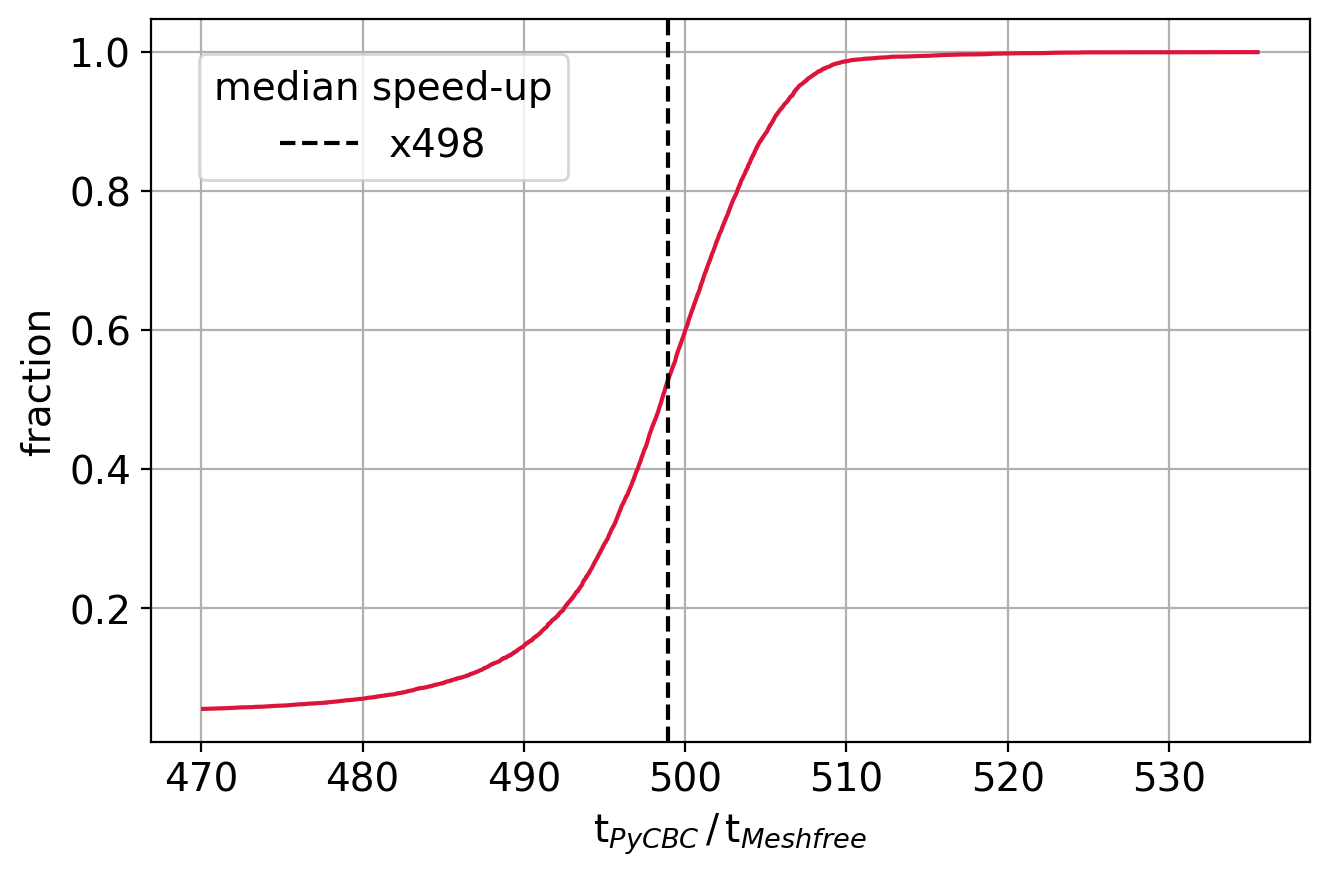}
    \caption{\texttt{IMRPhenomD} signal model}
    \label{figspeedup:IMR}
\end{subfigure}
\caption{
The plots compare the speed-up of the meshfree likelihood evaluation with \pycbc ~likelihood for two waveform models, \TaylorF~and \texttt{IMRPhenomD}. The meshfree method is around $49$ times faster for \TaylorF~and approximately $498$ times faster for \texttt{IMRPhenomD} (compared to the standard calculation). The median absolute error is approximately $\mathcal{O}(10^{-2})$ for both cases.
In the proposed meshfree method, one can trade-off speed in favor of better accuracy by retaining a larger number of SVD basis vectors (see Eq.~\eqref{eq:svdbasistimeseries}) and vice-versa. The optimal subjective choice is made from the spectrum of eigenvalues (Fig.~\ref{fig:sing_values}).
}
\label{fig:speedup}
\end{figure*}

We conducted a Bayesian PE study on the GW170817 binary neutron star (BNS) event to test our method. The strain data from the two advanced-LIGO and Virgo detectors were obtained from the open archival data sets available on GWOSC~\cite{GW170817data, RICHABBOTT2021100658}. We used a $360$ seconds data segment that was cleaned with a $4$th-order high-pass filter, with a cut-off frequency of $18$ Hz. The seismic cutoff frequency was set to $20$ Hz. The \TaylorF~waveform model~\cite{PhysRevD.49.1707, PhysRevD.59.124016, Faye_2012, PhysRevLett.74.3515} was employed to recover the source parameters. The noise PSD was estimated from the strain data using overlapping segments of length $2$ seconds, using the \texttt{median-mean} PSD estimation method by applying a \texttt{Hann} window as implemented in \pycbc. Subsequently, we generated $N = 800$ RBF nodes using the node placement algorithm described in section~\ref{subsec:startup-stage}.

Out of total RBF nodes, ${\ngauss = 200}$ nodes (number of Gaussian nodes) are selected from a multivariate Gaussian distribution, $\mathcal{N}(\vec \lambda_{\text{ref}}, \boldsymbol{\Sigma})$ where the mean is set as $\vec \lambda_{\text{ref}}$, and the covariance matrix, $\boldsymbol{\Sigma}$ is estimated by evaluating inverse of the Fisher matrix, $\boldsymbol{\Gamma}$~\cite{vallisneri2008use} at the reference point $\vec \lambda_{\text{ref}}$ (see Section~\ref{subsec:nodesplacement}). We utilize the \texttt{gwfast}~\cite{Iacovelli_2022} Python package to calculate the covariance matrix. The remaining nodes ${\nunif = 600}$ are uniformly sampled from the ranges specified in Table~\ref{tab:priordistr}. For generating the RBF interpolants, we employ the publicly available \texttt{RBF} python package~\cite{RBF_github}. A Gaussian RBF kernel ${\phi = \exp(-\varepsilon r^2)}$ is used as the basis function, where $\varepsilon$ serves as the shape parameter. We set $\varepsilon = 10$ through trial and error, but it can be optimized using the leave-one-out cross-validation~\cite{ROCHA20091573} method. The degree of the monomials is chosen as $\nu = 7$, and we retain the top $20$ basis vectors ($\ell=20$) for the reconstruction of ${\vec z^{(i)}(\vec \lambda^{q})}$.

\begin{figure}[hbtp]
\includegraphics[width=\columnwidth, clip=True]{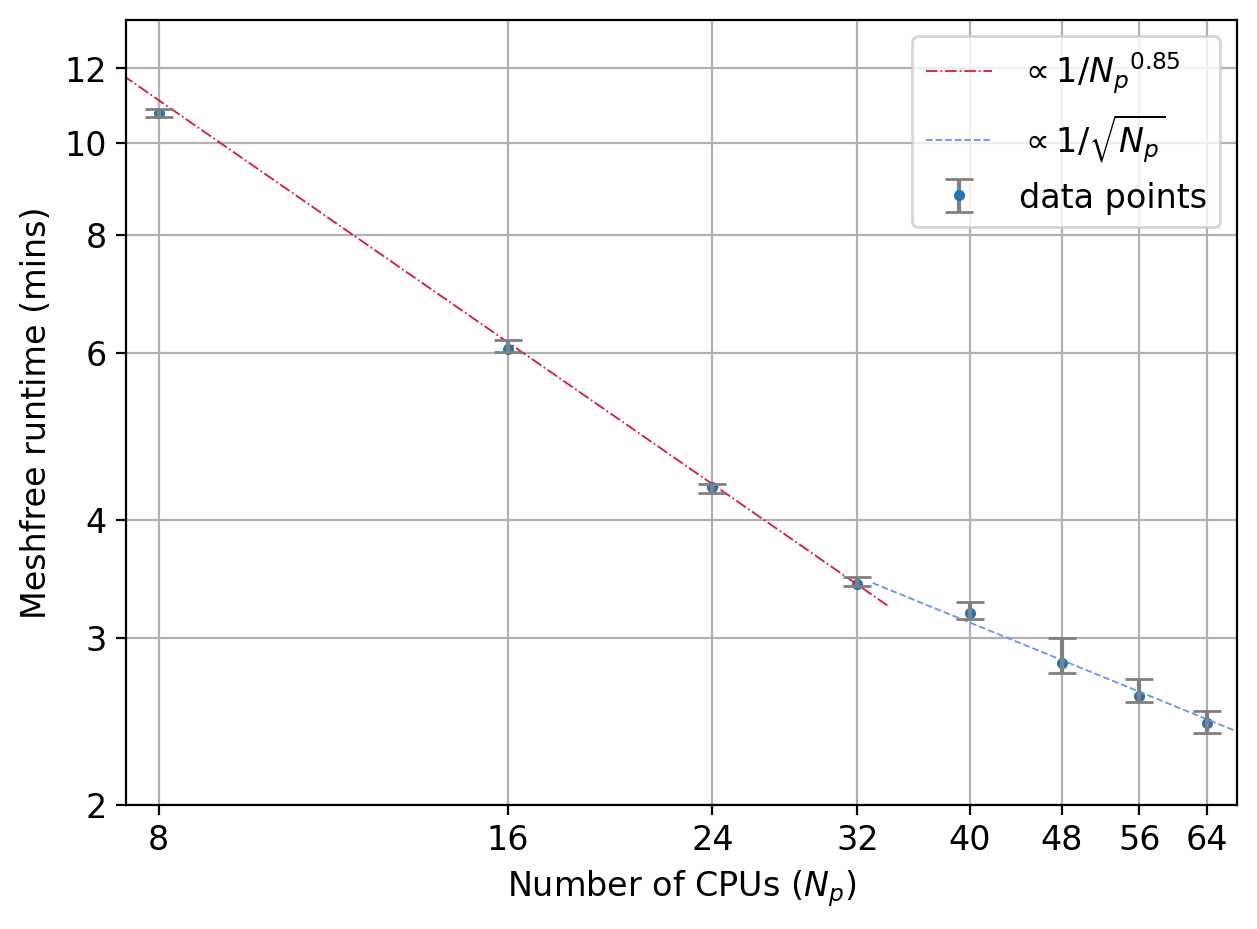}
\caption{The runtimes of the meshfree method have been plotted with varying the number of cores. We choose the GW170817 event for this specific example, and the \TaylorF~waveform model is taken for obtaining the posteriors. The plot shows a notable decrease in meshfree runtimes as the number of cores increased. We anticipate significant reductions in runtimes can be achieved with an even higher number of cores, though the gains will eventually saturate as we increase the number of cores.
}
\label{fig:timing_vs_cores}
\end{figure}

\begin{table}[hbt]
\def\arraystretch{1.5}
\begin{ruledtabular}
\begin{tabular}{lcl}
 Parameters     &   Range    &   Prior distribution \\
\hline
 $\mathcal{M}$  &   $[\mathcal{M}_{\text{cent}} \pm 0.0002]$ &   $\propto \mathcal{M}$ \\
 $q$            &   $[q_{\text{cent}} \pm 0.07]$             & $ \propto \left [ (1 + q)/q^3 \right ]^{2/5}$     \\
 $\chi_{1z, 2z}$   &   $[\chi_{1z \,\text{cent}} \pm 0.0025]$   & Uniform  \\
 $d_L$          & $[10, 60]$                & Uniform in volume\\
 $t_c$          & $t_{\text{trig}} \pm 0.12$& Uniform\\
 $\alpha$       & $[0, 2\pi]$               & Uniform\\
 $\delta$       & $\pm \pi/2$       & $\sin^{-1} \left [ {\text{Uniform}}[-1,1]\right ]$\\
 $\iota$        & $[0, \pi]$                & Uniform in $\cos \iota$\\
 $\psi$         & $[0, 2\pi]$               & Uniform angle\\
\end{tabular}
\end{ruledtabular}
\caption{Prior parameter space over the ten-dimensional parameter space $\vec \Lambda$.}
\label{tab:priordistr}
\end{table}

To sample the ten-dimensional parameter space $\vec \Lambda$, we utilize nested sampling implemented in the \texttt{dynesty} python package. The likelihood function used for the analysis is $\ln \mathcal{L}_{\text{RBF}}$. The prior distributions and their boundaries for all sampling parameters are provided in Table~\ref{tab:priordistr}. The sampler configuration is outlined as follows: \texttt{nlive} $=500$, \texttt{walks} $=100$, \texttt{sample} = ``rwalk'', and \texttt{dlogz} $=0.1$. These parameters crucially determine the accuracy and time taken by the nested sampling algorithm. Here, \texttt{nlive} represents the number of live points. Opting for a larger number of \texttt{nlive} points results in a more finely sampled posterior (and consequently, evidence), albeit at the expense of requiring more iterations for convergence. \texttt{walks} denotes the minimum number of points needed before proposing a new live point, \texttt{sample} indicates the chosen sampling approach for generating samples, and \texttt{dlogz} signifies the proportion of the remaining prior volume's contribution to the total evidence, which functions as a stopping criterion for terminating the sampling process. Further details on the dynesty's nested sampling algorithm and its implementation can be found in ~\cite{speagle2020dynesty, sergey_koposov_2023_7600689}. For comparison, we also employ the direct marginalized phase likelihood implemented in \pycbc ~ with the dynesty sampler, using the same sampler configuration mentioned above.

The PE run with the meshfree likelihood (for \TaylorF) completed in $\sim 3.4$ minutes (including the startup and sampling stages) when run on 32 cores. In a similar analysis with the \texttt{IMRPhenomD}~\cite{khan2016frequency} waveform model, we found good agreement between the posterior distributions obtained from the meshfree and \pycbc ~likelihood methods, and these results are broadly consistent with the LVK analysis of the same event. In this case, the meshfree method generated the posterior in $\sim 3.6$ minutes utilizing $32$ cores.

\begin{figure*}[htpb]
\label{fig:corner_plot_tay}
\centering
\includegraphics[width=0.85\linewidth]{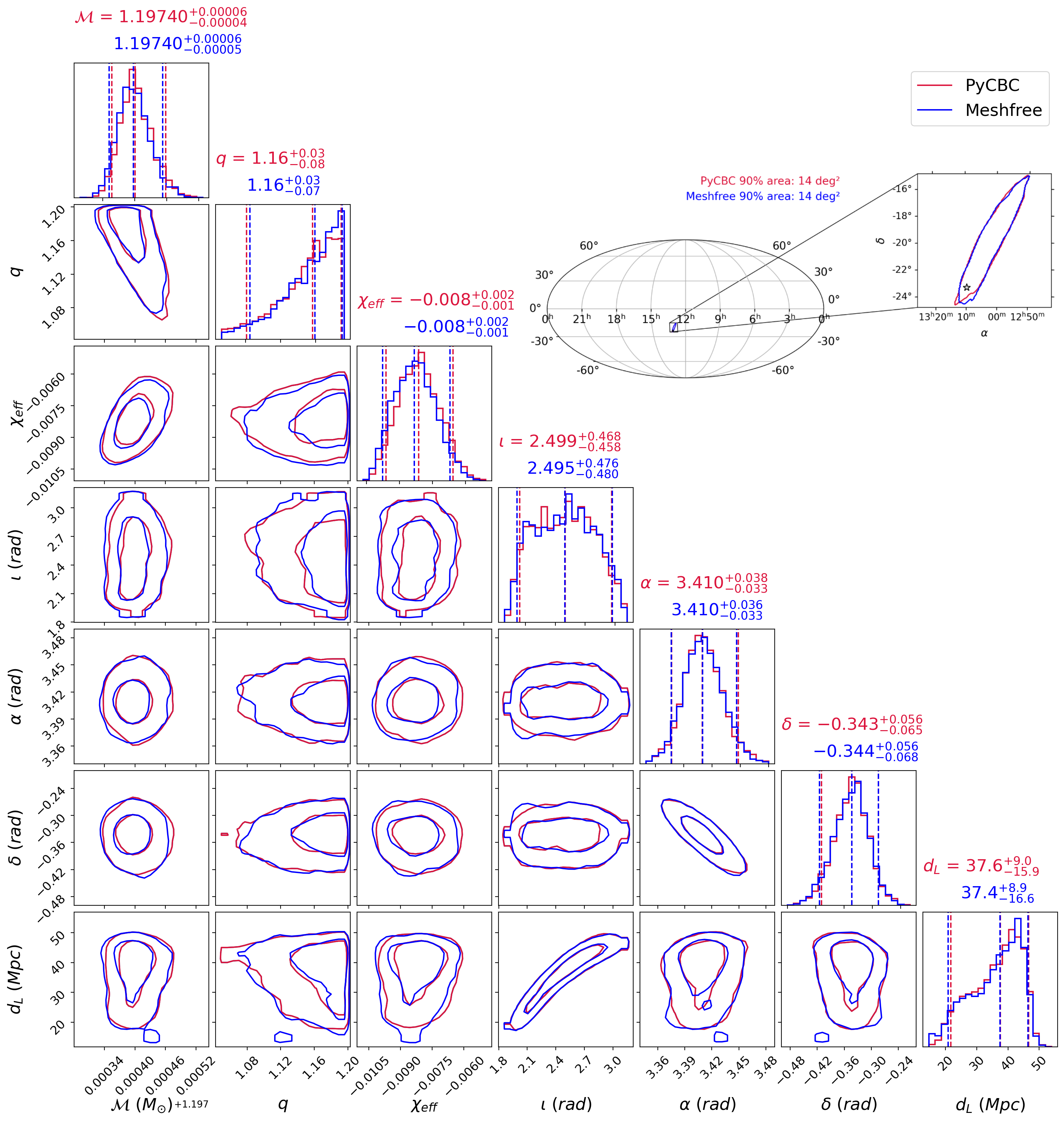}
\caption{\label{fig:cornerplot_20Hz} The corner plot shows the posterior distributions over a ten-dimensional parameter space using both the meshfree method and the standard \pycbc ~ likelihood. The seismic cutoff frequency is set to $20$ Hz, assuming the \TaylorF~waveform model. The estimated distributions from both methods exhibit good agreement, as indicated by the median values in the title of the 1-D marginalized posterior plots along the diagonal. The meshfree likelihood-based PE process was completed in $\sim 3.4$ minutes when performed on $32$ cores. 
 A skymap (inset) is also generated for both methods where the star denotes the actual location of the galaxy from which GW170817 is believed to have originated and is in agreement with~\cite{PhysRevX.9.011001}}. 
\end{figure*}

As shown in Fig.~\ref{fig:cornerplot_20Hz}, the posterior distributions estimated using the meshfree method and standard \pycbc ~ likelihood exhibit good agreement. Additionally, sky maps for GW170817 were generated using the \texttt{ligo-sky-map}~\cite{ligo_sky_map} utility for two methods, further demonstrating the effectiveness of the meshfree method in reconstructing source parameters. For the same analysis using $64$ cores, the number of likelihood evaluations for meshfree PE (standard \pycbc~PE) is $1396388$ $(1404588)$ whereas the number of posterior samples collected during sampling is $3633$ $(3546)$.

To estimate the speed-up achieved by using the meshfree likelihood compared to the \pycbc ~ likelihood, we generated $10^{4}$ random query points from the prior distribution specified in Table~\ref{tab:priordistr}. Subsequently, we calculated the log-likelihood at these points using both the meshfree and \pycbc ~ likelihood methods. The results show a speed-up of $\sim 49$ over standard \pycbc ~ likelihood, as shown in Fig.~\ref{fig:speedup}. In the case of \texttt{IMRPhenomD}, we observed a relative speed-up of  $\sim 498$ over standard \pycbc ~ likelihood. This difference in speed-up can be attributed to the difference in time taken to generate the template in the two different waveform models: {\TaylorF} waveforms are far less computationally expensive to generate than  \texttt{IMRPhenomD} model waveforms. The corresponding error plots are shown in Appendix~\ref{appendix:C}. 

Furthermore, we investigated the scaling behavior of PE run times (startup + online) using the meshfree method with respect to the number of cores employed. For the \TaylorF~analysis of the GW170817 event, we performed the ten-dimensional PE analysis with varying core counts ranging from $8$ to $64$. Our study indicates that the analysis can be completed in $\sim 2.4$ minutes using $64$ cores. Furthermore, we found that the current implementation scales linearly with the number of live points (keeping \texttt{dlogz} fixed). It is worth noting that further reductions in estimation times are expected if the number of cores is increased, as the scaling does not saturate at 64 cores. However, as depicted in Fig.~\ref{fig:timing_vs_cores}, the rate at which the estimation times decrease diminishes as the number of cores continues to increase, which may eventually result in run time saturation beyond a certain number of cores.

The total time shown in Fig.~\ref{fig:timing_vs_cores} includes the time taken by both the start-up and online stages of the meshfree algorithm, in addition to the time required for the dynamic nested sampling algorithm to compute the evidence integral up to a desired level of accuracy. To clarify further:
(a) The start-up stage of the meshfree method benefits from the availability of many CPU cores as it can be performed in parallel. (b) Conversely, the online stage is not designed for parallel processing. This is because estimating the likelihood is a relatively fast operation, typically taking $\sim \mathcal{O}(1)$ ms, achieved by evaluating the interpolating functions. This part of the computational cost remains nearly constant regardless of the specific sample point being processed.
As a result, the runtime scaling, as depicted in Fig.~\ref{fig:timing_vs_cores}, effectively combines the time consumed by the Dynesty sampler and the start-up stage of our method.

\subsection{Simulated Data}
\label{subsec:simulated Data}
To assess the performance of our method across a range of SNRs, we generated simulated data sets with different source parameters that mimic binary neutron star (BNS) systems with network-matched filter SNRs spanning from $10$ to $100$. In our approach, the number of Gaussian nodes, denoted as $\ngauss$, for a given source is intricately tied to its SNR. Specifically, the likelihood profile for events with lower SNRs exhibits a relatively broader spread over the parameter space compared to the more sharply peaked likelihood profiles observed for high SNR events. Consequently, for higher SNR events, we select a larger number of nodes from the Gaussian distribution in contrast to lower SNR events. Thus, we opt for different values of $\ngauss$ based on the SNR of each source. We will now detail the distribution of source parameters.  

The component masses were uniformly drawn from the interval ${[1.2, \, 2.35] M_{\odot}}$, while the magnitude of the dimensionless component spins (aligned with the orbital angular momentum) was uniformly drawn from the interval ${[-0.05, \, 0.05]}$. The sources were uniformly distributed in sky location and followed a uniform distribution with $d_{L}$ ranging from $10$ to $200$ Mpc. The inclination angle $\iota$ of the binary's plane with respect to the line of sight was chosen to be uniformly distributed in $\cos \iota$. The tidal deformability parameter $\Lambda_{\text{tide}}$ of the neutron stars was set to zero for all signals. These signals were simulated using the {\TaylorF}  waveform approximant, and the noise PSD was chosen from {\texttt{aLIGOZeroDetHighPower}}~\cite{aLIGO_ZDHP} for the Livingston and Hanford detectors, and \texttt{AdvVirgo}~\cite{AdVirgo} for the Virgo detector, as implemented in \pycbc.

It's important to recognize that the distribution of injection parameters we've chosen may not hold any astrophysical relevance. These choices have been made solely for the purpose of testing our method across various SNR scenarios, as our node placement algorithm is intricately linked to the SNRs of these events.

\begin{figure}[hbtp]
\includegraphics[width=\columnwidth, clip=True]{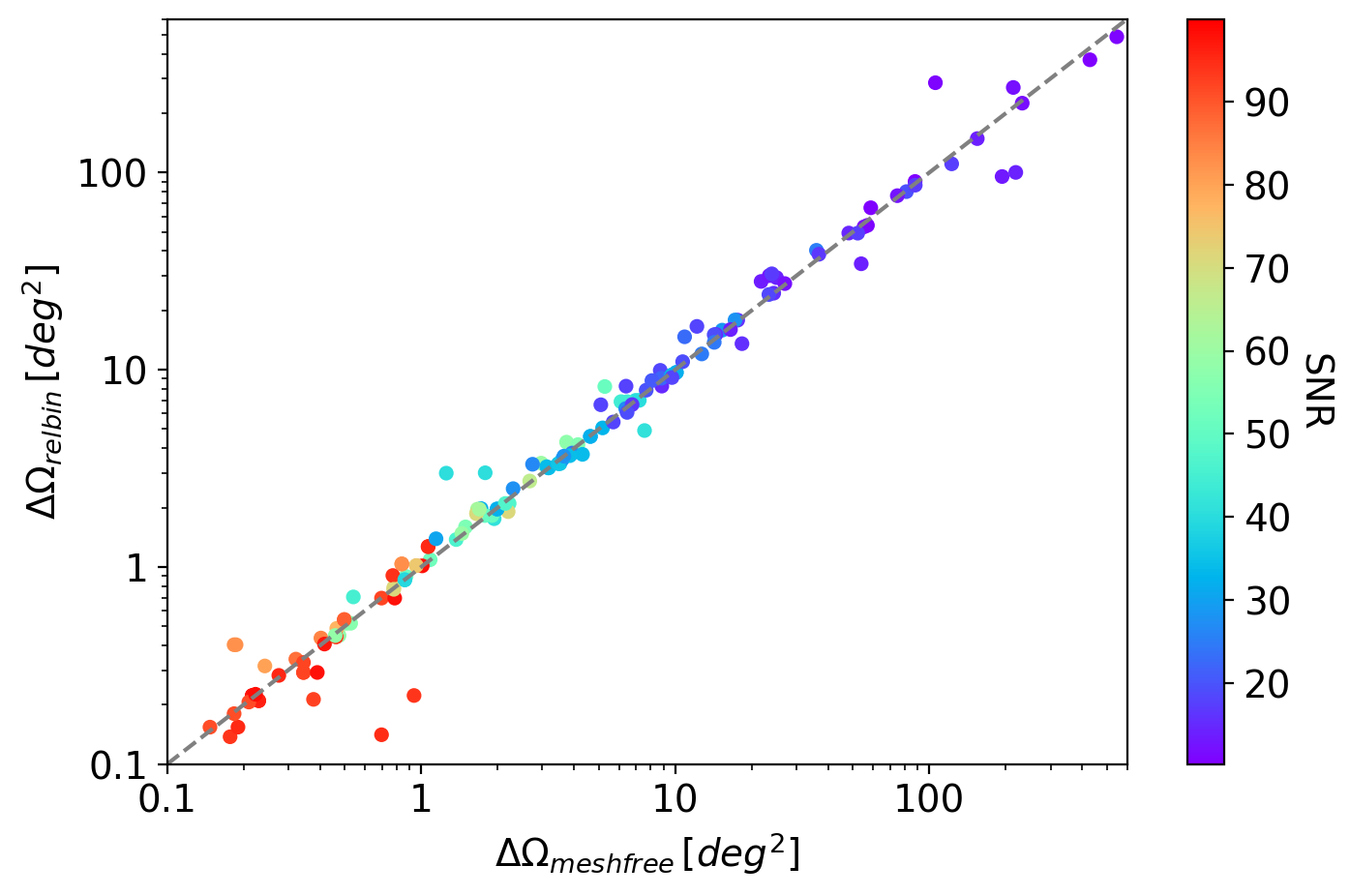}
    \caption{$90\%$ credible interval (CI) sky area as a function of SNR are shown. We compared the accuracy of meshfree with the {\relbin} method. The scattered points lie along the positive slope (dashed straight line), implying that the meshfree method has similar accuracy as the {\relbin} in estimating these parameters, i.e., sky location. It is observed that estimating these parameters with high SNR events is more accurate than with low SNR events. Note that there are some points that are away from the straight line. This is due to the fact that posteriors corresponding to those points are multimodal, giving rise to higher areas for one method in comparison to the other method.
}
\label{fig:simulated_error}
\end{figure}

\begin{figure}[hbtp]
\includegraphics[width=\columnwidth, clip=True]{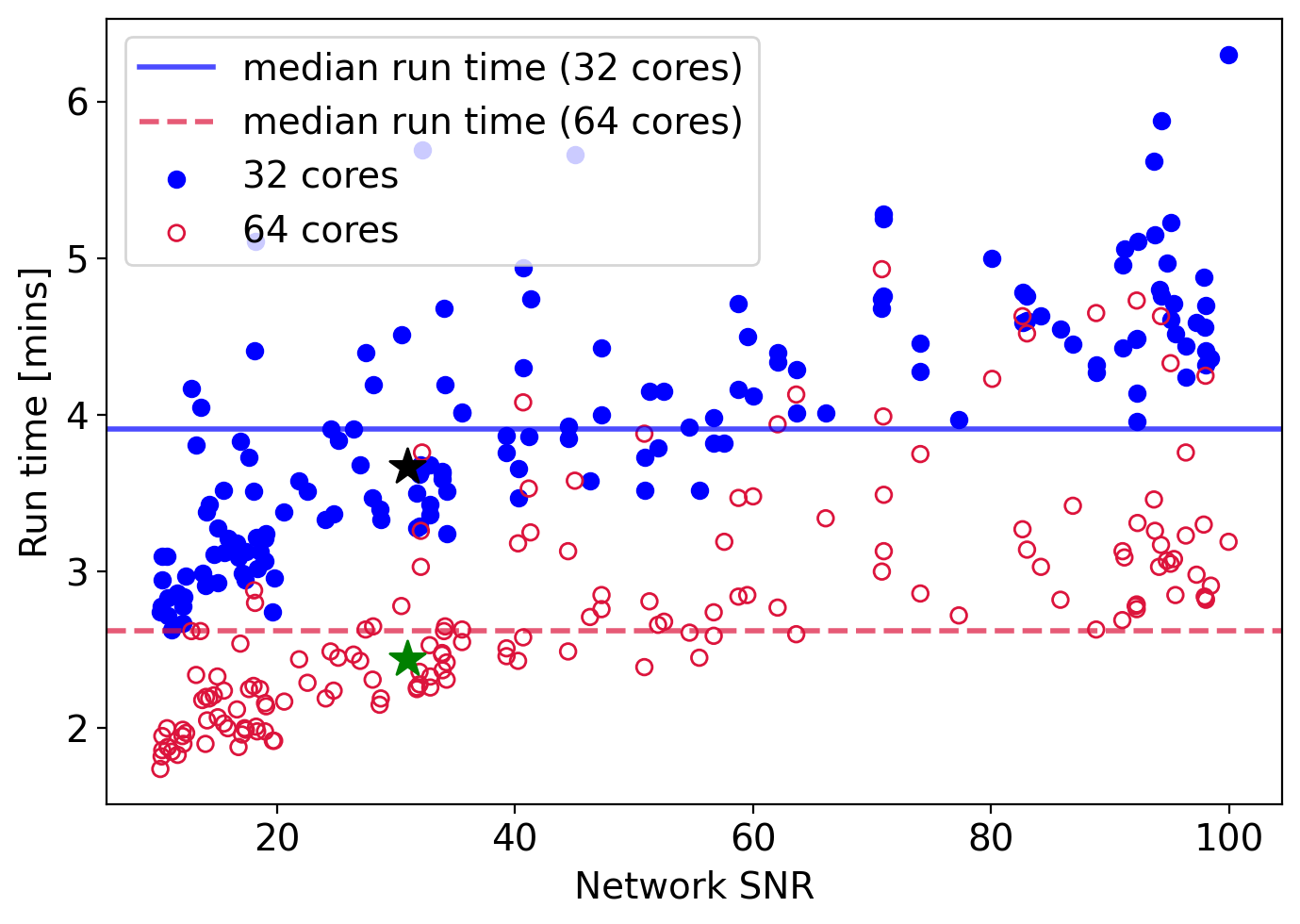}
\caption{We present the run times for reconstructing a number of simulated BNS events injected in Gaussian noise (Sec.~\ref{subsec:simulated Data}) using the meshfree method. Higher SNR injections took longer to reconstruct due to their narrowly peaked likelihood. The analysis utilized 32(64) cores, represented by blue dots (red circles). The black(green) star shows the GW170817 event's run time on 32(64) cores for reference. With 64 cores, the median run time decreased to around $\sim 2.6$ min compared to $\sim 3.9$ min with 32 cores. For \relbin~method, using $\epsilon = 0.000198$, the median run time is $\sim 22$ minutes on $32$ cores. The \relbin~PE run times crucially depend on the value $\epsilon$, and taking higher values of $\epsilon$ can result in lower run times.}
\label{fig:simulated_timings}
\end{figure}

At this point, we employ both the meshfree technique and the relative-binning method (\relbin) for estimating the BNS source parameters. In an ideal scenario, we would select the brute force likelihood method for comparison, but opting for such an approach might result in extended PE run times. Given that \relbin ~ is a well-established and widely used fast PE algorithm within the GW community, recognized for its combination of speed and accuracy, we have opted for this alternative over the standard approach. 

For all the simulated events, we chose $N = 800$ RBF nodes. However, $\ngauss$ was empirically chosen depending on the SNR of the events. Heuristically, we found that for SNR $\geq 50$, $\ngauss \in [0.4, 0.6]$ was a suitable choice. While for events with SNR $\leq 50$, $\ngauss$ was chosen from $[0, 0.2]$. There were a few exceptions to this prescription as well.

For the ~\relbin method, we set $\epsilon = 0.000198$, which determines the number of frequency bins in the \relbin ~ method (For more details, refer to ~\cite{Venumadhav2018, cornish2021heterodyned}). This parameter determines the accuracy and run-time of the relative binning method. We choose the reference parameters (or ``fiducial point'' in the case of ~\relbin) to be the injection parameters of each simulated event, even though in a realistic scenario, it would have been obtained by an optimization process starting from the best-matched search template.

The prior distributions over the extrinsic parameters ${(\alpha, \delta, \iota, \psi, d_L)}$ used in the PE are the same as those used for the GW170817 event in the earlier section. As for the intrinsic parameters, we select them according to the following distributions: (i) Uniform distribution in component masses $(m_1, m_2)$ such that ${\mathcal{M} \in \mathcal{M}_{\text{ref}} \pm 0.0002}$ and ${q \in q_{\text{ref}} \pm 0.07}$\footnote{In cases where $q_{\text{ref}} - 0.07 < 1$, we sample from $q \in [1, 1.14]$.} (ii) Uniformly in dimensionless component spins from the interval ${\chi_{1z/2z} \in {\chi_{1z/2z}}_{\text{ref}} \pm 0.0025}$ where $\mathcal{M}_{\text{ref}}, q_{\text{ref}}, {\chi_{1z/2z}}_{\text{ref}}$ are chirp mass, mass-ratio, and aligned spins components respectively, corresponding to the reference point found by the optimizing the network-SNR starting from the best-matched template. The prior over the time of coalescence $t_c$ is chosen to be uniform from the interval ${t_{\text{ref}} \pm 0.12}$ where $t_{\text{ref}}$ is the geocentric time of the trigger. 

As can be seen from the $90\%$ CI sky-area distribution in Fig.~\ref{fig:simulated_error}, the meshfree method accurately recovers the sky location with a similar level of accuracy as ~\relbin. The timing results for the simulated signals, shown in Fig.~\ref{fig:simulated_timings}, illustrate that the run-time increases with the SNR since signals with higher SNRs exhibit a more sharply peaked likelihood function that requires a longer time to accumulate a sufficient number of posterior samples to evaluate the evidence integral upto the desired accuracy.

All the tests were performed on AMD EPYC 7542 CPU@2.90GHz processors.

\section{Conclusion and Outlook}
\label{sec:concl_outl}
This work extends our previously prescribed meshfree method \cite{pathak2022rapid} to perform parameter estimation on a ten-dimensional parameter space for aligned spin waveform models using data from a multi-detector network framework. In the previous version of our method, we demonstrated its effectiveness on a simulated BNS event, as seen in a single detector, where we directly chose the injection parameters as our center for placing uniformly distributed RBF nodes. In this work, we start from the best-matched template parameters and optimize the network-SNR to reach a nearby point with a higher network-SNR value, which acts as a guide for placing RBF nodes. This node placement algorithm incorporates a blend of nodes from both multivariate Gaussian and uniform distributions. It enhances the accuracy of likelihood reconstruction (see Appendix~\ref{appendix:A} for a study of marginalized posterior profiles on different fractions of random interpolation nodes drawn from a multivariate Gaussian distribution). 

We tested our method for several simulated BNS events with varying SNRs and found good agreement between the meshfree and the relative binning methods. Further, we have demonstrated that our meshfree method can rapidly estimate the posteriors of the GW170817 binary neutron star (BNS) event and efficiently locate the electromagnetic (EM) counterpart in the sky within ${\sim}$2.4~(3.4) minutes after detection using 64~(32) CPU cores. 

These run times quoted above need to be put in the context of other fast-PE algorithms for which we chose the relative-binning method. Note that the time taken by the relative binning algorithm crucially depends on the choice of the $\epsilon$ parameter, which sets the frequency-bin resolution in the heterodyning process. We set $\epsilon=0.000198$ for which the time taken by the relative-binning method was found to take $\sim 20$ minutes on 32 cores to analyze the GW170817 event using data from a 3-detector (HLV) network. This choice of $\epsilon$ was taken to match the PE run times claimed by Finstad and Brown (2020)~\cite{Finstad_2020}. The sampler parameters were kept identical for both methods. Since we do not have the hardware (nor know the parameters) used by Finstad and Brown, this ad-hoc procedure was used to arrive at this reference time. As such, it won’t be prudent to arrive at any definite comparative conclusion based on these timing results. Nevertheless, we will continue to tune the meshfree method to explore further optimizations for speeding up the PE analysis. 

Note that the run time quoted here doesn't include the time spent in finding the center using optimization, which, when run on a single core, took $\sim 32$ seconds. However, it can be sped up by running in parallel over multiple CPU cores using multiprocessor versions of the optimization algorithm. For example, see the parallel version of the \texttt{scipy.optimize}~\cite{RJ-2019-030, florian_gerber_2020_3888570} Python routine. 

We also used our method for the NSBH event GW200115~\cite{NSBH_GW20015} (See Appendix~\ref{appendix:B}) and obtained a reasonably good match of posterior samples between our method and \pycbc ~standard likelihood.  

While our meshfree framework has the potential to contribute to future LIGO low-latency PE efforts, it is also essential to acknowledge some limitations of the current implementation:
Firstly, the RBF meshfree interpolating functions are seen to be accurate only over a relatively narrow domain in the intrinsic parameter space, which leads to using narrow prior boundaries to carry out the PE analysis. This constraint restricts the flexibility of the meshfree PE analysis compared to standard approaches. A potential solution to overcome this issue could be to divide the intrinsic parameter space into smaller ``patches'' and generate interpolants for each patch independently~\cite{Borne2019DomainDM}.  

In the present implementation, the prior bounds (albeit narrow) were obtained from the best-matched template as explained in Sec.~\ref{subsec:nodesplacement}. Since the PE pipeline is expected to be triggered by an upstream search pipeline, we have used the most significant trigger parameters ($\vec \lambda^\ast$) as the starting point. Using a suitable optimization algorithm,  this point ultimately leads us to a nearby reference point ($\vec \lambda_{\text{ref}}$), which serves as the center of the hyper-rectangle from which the samples are drawn. The dimensions of this hyper-rectangle along the different directions are heuristically set to get good accuracy in the likelihood reconstruction (i.e. we do not use the LVK results apriori to determine the prior bounds).
In the future, we would like to explore an automated way of determining the dimensions of this hyper-rectangle from the size of the $90\%$ overlap ellipsoid calculated at $\vec \lambda_{\text{ref}}$. Such an idea has been mooted in Pankow et al.~\cite{Pankow_2015}. However, since the axes of the overlap ellipsoid do not necessarily align with the eigen-directions of the data-driven covariance matrix, such an implementation would require careful thought. 

Secondly, it is worth noting that despite improvements in the placement of nodes compared to our previous work, it still needs investigation for a more generic prescription that can be applied to any CBC system. Incorporating an adaptive strategy~\cite{Zhang:2017aa} to identify significant nodes could enhance interpolation accuracy and strengthen the node placement strategy. 

In this analysis, we have focused exclusively on non-precessing signal models (\TaylorF~and \texttt{IMRPhenomD}), with a particular emphasis on the dominant harmonic of the GW signal. These models do not incorporate tidal terms~\cite{DelPozzo:2013ala, Dietrich_2019}. Since the tidal deformability parameters corresponding to the binary components are intrinsic in nature, we need to include them as interpolating parameters in our method, which increases the dimensionality of the intrinsic parameter space to six. Additionally, a greater number of nodes are required to ensure an accurate interpolation of the SVD coefficients and template norm square. We will explore these ideas in future follow-ups of this work. 

Although we anticipate that extending our method to include higher-order harmonics should be relatively straightforward, it may result in increased start-up costs. In the case of precessing-spin waveform models~\cite{IMRPhenomXPHM, IMRPhenomXHM}, a meshfree framework is particularly suitable since the number of RBF nodes does not exponentially increase with the dimensionality of the parameter space. In this regard, the relative binning algorithm has been recently implemented for gravitational-wave parameter estimation with higher-order modes and precession~\cite{narola2023relative}. In the future, we would like to overcome these limitations by enhancing the node placement algorithm further, expanding the boundaries of the sample space, and incorporating models that account for in-plane spins~\cite{IMRPhenomXPHM}.
%

\begin{acknowledgements}
We thank Vaibhav Tiwari for carefully going through our manuscript and giving helpful suggestions. We also thank Abhishek Sharma and Sachin Shukla for the helpful discussions at various stages of this work. We especially thank the anonymous referee for their careful review and helpful suggestions. 

L.~P. is supported by the Research Scholarship Program of Tata Consultancy Services (TCS). S.~M. acknowledges INSPIRE fellowship by the Department of Science and Technology (DST), India. A.~R is supported by the research program of the Netherlands Organisation for Scientific Research (NWO). A.~S. gratefully acknowledges the generous grant provided by the Department of Science and Technology, India, through the DST-ICPS (Interdisciplinary Cyber Physical Systems) cluster project funding. We thank the HPC support staff at IIT Gandhinagar for their help and cooperation. The authors are grateful for the computational resources provided by the LIGO Laboratory and supported by the National Science Foundation Grants No. PHY-0757058 and No. PHY-0823459. This material is based upon work supported by NSF's LIGO Laboratory, which is a major facility fully funded by the National Science Foundation. This research has made use of data or software obtained from the Gravitational Wave Open Science Center~\cite{gwosc_web}, a service of the LIGO Scientific Collaboration, the Virgo Collaboration, and KAGRA. This material is based upon work supported by NSF's LIGO Laboratory, which is a major facility fully funded by the National Science Foundation, as well as the Science and Technology Facilities Council (STFC) of the United Kingdom, the Max-Planck-Society (MPS), and the State of Niedersachsen/Germany for support of the construction of Advanced LIGO and construction and operation of the GEO600 detector. Additional support for Advanced LIGO was provided by the Australian Research Council. Virgo is funded through the European Gravitational Observatory (EGO), the French Centre National de Recherche Scientifique (CNRS), the Italian Istituto Nazionale di Fisica Nucleare (INFN), and the Dutch Nikhef, with contributions by institutions from Belgium, Germany, Greece, Hungary, Ireland, Japan, Monaco, Poland, Portugal, Spain. KAGRA is supported by the Ministry of Education, Culture, Sports, Science and Technology (MEXT), Japan Society for the Promotion of Science (JSPS) in Japan; National Research Foundation (NRF) and the Ministry of Science and ICT (MSIT) in Korea; Academia Sinica (AS) and National Science and Technology Council (NSTC) in Taiwan.

Code availability: Codes used in this analysis are publicly available in the following Github repository~\cite{meshfree_github}.

\end{acknowledgements}

\bibliography{references}

\begin{thebibliography}{92}%
\makeatletter
\providecommand \@ifxundefined [1]{%
 \@ifx{#1\undefined}
}%
\providecommand \@ifnum [1]{%
 \ifnum #1\expandafter \@firstoftwo
 \else \expandafter \@secondoftwo
 \fi
}%
\providecommand \@ifx [1]{%
 \ifx #1\expandafter \@firstoftwo
 \else \expandafter \@secondoftwo
 \fi
}%
\providecommand \natexlab [1]{#1}%
\providecommand \enquote  [1]{``#1''}%
\providecommand \bibnamefont  [1]{#1}%
\providecommand \bibfnamefont [1]{#1}%
\providecommand \citenamefont [1]{#1}%
\providecommand \href@noop [0]{\@secondoftwo}%
\providecommand \href [0]{\begingroup \@sanitize@url \@href}%
\providecommand \@href[1]{\@@startlink{#1}\@@href}%
\providecommand \@@href[1]{\endgroup#1\@@endlink}%
\providecommand \@sanitize@url [0]{\catcode `\\12\catcode `\$12\catcode
  `\&12\catcode `\#12\catcode `\^12\catcode `\_12\catcode `\%12\relax}%
\providecommand \@@startlink[1]{}%
\providecommand \@@endlink[0]{}%
\providecommand \url  [0]{\begingroup\@sanitize@url \@url }%
\providecommand \@url [1]{\endgroup\@href {#1}{\urlprefix }}%
\providecommand \urlprefix  [0]{URL }%
\providecommand \Eprint [0]{\href }%
\providecommand \doibase [0]{https://doi.org/}%
\providecommand \selectlanguage [0]{\@gobble}%
\providecommand \bibinfo  [0]{\@secondoftwo}%
\providecommand \bibfield  [0]{\@secondoftwo}%
\providecommand \translation [1]{[#1]}%
\providecommand \BibitemOpen [0]{}%
\providecommand \bibitemStop [0]{}%
\providecommand \bibitemNoStop [0]{.\EOS\space}%
\providecommand \EOS [0]{\spacefactor3000\relax}%
\providecommand \BibitemShut  [1]{\csname bibitem#1\endcsname}%
\let\auto@bib@innerbib\@empty
\bibitem [{\citenamefont {Eichler}\ \emph {et~al.}(1989)\citenamefont
  {Eichler}, \citenamefont {Livio}, \citenamefont {Piran},\ and\ \citenamefont
  {Schramm}}]{Eichler:1989ve}%
  \BibitemOpen
  \bibfield  {author} {\bibinfo {author} {\bibfnamefont {D.}~\bibnamefont
  {Eichler}}, \bibinfo {author} {\bibfnamefont {M.}~\bibnamefont {Livio}},
  \bibinfo {author} {\bibfnamefont {T.}~\bibnamefont {Piran}},\ and\ \bibinfo
  {author} {\bibfnamefont {D.~N.}\ \bibnamefont {Schramm}},\ }\bibfield
  {title} {\bibinfo {title} {{Nucleosynthesis, Neutrino Bursts and Gamma-Rays
  from Coalescing Neutron Stars}},\ }\href {https://doi.org/10.1038/340126a0}
  {\bibfield  {journal} {\bibinfo  {journal} {Nature}\ }\textbf {\bibinfo
  {volume} {340}},\ \bibinfo {pages} {126} (\bibinfo {year}
  {1989})}\BibitemShut {NoStop}%
\bibitem [{\citenamefont {{Goodman}}(1986)}]{1986ApJ_308L_47G}%
  \BibitemOpen
  \bibfield  {author} {\bibinfo {author} {\bibfnamefont {J.}~\bibnamefont
  {{Goodman}}},\ }\bibfield  {title} {\bibinfo {title} {{Are gamma-ray bursts
  optically thick?}},\ }\href {https://doi.org/10.1086/184741} {\bibfield
  {journal} {\bibinfo  {journal} {apjl}\ }\textbf {\bibinfo {volume} {308}},\
  \bibinfo {pages} {L47} (\bibinfo {year} {1986})}\BibitemShut {NoStop}%
\bibitem [{\citenamefont {Metzger}(2019)}]{kilonovae_metzger}%
  \BibitemOpen
  \bibfield  {author} {\bibinfo {author} {\bibfnamefont {B.~D.}\ \bibnamefont
  {Metzger}},\ }\bibfield  {title} {\bibinfo {title} {Kilonovae},\ }\href
  {https://doi.org/10.1007/s41114-019-0024-0} {\bibfield  {journal} {\bibinfo
  {journal} {Living Reviews in Relativity}\ }\textbf {\bibinfo {volume} {1}},\
  \bibinfo {pages} {1} (\bibinfo {year} {2019})}\BibitemShut {NoStop}%
\bibitem [{\citenamefont {Li}\ and\ \citenamefont
  {Paczyński}(1998)}]{Li_1998}%
  \BibitemOpen
  \bibfield  {author} {\bibinfo {author} {\bibfnamefont {L.-X.}\ \bibnamefont
  {Li}}\ and\ \bibinfo {author} {\bibfnamefont {B.}~\bibnamefont
  {Paczyński}},\ }\bibfield  {title} {\bibinfo {title} {Transient events from
  neutron star mergers},\ }\href {https://doi.org/10.1086/311680} {\bibfield
  {journal} {\bibinfo  {journal} {The Astrophysical Journal}\ }\textbf
  {\bibinfo {volume} {507}},\ \bibinfo {pages} {L59–L62} (\bibinfo {year}
  {1998})}\BibitemShut {NoStop}%
\bibitem [{\citenamefont {Gupta}\ \emph {et~al.}(2023)\citenamefont {Gupta},
  \citenamefont {Borhanian}, \citenamefont {Dhani}, \citenamefont
  {Chattopadhyay}, \citenamefont {Kashyap}, \citenamefont {Villar},\ and\
  \citenamefont {Sathyaprakash}}]{PhysRevD.107.124007}%
  \BibitemOpen
  \bibfield  {author} {\bibinfo {author} {\bibfnamefont {I.}~\bibnamefont
  {Gupta}}, \bibinfo {author} {\bibfnamefont {S.}~\bibnamefont {Borhanian}},
  \bibinfo {author} {\bibfnamefont {A.}~\bibnamefont {Dhani}}, \bibinfo
  {author} {\bibfnamefont {D.}~\bibnamefont {Chattopadhyay}}, \bibinfo {author}
  {\bibfnamefont {R.}~\bibnamefont {Kashyap}}, \bibinfo {author} {\bibfnamefont
  {V.~A.}\ \bibnamefont {Villar}},\ and\ \bibinfo {author} {\bibfnamefont
  {B.~S.}\ \bibnamefont {Sathyaprakash}},\ }\bibfield  {title} {\bibinfo
  {title} {Neutron star-black hole mergers in next generation
  gravitational-wave observatories},\ }\href
  {https://doi.org/10.1103/PhysRevD.107.124007} {\bibfield  {journal} {\bibinfo
   {journal} {Phys. Rev. D}\ }\textbf {\bibinfo {volume} {107}},\ \bibinfo
  {pages} {124007} (\bibinfo {year} {2023})}\BibitemShut {NoStop}%
\bibitem [{\citenamefont {Abbott}\ \emph {et~al.}(2017)\citenamefont {Abbott},
  \citenamefont {Abbott}, \citenamefont {Abbott}, \citenamefont {Acernese},
  \citenamefont {Ackley}, \citenamefont {Adams}, \citenamefont {Adams},
  \citenamefont {Addesso}, \citenamefont {Adhikari}, \citenamefont {Adya} \emph
  {et~al.}}]{abbott2017gw170817}%
  \BibitemOpen
  \bibfield  {author} {\bibinfo {author} {\bibfnamefont {B.~P.}\ \bibnamefont
  {Abbott}}, \bibinfo {author} {\bibfnamefont {R.}~\bibnamefont {Abbott}},
  \bibinfo {author} {\bibfnamefont {T.}~\bibnamefont {Abbott}}, \bibinfo
  {author} {\bibfnamefont {F.}~\bibnamefont {Acernese}}, \bibinfo {author}
  {\bibfnamefont {K.}~\bibnamefont {Ackley}}, \bibinfo {author} {\bibfnamefont
  {C.}~\bibnamefont {Adams}}, \bibinfo {author} {\bibfnamefont
  {T.}~\bibnamefont {Adams}}, \bibinfo {author} {\bibfnamefont
  {P.}~\bibnamefont {Addesso}}, \bibinfo {author} {\bibfnamefont
  {R.}~\bibnamefont {Adhikari}}, \bibinfo {author} {\bibfnamefont {V.~B.}\
  \bibnamefont {Adya}}, \emph {et~al.},\ }\bibfield  {title} {\bibinfo {title}
  {{GW170817}: observation of gravitational waves from a binary neutron star
  inspiral},\ }\href
  {https://doi.org/https://doi.org/10.1103/PhysRevLett.119.161101} {\bibfield
  {journal} {\bibinfo  {journal} {Phys. Rev. Lett.}\ }\textbf {\bibinfo
  {volume} {119}},\ \bibinfo {pages} {161101} (\bibinfo {year}
  {2017})}\BibitemShut {NoStop}%
\bibitem [{\citenamefont {Harry}(2010)}]{Harry:2010zz}%
  \BibitemOpen
  \bibfield  {author} {\bibinfo {author} {\bibfnamefont {G.~M.}\ \bibnamefont
  {Harry}} (\bibinfo {collaboration} {LIGO Scientific}),\ }\bibfield  {title}
  {\bibinfo {title} {{Advanced LIGO: The next generation of gravitational wave
  detectors}},\ }\href {https://doi.org/10.1088/0264-9381/27/8/084006}
  {\bibfield  {journal} {\bibinfo  {journal} {Class. Quant. Grav.}\ }\textbf
  {\bibinfo {volume} {27}},\ \bibinfo {pages} {084006} (\bibinfo {year}
  {2010})}\BibitemShut {NoStop}%
\bibitem [{\citenamefont {Abbott}\ \emph
  {et~al.}(2020{\natexlab{a}})\citenamefont {Abbott}, \citenamefont {Abbott},
  \citenamefont {Abbott}, \citenamefont {Abraham}, \citenamefont {Acernese},
  \citenamefont {Ackley}, \citenamefont {Adams}, \citenamefont {Adya},
  \citenamefont {Affeldt}, \citenamefont {Agathos} \emph
  {et~al.}}]{Abbott_2020}%
  \BibitemOpen
  \bibfield  {author} {\bibinfo {author} {\bibfnamefont {B.~P.}\ \bibnamefont
  {Abbott}}, \bibinfo {author} {\bibfnamefont {R.}~\bibnamefont {Abbott}},
  \bibinfo {author} {\bibfnamefont {T.~D.}\ \bibnamefont {Abbott}}, \bibinfo
  {author} {\bibfnamefont {S.}~\bibnamefont {Abraham}}, \bibinfo {author}
  {\bibfnamefont {F.}~\bibnamefont {Acernese}}, \bibinfo {author}
  {\bibfnamefont {K.}~\bibnamefont {Ackley}}, \bibinfo {author} {\bibfnamefont
  {C.}~\bibnamefont {Adams}}, \bibinfo {author} {\bibfnamefont {V.~B.}\
  \bibnamefont {Adya}}, \bibinfo {author} {\bibfnamefont {C.}~\bibnamefont
  {Affeldt}}, \bibinfo {author} {\bibfnamefont {M.}~\bibnamefont {Agathos}},
  \emph {et~al.},\ }\bibfield  {title} {\bibinfo {title} {Prospects for
  observing and localizing gravitational-wave transients with advanced {LIGO},
  advanced virgo and {KAGRA}},\ }\bibfield  {journal} {\bibinfo  {journal}
  {Living Reviews in Relativity}\ }\textbf {\bibinfo {volume} {23}},\ \href
  {https://doi.org/10.1007/s41114-020-00026-9} {10.1007/s41114-020-00026-9}
  (\bibinfo {year} {2020}{\natexlab{a}})\BibitemShut {NoStop}%
\bibitem [{\citenamefont {Acernese}\ \emph {et~al.}(2014)\citenamefont
  {Acernese}, \citenamefont {Agathos}, \citenamefont {Agatsuma}, \citenamefont
  {Aisa}, \citenamefont {Allemandou}, \citenamefont {Allocca}, \citenamefont
  {Amarni}, \citenamefont {Astone}, \citenamefont {Balestri}, \citenamefont
  {Ballardin} \emph {et~al.}}]{Acernese_2014}%
  \BibitemOpen
  \bibfield  {author} {\bibinfo {author} {\bibfnamefont {F.}~\bibnamefont
  {Acernese}}, \bibinfo {author} {\bibfnamefont {M.}~\bibnamefont {Agathos}},
  \bibinfo {author} {\bibfnamefont {K.}~\bibnamefont {Agatsuma}}, \bibinfo
  {author} {\bibfnamefont {D.}~\bibnamefont {Aisa}}, \bibinfo {author}
  {\bibfnamefont {N.}~\bibnamefont {Allemandou}}, \bibinfo {author}
  {\bibfnamefont {A.}~\bibnamefont {Allocca}}, \bibinfo {author} {\bibfnamefont
  {J.}~\bibnamefont {Amarni}}, \bibinfo {author} {\bibfnamefont
  {P.}~\bibnamefont {Astone}}, \bibinfo {author} {\bibfnamefont
  {G.}~\bibnamefont {Balestri}}, \bibinfo {author} {\bibfnamefont
  {G.}~\bibnamefont {Ballardin}}, \emph {et~al.},\ }\bibfield  {title}
  {\bibinfo {title} {Advanced virgo: a second-generation interferometric
  gravitational wave detector},\ }\href
  {https://doi.org/10.1088/0264-9381/32/2/024001} {\bibfield  {journal}
  {\bibinfo  {journal} {Classical and Quantum Gravity}\ }\textbf {\bibinfo
  {volume} {32}},\ \bibinfo {pages} {024001} (\bibinfo {year}
  {2014})}\BibitemShut {NoStop}%
\bibitem [{\citenamefont {Aasi}\ \emph {et~al.}(2015)\citenamefont {Aasi},
  \citenamefont {Abbott}, \citenamefont {Abbott}, \citenamefont {Abbott},
  \citenamefont {Abernathy}, \citenamefont {Ackley}, \citenamefont {Adams},
  \citenamefont {Adams}, \citenamefont {Addesso}, \citenamefont {Adhikari}
  \emph {et~al.}}]{2015}%
  \BibitemOpen
  \bibfield  {author} {\bibinfo {author} {\bibfnamefont {J.}~\bibnamefont
  {Aasi}}, \bibinfo {author} {\bibfnamefont {B.~P.}\ \bibnamefont {Abbott}},
  \bibinfo {author} {\bibfnamefont {R.}~\bibnamefont {Abbott}}, \bibinfo
  {author} {\bibfnamefont {T.}~\bibnamefont {Abbott}}, \bibinfo {author}
  {\bibfnamefont {M.~R.}\ \bibnamefont {Abernathy}}, \bibinfo {author}
  {\bibfnamefont {K.}~\bibnamefont {Ackley}}, \bibinfo {author} {\bibfnamefont
  {C.}~\bibnamefont {Adams}}, \bibinfo {author} {\bibfnamefont
  {T.}~\bibnamefont {Adams}}, \bibinfo {author} {\bibfnamefont
  {P.}~\bibnamefont {Addesso}}, \bibinfo {author} {\bibfnamefont {R.~X.}\
  \bibnamefont {Adhikari}}, \emph {et~al.},\ }\bibfield  {title} {\bibinfo
  {title} {Advanced {LIGO}},\ }\href
  {https://doi.org/10.1088/0264-9381/32/7/074001} {\bibfield  {journal}
  {\bibinfo  {journal} {Classical and Quantum Gravity}\ }\textbf {\bibinfo
  {volume} {32}},\ \bibinfo {pages} {074001} (\bibinfo {year}
  {2015})}\BibitemShut {NoStop}%
\bibitem [{\citenamefont {Martynov}\ \emph {et~al.}(2016)\citenamefont
  {Martynov}, \citenamefont {Hall}, \citenamefont {Abbott}, \citenamefont
  {Abbott}, \citenamefont {Abbott}, \citenamefont {Adams}, \citenamefont
  {Adhikari}, \citenamefont {Anderson}, \citenamefont {Anderson}, \citenamefont
  {Arai} \emph {et~al.}}]{PhysRevD.93.112004}%
  \BibitemOpen
  \bibfield  {author} {\bibinfo {author} {\bibfnamefont {D.~V.}\ \bibnamefont
  {Martynov}}, \bibinfo {author} {\bibfnamefont {E.~D.}\ \bibnamefont {Hall}},
  \bibinfo {author} {\bibfnamefont {B.~P.}\ \bibnamefont {Abbott}}, \bibinfo
  {author} {\bibfnamefont {R.}~\bibnamefont {Abbott}}, \bibinfo {author}
  {\bibfnamefont {T.~D.}\ \bibnamefont {Abbott}}, \bibinfo {author}
  {\bibfnamefont {C.}~\bibnamefont {Adams}}, \bibinfo {author} {\bibfnamefont
  {R.~X.}\ \bibnamefont {Adhikari}}, \bibinfo {author} {\bibfnamefont {R.~A.}\
  \bibnamefont {Anderson}}, \bibinfo {author} {\bibfnamefont {S.~B.}\
  \bibnamefont {Anderson}}, \bibinfo {author} {\bibfnamefont {K.}~\bibnamefont
  {Arai}}, \emph {et~al.},\ }\bibfield  {title} {\bibinfo {title} {Sensitivity
  of the advanced ligo detectors at the beginning of gravitational wave
  astronomy},\ }\href {https://doi.org/10.1103/PhysRevD.93.112004} {\bibfield
  {journal} {\bibinfo  {journal} {Phys. Rev. D}\ }\textbf {\bibinfo {volume}
  {93}},\ \bibinfo {pages} {112004} (\bibinfo {year} {2016})}\BibitemShut
  {NoStop}%
\bibitem [{\citenamefont {Lamb}\ and\ \citenamefont
  {Kobayashi}(2018)}]{lamb2018grb}%
  \BibitemOpen
  \bibfield  {author} {\bibinfo {author} {\bibfnamefont {G.~P.}\ \bibnamefont
  {Lamb}}\ and\ \bibinfo {author} {\bibfnamefont {S.}~\bibnamefont
  {Kobayashi}},\ }\bibfield  {title} {\bibinfo {title} {{GRB 170817A} as a jet
  counterpart to gravitational wave trigger {GW 170817}},\ }\href
  {https://doi.org/https://doi.org/10.1093/mnras/sty1108} {\bibfield  {journal}
  {\bibinfo  {journal} {Mon. Notices of the Royal Astron. Soc.}\ }\textbf
  {\bibinfo {volume} {478}},\ \bibinfo {pages} {733} (\bibinfo {year}
  {2018})}\BibitemShut {NoStop}%
\bibitem [{\citenamefont {Abbott}\ \emph
  {et~al.}(2020{\natexlab{b}})\citenamefont {Abbott} \emph
  {et~al.}}]{abbott2020prospects}%
  \BibitemOpen
  \bibfield  {author} {\bibinfo {author} {\bibfnamefont {B.~P.}\ \bibnamefont
  {Abbott}} \emph {et~al.} (\bibinfo {collaboration} {{KAGRA, LIGO Scientific
  and Virgo Collaboration}}),\ }\bibfield  {title} {\bibinfo {title}
  {{Prospects for observing and localizing gravitational-wave transients with
  Advanced LIGO, Advanced Virgo and KAGRA}},\ }\href
  {https://doi.org/10.1007/s41114-020-00026-9} {\bibfield  {journal} {\bibinfo
  {journal} {Living Rev. Relativ.}\ }\textbf {\bibinfo {volume} {23}},\
  \bibinfo {pages} {1} (\bibinfo {year} {2020}{\natexlab{b}})}\BibitemShut
  {NoStop}%
\bibitem [{\citenamefont {Margalit}\ and\ \citenamefont
  {Metzger}(2019)}]{Margalit_2019}%
  \BibitemOpen
  \bibfield  {author} {\bibinfo {author} {\bibfnamefont {B.}~\bibnamefont
  {Margalit}}\ and\ \bibinfo {author} {\bibfnamefont {B.~D.}\ \bibnamefont
  {Metzger}},\ }\bibfield  {title} {\bibinfo {title} {The multi-messenger
  matrix: The future of neutron star merger constraints on the nuclear equation
  of state},\ }\href {https://doi.org/10.3847/2041-8213/ab2ae2} {\bibfield
  {journal} {\bibinfo  {journal} {Astrophys. J. Lett.}\ }\textbf {\bibinfo
  {volume} {880}},\ \bibinfo {pages} {L15} (\bibinfo {year}
  {2019})}\BibitemShut {NoStop}%
\bibitem [{\citenamefont {Lenon}\ \emph {et~al.}(2020)\citenamefont {Lenon},
  \citenamefont {Nitz},\ and\ \citenamefont {Brown}}]{Lenon_2020}%
  \BibitemOpen
  \bibfield  {author} {\bibinfo {author} {\bibfnamefont {A.~K.}\ \bibnamefont
  {Lenon}}, \bibinfo {author} {\bibfnamefont {A.~H.}\ \bibnamefont {Nitz}},\
  and\ \bibinfo {author} {\bibfnamefont {D.~A.}\ \bibnamefont {Brown}},\
  }\bibfield  {title} {\bibinfo {title} {Measuring the eccentricity of
  {GW}170817 and {GW}190425},\ }\href {https://doi.org/10.1093/mnras/staa2120}
  {\bibfield  {journal} {\bibinfo  {journal} {Monthly Notices of the Royal
  Astronomical Society}\ }\textbf {\bibinfo {volume} {497}},\ \bibinfo {pages}
  {1966} (\bibinfo {year} {2020})}\BibitemShut {NoStop}%
\bibitem [{\citenamefont {Apostolatos}\ \emph {et~al.}(1994)\citenamefont
  {Apostolatos}, \citenamefont {Cutler}, \citenamefont {Sussman},\ and\
  \citenamefont {Thorne}}]{PhysRevD.49.6274}%
  \BibitemOpen
  \bibfield  {author} {\bibinfo {author} {\bibfnamefont {T.~A.}\ \bibnamefont
  {Apostolatos}}, \bibinfo {author} {\bibfnamefont {C.}~\bibnamefont {Cutler}},
  \bibinfo {author} {\bibfnamefont {G.~J.}\ \bibnamefont {Sussman}},\ and\
  \bibinfo {author} {\bibfnamefont {K.~S.}\ \bibnamefont {Thorne}},\ }\bibfield
   {title} {\bibinfo {title} {Spin-induced orbital precession and its
  modulation of the gravitational waveforms from merging binaries},\ }\href
  {https://doi.org/10.1103/PhysRevD.49.6274} {\bibfield  {journal} {\bibinfo
  {journal} {Phys. Rev. D}\ }\textbf {\bibinfo {volume} {49}},\ \bibinfo
  {pages} {6274} (\bibinfo {year} {1994})}\BibitemShut {NoStop}%
\bibitem [{\citenamefont {Hannam}\ \emph {et~al.}(2022)\citenamefont {Hannam},
  \citenamefont {Hoy}, \citenamefont {Thompson}, \citenamefont {Fairhurst},
  \citenamefont {Raymond}, \citenamefont {Colleoni}, \citenamefont {Davis},
  \citenamefont {Estellés}, \citenamefont {Haster}, \citenamefont
  {Helmling-Cornell}, \citenamefont {Husa}, \citenamefont {Keitel},
  \citenamefont {Massinger}, \citenamefont {Menéndez-Vázquez}, \citenamefont
  {Mogushi}, \citenamefont {Ossokine}, \citenamefont {Payne}, \citenamefont
  {Pratten}, \citenamefont {Romero-Shaw}, \citenamefont {Sadiq}, \citenamefont
  {Schmidt}, \citenamefont {Tenorio}, \citenamefont {Udall}, \citenamefont
  {Veitch}, \citenamefont {Williams}, \citenamefont {Yelikar},\ and\
  \citenamefont {Zimmerman}}]{Hannam_2022}%
  \BibitemOpen
  \bibfield  {author} {\bibinfo {author} {\bibfnamefont {M.}~\bibnamefont
  {Hannam}}, \bibinfo {author} {\bibfnamefont {C.}~\bibnamefont {Hoy}},
  \bibinfo {author} {\bibfnamefont {J.~E.}\ \bibnamefont {Thompson}}, \bibinfo
  {author} {\bibfnamefont {S.}~\bibnamefont {Fairhurst}}, \bibinfo {author}
  {\bibfnamefont {V.}~\bibnamefont {Raymond}}, \bibinfo {author} {\bibfnamefont
  {M.}~\bibnamefont {Colleoni}}, \bibinfo {author} {\bibfnamefont
  {D.}~\bibnamefont {Davis}}, \bibinfo {author} {\bibfnamefont
  {H.}~\bibnamefont {Estellés}}, \bibinfo {author} {\bibfnamefont {C.-J.}\
  \bibnamefont {Haster}}, \bibinfo {author} {\bibfnamefont {A.}~\bibnamefont
  {Helmling-Cornell}}, \bibinfo {author} {\bibfnamefont {S.}~\bibnamefont
  {Husa}}, \bibinfo {author} {\bibfnamefont {D.}~\bibnamefont {Keitel}},
  \bibinfo {author} {\bibfnamefont {T.~J.}\ \bibnamefont {Massinger}}, \bibinfo
  {author} {\bibfnamefont {A.}~\bibnamefont {Menéndez-Vázquez}}, \bibinfo
  {author} {\bibfnamefont {K.}~\bibnamefont {Mogushi}}, \bibinfo {author}
  {\bibfnamefont {S.}~\bibnamefont {Ossokine}}, \bibinfo {author}
  {\bibfnamefont {E.}~\bibnamefont {Payne}}, \bibinfo {author} {\bibfnamefont
  {G.}~\bibnamefont {Pratten}}, \bibinfo {author} {\bibfnamefont
  {I.}~\bibnamefont {Romero-Shaw}}, \bibinfo {author} {\bibfnamefont
  {J.}~\bibnamefont {Sadiq}}, \bibinfo {author} {\bibfnamefont
  {P.}~\bibnamefont {Schmidt}}, \bibinfo {author} {\bibfnamefont
  {R.}~\bibnamefont {Tenorio}}, \bibinfo {author} {\bibfnamefont
  {R.}~\bibnamefont {Udall}}, \bibinfo {author} {\bibfnamefont
  {J.}~\bibnamefont {Veitch}}, \bibinfo {author} {\bibfnamefont
  {D.}~\bibnamefont {Williams}}, \bibinfo {author} {\bibfnamefont {A.~B.}\
  \bibnamefont {Yelikar}},\ and\ \bibinfo {author} {\bibfnamefont
  {A.}~\bibnamefont {Zimmerman}},\ }\bibfield  {title} {\bibinfo {title}
  {General-relativistic precession in a black-hole binary},\ }\href
  {https://doi.org/10.1038/s41586-022-05212-z} {\bibfield  {journal} {\bibinfo
  {journal} {Nature}\ }\textbf {\bibinfo {volume} {610}},\ \bibinfo {pages}
  {652–655} (\bibinfo {year} {2022})}\BibitemShut {NoStop}%
\bibitem [{\citenamefont {Wolfe}\ \emph {et~al.}(2023)\citenamefont {Wolfe},
  \citenamefont {Vitale},\ and\ \citenamefont {Talbot}}]{wolfe2023small}%
  \BibitemOpen
  \bibfield  {author} {\bibinfo {author} {\bibfnamefont {N.~E.}\ \bibnamefont
  {Wolfe}}, \bibinfo {author} {\bibfnamefont {S.}~\bibnamefont {Vitale}},\ and\
  \bibinfo {author} {\bibfnamefont {C.}~\bibnamefont {Talbot}},\ }\href@noop {}
  {\bibinfo {title} {Too small to fail: characterizing sub-solar mass black
  hole mergers with gravitational waves}} (\bibinfo {year} {2023}),\ \Eprint
  {https://arxiv.org/abs/2305.19907} {arXiv:2305.19907 [astro-ph.HE]}
  \BibitemShut {NoStop}%
\bibitem [{\citenamefont {Pratten}\ \emph
  {et~al.}(2021{\natexlab{a}})\citenamefont {Pratten}, \citenamefont
  {Garc\'{\i}a-Quir\'os}, \citenamefont {Colleoni}, \citenamefont
  {Ramos-Buades}, \citenamefont {Estell\'es}, \citenamefont {Mateu-Lucena},
  \citenamefont {Jaume}, \citenamefont {Haney}, \citenamefont {Keitel},
  \citenamefont {Thompson},\ and\ \citenamefont {Husa}}]{PhysRevD.103.104056}%
  \BibitemOpen
  \bibfield  {author} {\bibinfo {author} {\bibfnamefont {G.}~\bibnamefont
  {Pratten}}, \bibinfo {author} {\bibfnamefont {C.}~\bibnamefont
  {Garc\'{\i}a-Quir\'os}}, \bibinfo {author} {\bibfnamefont {M.}~\bibnamefont
  {Colleoni}}, \bibinfo {author} {\bibfnamefont {A.}~\bibnamefont
  {Ramos-Buades}}, \bibinfo {author} {\bibfnamefont {H.}~\bibnamefont
  {Estell\'es}}, \bibinfo {author} {\bibfnamefont {M.}~\bibnamefont
  {Mateu-Lucena}}, \bibinfo {author} {\bibfnamefont {R.}~\bibnamefont {Jaume}},
  \bibinfo {author} {\bibfnamefont {M.}~\bibnamefont {Haney}}, \bibinfo
  {author} {\bibfnamefont {D.}~\bibnamefont {Keitel}}, \bibinfo {author}
  {\bibfnamefont {J.~E.}\ \bibnamefont {Thompson}},\ and\ \bibinfo {author}
  {\bibfnamefont {S.}~\bibnamefont {Husa}},\ }\bibfield  {title} {\bibinfo
  {title} {Computationally efficient models for the dominant and subdominant
  harmonic modes of precessing binary black holes},\ }\href
  {https://doi.org/10.1103/PhysRevD.103.104056} {\bibfield  {journal} {\bibinfo
   {journal} {Phys. Rev. D}\ }\textbf {\bibinfo {volume} {103}},\ \bibinfo
  {pages} {104056} (\bibinfo {year} {2021}{\natexlab{a}})}\BibitemShut
  {NoStop}%
\bibitem [{\citenamefont {Varma}\ \emph {et~al.}(2019)\citenamefont {Varma},
  \citenamefont {Field}, \citenamefont {Scheel}, \citenamefont {Blackman},
  \citenamefont {Gerosa}, \citenamefont {Stein}, \citenamefont {Kidder},\ and\
  \citenamefont {Pfeiffer}}]{PhysRevResearch.1.033015}%
  \BibitemOpen
  \bibfield  {author} {\bibinfo {author} {\bibfnamefont {V.}~\bibnamefont
  {Varma}}, \bibinfo {author} {\bibfnamefont {S.~E.}\ \bibnamefont {Field}},
  \bibinfo {author} {\bibfnamefont {M.~A.}\ \bibnamefont {Scheel}}, \bibinfo
  {author} {\bibfnamefont {J.}~\bibnamefont {Blackman}}, \bibinfo {author}
  {\bibfnamefont {D.}~\bibnamefont {Gerosa}}, \bibinfo {author} {\bibfnamefont
  {L.~C.}\ \bibnamefont {Stein}}, \bibinfo {author} {\bibfnamefont {L.~E.}\
  \bibnamefont {Kidder}},\ and\ \bibinfo {author} {\bibfnamefont {H.~P.}\
  \bibnamefont {Pfeiffer}},\ }\bibfield  {title} {\bibinfo {title} {Surrogate
  models for precessing binary black hole simulations with unequal masses},\
  }\href {https://doi.org/10.1103/PhysRevResearch.1.033015} {\bibfield
  {journal} {\bibinfo  {journal} {Phys. Rev. Res.}\ }\textbf {\bibinfo {volume}
  {1}},\ \bibinfo {pages} {033015} (\bibinfo {year} {2019})}\BibitemShut
  {NoStop}%
\bibitem [{\citenamefont {Ramos-Buades}\ \emph {et~al.}(2023)\citenamefont
  {Ramos-Buades}, \citenamefont {Buonanno}, \citenamefont {Estellés},
  \citenamefont {Khalil}, \citenamefont {Mihaylov}, \citenamefont {Ossokine},
  \citenamefont {Pompili},\ and\ \citenamefont
  {Shiferaw}}]{ramosbuades2023seobnrv5phm}%
  \BibitemOpen
  \bibfield  {author} {\bibinfo {author} {\bibfnamefont {A.}~\bibnamefont
  {Ramos-Buades}}, \bibinfo {author} {\bibfnamefont {A.}~\bibnamefont
  {Buonanno}}, \bibinfo {author} {\bibfnamefont {H.}~\bibnamefont {Estellés}},
  \bibinfo {author} {\bibfnamefont {M.}~\bibnamefont {Khalil}}, \bibinfo
  {author} {\bibfnamefont {D.~P.}\ \bibnamefont {Mihaylov}}, \bibinfo {author}
  {\bibfnamefont {S.}~\bibnamefont {Ossokine}}, \bibinfo {author}
  {\bibfnamefont {L.}~\bibnamefont {Pompili}},\ and\ \bibinfo {author}
  {\bibfnamefont {M.}~\bibnamefont {Shiferaw}},\ }\href@noop {} {\bibinfo
  {title} {Seobnrv5phm: Next generation of accurate and efficient multipolar
  precessing-spin effective-one-body waveforms for binary black holes}}
  (\bibinfo {year} {2023}),\ \Eprint {https://arxiv.org/abs/2303.18046}
  {arXiv:2303.18046 [gr-qc]} \BibitemShut {NoStop}%
\bibitem [{\citenamefont {Islam}\ \emph {et~al.}(2022)\citenamefont {Islam},
  \citenamefont {Roulet},\ and\ \citenamefont
  {Venumadhav}}]{islam2022factorized}%
  \BibitemOpen
  \bibfield  {author} {\bibinfo {author} {\bibfnamefont {T.}~\bibnamefont
  {Islam}}, \bibinfo {author} {\bibfnamefont {J.}~\bibnamefont {Roulet}},\ and\
  \bibinfo {author} {\bibfnamefont {T.}~\bibnamefont {Venumadhav}},\
  }\href@noop {} {\bibinfo {title} {Factorized parameter estimation for
  real-time gravitational wave inference}} (\bibinfo {year} {2022}),\ \Eprint
  {https://arxiv.org/abs/2210.16278} {arXiv:2210.16278 [gr-qc]} \BibitemShut
  {NoStop}%
\bibitem [{\citenamefont {Singer}\ and\ \citenamefont
  {Price}(2016{\natexlab{a}})}]{singer2016rapid}%
  \BibitemOpen
  \bibfield  {author} {\bibinfo {author} {\bibfnamefont {L.~P.}\ \bibnamefont
  {Singer}}\ and\ \bibinfo {author} {\bibfnamefont {L.~R.}\ \bibnamefont
  {Price}},\ }\bibfield  {title} {\bibinfo {title} {Rapid {B}ayesian position
  reconstruction for gravitational-wave transients},\ }\href
  {https://doi.org/https://doi.org/10.1103/PhysRevD.93.024013} {\bibfield
  {journal} {\bibinfo  {journal} {Phys. Rev. D}\ }\textbf {\bibinfo {volume}
  {93}},\ \bibinfo {pages} {024013} (\bibinfo {year}
  {2016}{\natexlab{a}})}\BibitemShut {NoStop}%
\bibitem [{\citenamefont {Finstad}\ and\ \citenamefont
  {Brown}(2020)}]{Finstad_2020}%
  \BibitemOpen
  \bibfield  {author} {\bibinfo {author} {\bibfnamefont {D.}~\bibnamefont
  {Finstad}}\ and\ \bibinfo {author} {\bibfnamefont {D.~A.}\ \bibnamefont
  {Brown}},\ }\bibfield  {title} {\bibinfo {title} {Fast parameter estimation
  of binary mergers for multimessenger follow-up},\ }\href
  {https://doi.org/10.3847/2041-8213/abca9e} {\bibfield  {journal} {\bibinfo
  {journal} {Astrophys. J. Lett.}\ }\textbf {\bibinfo {volume} {905}},\
  \bibinfo {pages} {L9} (\bibinfo {year} {2020})}\BibitemShut {NoStop}%
\bibitem [{\citenamefont {Chatterjee}\ and\ \citenamefont
  {Wen}(2022)}]{chatterjee2022premerger}%
  \BibitemOpen
  \bibfield  {author} {\bibinfo {author} {\bibfnamefont {C.}~\bibnamefont
  {Chatterjee}}\ and\ \bibinfo {author} {\bibfnamefont {L.}~\bibnamefont
  {Wen}},\ }\href@noop {} {\bibinfo {title} {Pre-merger sky localization of
  gravitational waves from binary neutron star mergers using deep learning}}
  (\bibinfo {year} {2022}),\ \Eprint {https://arxiv.org/abs/2301.03558}
  {arXiv:2301.03558 [astro-ph.HE]} \BibitemShut {NoStop}%
\bibitem [{\citenamefont {Canizares}\ \emph {et~al.}(2013)\citenamefont
  {Canizares}, \citenamefont {Field}, \citenamefont {Gair},\ and\ \citenamefont
  {Tiglio}}]{canizares2013gravitational}%
  \BibitemOpen
  \bibfield  {author} {\bibinfo {author} {\bibfnamefont {P.}~\bibnamefont
  {Canizares}}, \bibinfo {author} {\bibfnamefont {S.~E.}\ \bibnamefont
  {Field}}, \bibinfo {author} {\bibfnamefont {J.~R.}\ \bibnamefont {Gair}},\
  and\ \bibinfo {author} {\bibfnamefont {M.}~\bibnamefont {Tiglio}},\
  }\bibfield  {title} {\bibinfo {title} {Gravitational wave parameter
  estimation with compressed likelihood evaluations},\ }\href
  {https://doi.org/https://doi.org/10.1103/PhysRevD.87.124005} {\bibfield
  {journal} {\bibinfo  {journal} {Phys. Rev. D}\ }\textbf {\bibinfo {volume}
  {87}},\ \bibinfo {pages} {124005} (\bibinfo {year} {2013})}\BibitemShut
  {NoStop}%
\bibitem [{\citenamefont {Canizares}\ \emph {et~al.}(2015)\citenamefont
  {Canizares}, \citenamefont {Field}, \citenamefont {Gair}, \citenamefont
  {Raymond}, \citenamefont {Smith},\ and\ \citenamefont
  {Tiglio}}]{canizares2015accelerated}%
  \BibitemOpen
  \bibfield  {author} {\bibinfo {author} {\bibfnamefont {P.}~\bibnamefont
  {Canizares}}, \bibinfo {author} {\bibfnamefont {S.~E.}\ \bibnamefont
  {Field}}, \bibinfo {author} {\bibfnamefont {J.}~\bibnamefont {Gair}},
  \bibinfo {author} {\bibfnamefont {V.}~\bibnamefont {Raymond}}, \bibinfo
  {author} {\bibfnamefont {R.}~\bibnamefont {Smith}},\ and\ \bibinfo {author}
  {\bibfnamefont {M.}~\bibnamefont {Tiglio}},\ }\bibfield  {title} {\bibinfo
  {title} {Accelerated gravitational wave parameter estimation with reduced
  order modeling},\ }\href
  {https://doi.org/https://doi.org/10.1103/PhysRevLett.114.071104} {\bibfield
  {journal} {\bibinfo  {journal} {Phy. Rev. Lett.}\ }\textbf {\bibinfo {volume}
  {114}},\ \bibinfo {pages} {071104} (\bibinfo {year} {2015})}\BibitemShut
  {NoStop}%
\bibitem [{\citenamefont {Smith}\ \emph {et~al.}(2016)\citenamefont {Smith},
  \citenamefont {Field}, \citenamefont {Blackburn}, \citenamefont {Haster},
  \citenamefont {P{\"u}rrer}, \citenamefont {Raymond},\ and\ \citenamefont
  {Schmidt}}]{smith2016fast}%
  \BibitemOpen
  \bibfield  {author} {\bibinfo {author} {\bibfnamefont {R.}~\bibnamefont
  {Smith}}, \bibinfo {author} {\bibfnamefont {S.~E.}\ \bibnamefont {Field}},
  \bibinfo {author} {\bibfnamefont {K.}~\bibnamefont {Blackburn}}, \bibinfo
  {author} {\bibfnamefont {C.-J.}\ \bibnamefont {Haster}}, \bibinfo {author}
  {\bibfnamefont {M.}~\bibnamefont {P{\"u}rrer}}, \bibinfo {author}
  {\bibfnamefont {V.}~\bibnamefont {Raymond}},\ and\ \bibinfo {author}
  {\bibfnamefont {P.}~\bibnamefont {Schmidt}},\ }\bibfield  {title} {\bibinfo
  {title} {Fast and accurate inference on gravitational waves from precessing
  compact binaries},\ }\href
  {https://doi.org/https://doi.org/10.1103/PhysRevD.94.044031} {\bibfield
  {journal} {\bibinfo  {journal} {Phys. Rev. D}\ }\textbf {\bibinfo {volume}
  {94}},\ \bibinfo {pages} {044031} (\bibinfo {year} {2016})}\BibitemShut
  {NoStop}%
\bibitem [{\citenamefont {Morisaki}\ and\ \citenamefont
  {Raymond}(2020)}]{Morisaki_2020}%
  \BibitemOpen
  \bibfield  {author} {\bibinfo {author} {\bibfnamefont {S.}~\bibnamefont
  {Morisaki}}\ and\ \bibinfo {author} {\bibfnamefont {V.}~\bibnamefont
  {Raymond}},\ }\bibfield  {title} {\bibinfo {title} {{Rapid parameter
  estimation of gravitational waves from binary neutron star coalescence using
  focused reduced order quadrature}},\ }\bibfield  {journal} {\bibinfo
  {journal} {Physical Review D}\ }\textbf {\bibinfo {volume} {102}},\ \href
  {https://doi.org/10.1103/physrevd.102.104020} {10.1103/physrevd.102.104020}
  (\bibinfo {year} {2020})\BibitemShut {NoStop}%
\bibitem [{\citenamefont {Morisaki}\ \emph {et~al.}(2023)\citenamefont
  {Morisaki}, \citenamefont {Smith}, \citenamefont {Tsukada}, \citenamefont
  {Sachdev}, \citenamefont {Stevenson}, \citenamefont {Talbot},\ and\
  \citenamefont {Zimmerman}}]{morisaki2023rapid}%
  \BibitemOpen
  \bibfield  {author} {\bibinfo {author} {\bibfnamefont {S.}~\bibnamefont
  {Morisaki}}, \bibinfo {author} {\bibfnamefont {R.}~\bibnamefont {Smith}},
  \bibinfo {author} {\bibfnamefont {L.}~\bibnamefont {Tsukada}}, \bibinfo
  {author} {\bibfnamefont {S.}~\bibnamefont {Sachdev}}, \bibinfo {author}
  {\bibfnamefont {S.}~\bibnamefont {Stevenson}}, \bibinfo {author}
  {\bibfnamefont {C.}~\bibnamefont {Talbot}},\ and\ \bibinfo {author}
  {\bibfnamefont {A.}~\bibnamefont {Zimmerman}},\ }\href
  {https://doi.org/https://arxiv.org/abs/2307.13380} {\bibinfo {title} {{Rapid
  localization and inference on compact binary coalescences with the Advanced
  LIGO-Virgo-KAGRA gravitational-wave detector network}}} (\bibinfo {year}
  {2023}),\ \Eprint {https://arxiv.org/abs/2307.13380} {arXiv:2307.13380
  [gr-qc]} \BibitemShut {NoStop}%
\bibitem [{\citenamefont {Zackay}\ \emph {et~al.}(2018)\citenamefont {Zackay},
  \citenamefont {Dai},\ and\ \citenamefont {Venumadhav}}]{Venumadhav2018}%
  \BibitemOpen
  \bibfield  {author} {\bibinfo {author} {\bibfnamefont {B.}~\bibnamefont
  {Zackay}}, \bibinfo {author} {\bibfnamefont {L.}~\bibnamefont {Dai}},\ and\
  \bibinfo {author} {\bibfnamefont {T.}~\bibnamefont {Venumadhav}},\ }\href
  {https://doi.org/10.48550/ARXIV.1806.08792} {\bibinfo {title} {Relative
  binning and fast likelihood evaluation for gravitational wave parameter
  estimation}} (\bibinfo {year} {2018}),\ \Eprint
  {https://arxiv.org/abs/1806.08792} {arXiv:1806.08792 [astro-ph.IM]}
  \BibitemShut {NoStop}%
\bibitem [{\citenamefont {Cornish}(2021)}]{cornish2021heterodyned}%
  \BibitemOpen
  \bibfield  {author} {\bibinfo {author} {\bibfnamefont {N.~J.}\ \bibnamefont
  {Cornish}},\ }\bibfield  {title} {\bibinfo {title} {Heterodyned likelihood
  for rapid gravitational wave parameter inference},\ }\href
  {https://doi.org/10.1103/PhysRevD.104.104054} {\bibfield  {journal} {\bibinfo
   {journal} {Phys. Rev. D}\ }\textbf {\bibinfo {volume} {104}},\ \bibinfo
  {pages} {104054} (\bibinfo {year} {2021})}\BibitemShut {NoStop}%
\bibitem [{\citenamefont {Fairhurst}\ \emph {et~al.}(2023)\citenamefont
  {Fairhurst}, \citenamefont {Hoy}, \citenamefont {Green}, \citenamefont
  {Mills},\ and\ \citenamefont {Usman}}]{fairhurst2023fast}%
  \BibitemOpen
  \bibfield  {author} {\bibinfo {author} {\bibfnamefont {S.}~\bibnamefont
  {Fairhurst}}, \bibinfo {author} {\bibfnamefont {C.}~\bibnamefont {Hoy}},
  \bibinfo {author} {\bibfnamefont {R.}~\bibnamefont {Green}}, \bibinfo
  {author} {\bibfnamefont {C.}~\bibnamefont {Mills}},\ and\ \bibinfo {author}
  {\bibfnamefont {S.~A.}\ \bibnamefont {Usman}},\ }\href@noop {} {\bibinfo
  {title} {Fast inference of binary merger properties using the information
  encoded in the gravitational-wave signal}} (\bibinfo {year} {2023}),\ \Eprint
  {https://arxiv.org/abs/2304.03731} {arXiv:2304.03731 [gr-qc]} \BibitemShut
  {NoStop}%
\bibitem [{\citenamefont {Rasmussen}\ and\ \citenamefont
  {Williams}(2006)}]{rasmussen2006gaussian}%
  \BibitemOpen
  \bibfield  {author} {\bibinfo {author} {\bibfnamefont {C.~E.}\ \bibnamefont
  {Rasmussen}}\ and\ \bibinfo {author} {\bibfnamefont {C.~K.}\ \bibnamefont
  {Williams}},\ }\href@noop {} {\emph {\bibinfo {title} {{Gaussian Processes
  for Machine Learning}}}}\ (\bibinfo  {publisher} {The MIT Press},\ \bibinfo
  {year} {2006})\BibitemShut {NoStop}%
\bibitem [{\citenamefont {Lange}\ \emph {et~al.}(2018)\citenamefont {Lange},
  \citenamefont {O'Shaughnessy},\ and\ \citenamefont
  {Rizzo}}]{https://doi.org/10.48550/arxiv.1805.10457}%
  \BibitemOpen
  \bibfield  {author} {\bibinfo {author} {\bibfnamefont {J.}~\bibnamefont
  {Lange}}, \bibinfo {author} {\bibfnamefont {R.}~\bibnamefont
  {O'Shaughnessy}},\ and\ \bibinfo {author} {\bibfnamefont {M.}~\bibnamefont
  {Rizzo}},\ }\href {https://doi.org/10.48550/ARXIV.1805.10457} {\bibinfo
  {title} {Rapid and accurate parameter inference for coalescing, precessing
  compact binaries}} (\bibinfo {year} {2018})\BibitemShut {NoStop}%
\bibitem [{\citenamefont {Morisaki}(2021)}]{Morisaki:2021ngj}%
  \BibitemOpen
  \bibfield  {author} {\bibinfo {author} {\bibfnamefont {S.}~\bibnamefont
  {Morisaki}},\ }\bibfield  {title} {\bibinfo {title} {{{Accelerating parameter
  estimation of gravitational waves from compact binary coalescence using
  adaptive frequency resolutions}}},\ }\href
  {https://doi.org/10.1103/PhysRevD.104.044062} {\bibfield  {journal} {\bibinfo
   {journal} {Phys. Rev. D}\ }\textbf {\bibinfo {volume} {104}},\ \bibinfo
  {pages} {044062} (\bibinfo {year} {2021})},\ \Eprint
  {https://arxiv.org/abs/2104.07813} {arXiv:2104.07813 [gr-qc]} \BibitemShut
  {NoStop}%
\bibitem [{\citenamefont {Lee}\ \emph {et~al.}(2022)\citenamefont {Lee},
  \citenamefont {Morisaki},\ and\ \citenamefont {Tagoshi}}]{Lee:2022jpn}%
  \BibitemOpen
  \bibfield  {author} {\bibinfo {author} {\bibfnamefont {E.}~\bibnamefont
  {Lee}}, \bibinfo {author} {\bibfnamefont {S.}~\bibnamefont {Morisaki}},\ and\
  \bibinfo {author} {\bibfnamefont {H.}~\bibnamefont {Tagoshi}},\ }\bibfield
  {title} {\bibinfo {title} {{{Mass-spin reparametrization for a rapid
  parameter estimation of inspiral gravitational-wave signals}}},\ }\href
  {https://doi.org/10.1103/PhysRevD.105.124057} {\bibfield  {journal} {\bibinfo
   {journal} {Phys. Rev. D}\ }\textbf {\bibinfo {volume} {105}},\ \bibinfo
  {pages} {124057} (\bibinfo {year} {2022})},\ \Eprint
  {https://arxiv.org/abs/2203.05216} {arXiv:2203.05216 [gr-qc]} \BibitemShut
  {NoStop}%
\bibitem [{\citenamefont {Morras}\ \emph {et~al.}(2023)\citenamefont {Morras},
  \citenamefont {Siles},\ and\ \citenamefont {Garcia-Bellido}}]{effroqs23}%
  \BibitemOpen
  \bibfield  {author} {\bibinfo {author} {\bibfnamefont {G.}~\bibnamefont
  {Morras}}, \bibinfo {author} {\bibfnamefont {J.~F.~N.}\ \bibnamefont
  {Siles}},\ and\ \bibinfo {author} {\bibfnamefont {J.}~\bibnamefont
  {Garcia-Bellido}},\ }\href {https://doi.org/https://arxiv.org/abs/2307.16610}
  {\bibinfo {title} {Efficient reduced order quadrature construction algorithms
  for fast gravitational wave inference}} (\bibinfo {year} {2023}),\ \Eprint
  {https://arxiv.org/abs/2307.16610} {arXiv:2307.16610 [gr-qc]} \BibitemShut
  {NoStop}%
\bibitem [{\citenamefont {Chua}\ and\ \citenamefont
  {Vallisneri}(2020)}]{Chua_2020}%
  \BibitemOpen
  \bibfield  {author} {\bibinfo {author} {\bibfnamefont {A.~J.}\ \bibnamefont
  {Chua}}\ and\ \bibinfo {author} {\bibfnamefont {M.}~\bibnamefont
  {Vallisneri}},\ }\bibfield  {title} {\bibinfo {title} {Learning bayesian
  posteriors with neural networks for gravitational-wave inference},\
  }\bibfield  {journal} {\bibinfo  {journal} {Physical Review Letters}\
  }\textbf {\bibinfo {volume} {124}},\ \href
  {https://doi.org/10.1103/physrevlett.124.041102}
  {10.1103/physrevlett.124.041102} (\bibinfo {year} {2020})\BibitemShut
  {NoStop}%
\bibitem [{\citenamefont {Green}\ \emph {et~al.}(2020)\citenamefont {Green},
  \citenamefont {Simpson},\ and\ \citenamefont {Gair}}]{Green_2020}%
  \BibitemOpen
  \bibfield  {author} {\bibinfo {author} {\bibfnamefont {S.~R.}\ \bibnamefont
  {Green}}, \bibinfo {author} {\bibfnamefont {C.}~\bibnamefont {Simpson}},\
  and\ \bibinfo {author} {\bibfnamefont {J.}~\bibnamefont {Gair}},\ }\bibfield
  {title} {\bibinfo {title} {Gravitational-wave parameter estimation with
  autoregressive neural network flows},\ }\bibfield  {journal} {\bibinfo
  {journal} {Physical Review D}\ }\textbf {\bibinfo {volume} {102}},\ \href
  {https://doi.org/10.1103/physrevd.102.104057} {10.1103/physrevd.102.104057}
  (\bibinfo {year} {2020})\BibitemShut {NoStop}%
\bibitem [{\citenamefont {Green}\ and\ \citenamefont
  {Gair}(2020)}]{green2020complete}%
  \BibitemOpen
  \bibfield  {author} {\bibinfo {author} {\bibfnamefont {S.~R.}\ \bibnamefont
  {Green}}\ and\ \bibinfo {author} {\bibfnamefont {J.}~\bibnamefont {Gair}},\
  }\href@noop {} {\bibinfo {title} {Complete parameter inference for gw150914
  using deep learning}} (\bibinfo {year} {2020}),\ \Eprint
  {https://arxiv.org/abs/2008.03312} {arXiv:2008.03312 [astro-ph.IM]}
  \BibitemShut {NoStop}%
\bibitem [{\citenamefont {Gabbard}\ \emph {et~al.}(2021)\citenamefont
  {Gabbard}, \citenamefont {Messenger}, \citenamefont {Heng}, \citenamefont
  {Tonolini},\ and\ \citenamefont {Murray-Smith}}]{gabbard2022bayesian}%
  \BibitemOpen
  \bibfield  {author} {\bibinfo {author} {\bibfnamefont {H.}~\bibnamefont
  {Gabbard}}, \bibinfo {author} {\bibfnamefont {C.}~\bibnamefont {Messenger}},
  \bibinfo {author} {\bibfnamefont {I.~S.}\ \bibnamefont {Heng}}, \bibinfo
  {author} {\bibfnamefont {F.}~\bibnamefont {Tonolini}},\ and\ \bibinfo
  {author} {\bibfnamefont {R.}~\bibnamefont {Murray-Smith}},\ }\bibfield
  {title} {\bibinfo {title} {Bayesian parameter estimation using conditional
  variational autoencoders for gravitational-wave astronomy},\ }\href
  {https://doi.org/10.1038/s41567-021-01425-7} {\bibfield  {journal} {\bibinfo
  {journal} {Nature Physics}\ }\textbf {\bibinfo {volume} {18}},\ \bibinfo
  {pages} {112} (\bibinfo {year} {2021})}\BibitemShut {NoStop}%
\bibitem [{\citenamefont {Wong}\ \emph {et~al.}(2023)\citenamefont {Wong},
  \citenamefont {Isi},\ and\ \citenamefont {Edwards}}]{wong2023fast}%
  \BibitemOpen
  \bibfield  {author} {\bibinfo {author} {\bibfnamefont {K.~W.~K.}\
  \bibnamefont {Wong}}, \bibinfo {author} {\bibfnamefont {M.}~\bibnamefont
  {Isi}},\ and\ \bibinfo {author} {\bibfnamefont {T.~D.~P.}\ \bibnamefont
  {Edwards}},\ }\href@noop {} {\bibinfo {title} {Fast gravitational wave
  parameter estimation without compromises}} (\bibinfo {year} {2023}),\ \Eprint
  {https://arxiv.org/abs/2302.05333} {arXiv:2302.05333 [astro-ph.IM]}
  \BibitemShut {NoStop}%
\bibitem [{\citenamefont {Legin}\ \emph {et~al.}(2023)\citenamefont {Legin},
  \citenamefont {Isi}, \citenamefont {Wong}, \citenamefont {Adam},
  \citenamefont {Perreault-Levasseur},\ and\ \citenamefont
  {Hezaveh}}]{score_based_likelihood}%
  \BibitemOpen
  \bibfield  {author} {\bibinfo {author} {\bibfnamefont {R.}~\bibnamefont
  {Legin}}, \bibinfo {author} {\bibfnamefont {M.}~\bibnamefont {Isi}}, \bibinfo
  {author} {\bibfnamefont {K.}~\bibnamefont {Wong}}, \bibinfo {author}
  {\bibfnamefont {A.}~\bibnamefont {Adam}}, \bibinfo {author} {\bibfnamefont
  {L.}~\bibnamefont {Perreault-Levasseur}},\ and\ \bibinfo {author}
  {\bibfnamefont {Y.}~\bibnamefont {Hezaveh}},\ }\bibfield  {title} {\bibinfo
  {title} {Towards unbiased gravitational-wave parameter estimation using
  score-based likelihood characterization},\ }in\ \href
  {https://ml4astro.github.io/icml2023/assets/70.pdf} {\emph {\bibinfo
  {booktitle} {2nd ICML Workshop on Machine Learning for Astrophysics}}}\
  (\bibinfo {address} {Honolulu, Hawai'i},\ \bibinfo {year} {2023})\BibitemShut
  {NoStop}%
\bibitem [{\citenamefont {Pathak}\ \emph {et~al.}(2023)\citenamefont {Pathak},
  \citenamefont {Reza},\ and\ \citenamefont {Sengupta}}]{pathak2022rapid}%
  \BibitemOpen
  \bibfield  {author} {\bibinfo {author} {\bibfnamefont {L.}~\bibnamefont
  {Pathak}}, \bibinfo {author} {\bibfnamefont {A.}~\bibnamefont {Reza}},\ and\
  \bibinfo {author} {\bibfnamefont {A.~S.}\ \bibnamefont {Sengupta}},\
  }\bibfield  {title} {\bibinfo {title} {Fast likelihood evaluation using
  meshfree approximations for reconstructing compact binary sources},\ }\href
  {https://doi.org/10.1103/PhysRevD.108.064055} {\bibfield  {journal} {\bibinfo
   {journal} {Phys. Rev. D}\ }\textbf {\bibinfo {volume} {108}},\ \bibinfo
  {pages} {064055} (\bibinfo {year} {2023})}\BibitemShut {NoStop}%
\bibitem [{\citenamefont {Nitz}\ \emph {et~al.}(2022)\citenamefont {Nitz},
  \citenamefont {Harry}, \citenamefont {Brown}, \citenamefont {Biwer},
  \citenamefont {Willis}, \citenamefont {Canton}, \citenamefont {Capano},
  \citenamefont {Dent}, \citenamefont {Pekowsky}, \citenamefont {Williamson}
  \emph {et~al.}}]{alex_nitz_2022_6912865}%
  \BibitemOpen
  \bibfield  {author} {\bibinfo {author} {\bibfnamefont {A.}~\bibnamefont
  {Nitz}}, \bibinfo {author} {\bibfnamefont {I.}~\bibnamefont {Harry}},
  \bibinfo {author} {\bibfnamefont {D.}~\bibnamefont {Brown}}, \bibinfo
  {author} {\bibfnamefont {C.~M.}\ \bibnamefont {Biwer}}, \bibinfo {author}
  {\bibfnamefont {J.}~\bibnamefont {Willis}}, \bibinfo {author} {\bibfnamefont
  {T.~D.}\ \bibnamefont {Canton}}, \bibinfo {author} {\bibfnamefont
  {C.}~\bibnamefont {Capano}}, \bibinfo {author} {\bibfnamefont
  {T.}~\bibnamefont {Dent}}, \bibinfo {author} {\bibfnamefont {L.}~\bibnamefont
  {Pekowsky}}, \bibinfo {author} {\bibfnamefont {A.~R.}\ \bibnamefont
  {Williamson}}, \emph {et~al.},\ }\href
  {https://doi.org/10.5281/zenodo.6912865} {\bibinfo {title} {gwastro/pycbc:
  v2.0.5 release of {PyCBC}}} (\bibinfo {year} {2022})\BibitemShut {NoStop}%
\bibitem [{\citenamefont {Foreman-Mackey}\ \emph {et~al.}(2013)\citenamefont
  {Foreman-Mackey}, \citenamefont {Hogg}, \citenamefont {Lang},\ and\
  \citenamefont {Goodman}}]{Foreman_Mackey_2013}%
  \BibitemOpen
  \bibfield  {author} {\bibinfo {author} {\bibfnamefont {D.}~\bibnamefont
  {Foreman-Mackey}}, \bibinfo {author} {\bibfnamefont {D.~W.}\ \bibnamefont
  {Hogg}}, \bibinfo {author} {\bibfnamefont {D.}~\bibnamefont {Lang}},\ and\
  \bibinfo {author} {\bibfnamefont {J.}~\bibnamefont {Goodman}},\ }\bibfield
  {title} {\bibinfo {title} {{\texttt{emcee}}: The {MCMC} {H}ammer},\ }\href
  {https://doi.org/10.1086/670067} {\bibfield  {journal} {\bibinfo  {journal}
  {Publ. Astron. Soc. Pac.}\ }\textbf {\bibinfo {volume} {125}},\ \bibinfo
  {pages} {306} (\bibinfo {year} {2013})}\BibitemShut {NoStop}%
\bibitem [{\citenamefont {Skilling}(2006)}]{skilling2006nested}%
  \BibitemOpen
  \bibfield  {author} {\bibinfo {author} {\bibfnamefont {J.}~\bibnamefont
  {Skilling}},\ }\bibfield  {title} {\bibinfo {title} {{Nested sampling for
  general Bayesian computation}},\ }\href {https://doi.org/10.1214/06-BA127}
  {\bibfield  {journal} {\bibinfo  {journal} {{Bayesian Anal.}}\ }\textbf
  {\bibinfo {volume} {1}},\ \bibinfo {pages} {833} (\bibinfo {year}
  {2006})}\BibitemShut {NoStop}%
\bibitem [{\citenamefont {Speagle}(2020)}]{speagle2020dynesty}%
  \BibitemOpen
  \bibfield  {author} {\bibinfo {author} {\bibfnamefont {J.~S.}\ \bibnamefont
  {Speagle}},\ }\bibfield  {title} {\bibinfo {title} {{\sc{DYNESTY}}: a dynamic
  nested sampling package for estimating {Bayesian} posteriors and evidences},\
  }\href {https://doi.org/10.1093/mnras/staa278} {\bibfield  {journal}
  {\bibinfo  {journal} {Mon. Not. R. Astron. Soc.}\ }\textbf {\bibinfo {volume}
  {493}},\ \bibinfo {pages} {3132} (\bibinfo {year} {2020})}\BibitemShut
  {NoStop}%
\bibitem [{\citenamefont {Koposov}\ \emph {et~al.}(2023)\citenamefont
  {Koposov}, \citenamefont {Speagle}, \citenamefont {Barbary}, \citenamefont
  {Ashton}, \citenamefont {Bennett}, \citenamefont {Buchner}, \citenamefont
  {Scheffler}, \citenamefont {Cook}, \citenamefont {Talbot}, \citenamefont
  {Guillochon} \emph {et~al.}}]{sergey_koposov_2023_7600689}%
  \BibitemOpen
  \bibfield  {author} {\bibinfo {author} {\bibfnamefont {S.}~\bibnamefont
  {Koposov}}, \bibinfo {author} {\bibfnamefont {J.}~\bibnamefont {Speagle}},
  \bibinfo {author} {\bibfnamefont {K.}~\bibnamefont {Barbary}}, \bibinfo
  {author} {\bibfnamefont {G.}~\bibnamefont {Ashton}}, \bibinfo {author}
  {\bibfnamefont {E.}~\bibnamefont {Bennett}}, \bibinfo {author} {\bibfnamefont
  {J.}~\bibnamefont {Buchner}}, \bibinfo {author} {\bibfnamefont
  {C.}~\bibnamefont {Scheffler}}, \bibinfo {author} {\bibfnamefont
  {B.}~\bibnamefont {Cook}}, \bibinfo {author} {\bibfnamefont {C.}~\bibnamefont
  {Talbot}}, \bibinfo {author} {\bibfnamefont {J.}~\bibnamefont {Guillochon}},
  \emph {et~al.},\ }\href {https://doi.org/10.5281/zenodo.7600689} {\bibinfo
  {title} {joshspeagle/dynesty: v2.1.0}} (\bibinfo {year} {2023})\BibitemShut
  {NoStop}%
\bibitem [{\citenamefont {Tiwari}\ \emph {et~al.}(2023)\citenamefont {Tiwari},
  \citenamefont {Hoy}, \citenamefont {Fairhurst},\ and\ \citenamefont
  {MacLeod}}]{varaha}%
  \BibitemOpen
  \bibfield  {author} {\bibinfo {author} {\bibfnamefont {V.}~\bibnamefont
  {Tiwari}}, \bibinfo {author} {\bibfnamefont {C.}~\bibnamefont {Hoy}},
  \bibinfo {author} {\bibfnamefont {S.}~\bibnamefont {Fairhurst}},\ and\
  \bibinfo {author} {\bibfnamefont {D.}~\bibnamefont {MacLeod}},\ }\bibfield
  {title} {\bibinfo {title} {{Fast non-Markovian sampler for estimating
  gravitational-wave posteriors}},\ }\href
  {https://doi.org/10.1103/PhysRevD.108.023001} {\bibfield  {journal} {\bibinfo
   {journal} {Phys. Rev. D}\ }\textbf {\bibinfo {volume} {108}},\ \bibinfo
  {pages} {023001} (\bibinfo {year} {2023})}\BibitemShut {NoStop}%
\bibitem [{\citenamefont {Veitch}\ \emph {et~al.}(2015)\citenamefont {Veitch},
  \citenamefont {Raymond}, \citenamefont {Farr}, \citenamefont {Farr},
  \citenamefont {Graff}, \citenamefont {Vitale}, \citenamefont {Aylott},
  \citenamefont {Blackburn}, \citenamefont {Christensen}, \citenamefont
  {Coughlin} \emph {et~al.}}]{lalinference}%
  \BibitemOpen
  \bibfield  {author} {\bibinfo {author} {\bibfnamefont {J.}~\bibnamefont
  {Veitch}}, \bibinfo {author} {\bibfnamefont {V.}~\bibnamefont {Raymond}},
  \bibinfo {author} {\bibfnamefont {B.}~\bibnamefont {Farr}}, \bibinfo {author}
  {\bibfnamefont {W.}~\bibnamefont {Farr}}, \bibinfo {author} {\bibfnamefont
  {P.}~\bibnamefont {Graff}}, \bibinfo {author} {\bibfnamefont
  {S.}~\bibnamefont {Vitale}}, \bibinfo {author} {\bibfnamefont
  {B.}~\bibnamefont {Aylott}}, \bibinfo {author} {\bibfnamefont
  {K.}~\bibnamefont {Blackburn}}, \bibinfo {author} {\bibfnamefont
  {N.}~\bibnamefont {Christensen}}, \bibinfo {author} {\bibfnamefont
  {M.}~\bibnamefont {Coughlin}}, \emph {et~al.},\ }\bibfield  {title} {\bibinfo
  {title} {Parameter estimation for compact binaries with ground-based
  gravitational-wave observations using the lalinference software library},\
  }\href {https://doi.org/10.1103/PhysRevD.91.042003} {\bibfield  {journal}
  {\bibinfo  {journal} {Phys. Rev. D}\ }\textbf {\bibinfo {volume} {91}},\
  \bibinfo {pages} {042003} (\bibinfo {year} {2015})}\BibitemShut {NoStop}%
\bibitem [{\citenamefont {Whelan}(2013)}]{detectorTensor}%
  \BibitemOpen
  \bibfield  {author} {\bibinfo {author} {\bibfnamefont {J.~T.}\ \bibnamefont
  {Whelan}},\ }\href
  {https://dcc.ligo.org/public/0106/T1300666/003/Whelan_notes.pdf} {\emph
  {\bibinfo {title} {Notes on The Geometry of Gravitational Wave Detection}}},\
  \bibinfo {type} {Tech. Rep.}\ \bibinfo {number} {LIGO-T1300666-v3}\ (\bibinfo
  {year} {2013})\ \bibinfo {note} {notes for lectures at the Banach Centre
  School of Gravitational Waves, Warsaw, Poland, 2013 July 1-5.}\BibitemShut
  {Stop}%
\bibitem [{\citenamefont {Pankow}\ \emph
  {et~al.}(2015{\natexlab{a}})\citenamefont {Pankow}, \citenamefont {Brady},
  \citenamefont {Ochsner},\ and\ \citenamefont
  {O'Shaughnessy}}]{PhysRevD.92.023002}%
  \BibitemOpen
  \bibfield  {author} {\bibinfo {author} {\bibfnamefont {C.}~\bibnamefont
  {Pankow}}, \bibinfo {author} {\bibfnamefont {P.}~\bibnamefont {Brady}},
  \bibinfo {author} {\bibfnamefont {E.}~\bibnamefont {Ochsner}},\ and\ \bibinfo
  {author} {\bibfnamefont {R.}~\bibnamefont {O'Shaughnessy}},\ }\bibfield
  {title} {\bibinfo {title} {Novel scheme for rapid parallel parameter
  estimation of gravitational waves from compact binary coalescences},\ }\href
  {https://doi.org/10.1103/PhysRevD.92.023002} {\bibfield  {journal} {\bibinfo
  {journal} {Phys. Rev. D}\ }\textbf {\bibinfo {volume} {92}},\ \bibinfo
  {pages} {023002} (\bibinfo {year} {2015}{\natexlab{a}})}\BibitemShut
  {NoStop}%
\bibitem [{\citenamefont {Thrane}\ and\ \citenamefont
  {Talbot}(2019)}]{thrane_2019}%
  \BibitemOpen
  \bibfield  {author} {\bibinfo {author} {\bibfnamefont {E.}~\bibnamefont
  {Thrane}}\ and\ \bibinfo {author} {\bibfnamefont {C.}~\bibnamefont
  {Talbot}},\ }\bibfield  {title} {\bibinfo {title} {An introduction to
  {Bayesian} inference in gravitational-wave astronomy: parameter estimation,
  model selection, and hierarchical models},\ }\bibfield  {journal} {\bibinfo
  {journal} {Publications of the Astronomical Society of Australia}\ }\textbf
  {\bibinfo {volume} {36}},\ \href
  {https://doi.org/https://doi.org/10.1017/pasa.2019.2}
  {https://doi.org/10.1017/pasa.2019.2} (\bibinfo {year} {2019})\BibitemShut
  {NoStop}%
\bibitem [{\citenamefont {Cannon}\ \emph {et~al.}(2010)\citenamefont {Cannon},
  \citenamefont {Chapman}, \citenamefont {Hanna}, \citenamefont {Keppel},
  \citenamefont {Searle},\ and\ \citenamefont {Weinstein}}]{GstLAL_2010}%
  \BibitemOpen
  \bibfield  {author} {\bibinfo {author} {\bibfnamefont {K.}~\bibnamefont
  {Cannon}}, \bibinfo {author} {\bibfnamefont {A.}~\bibnamefont {Chapman}},
  \bibinfo {author} {\bibfnamefont {C.}~\bibnamefont {Hanna}}, \bibinfo
  {author} {\bibfnamefont {D.}~\bibnamefont {Keppel}}, \bibinfo {author}
  {\bibfnamefont {A.~C.}\ \bibnamefont {Searle}},\ and\ \bibinfo {author}
  {\bibfnamefont {A.~J.}\ \bibnamefont {Weinstein}},\ }\bibfield  {title}
  {\bibinfo {title} {Singular value decomposition applied to compact binary
  coalescence gravitational-wave signals},\ }\href
  {https://doi.org/10.1103/PhysRevD.82.044025} {\bibfield  {journal} {\bibinfo
  {journal} {Phys. Rev. D}\ }\textbf {\bibinfo {volume} {82}},\ \bibinfo
  {pages} {044025} (\bibinfo {year} {2010})}\BibitemShut {NoStop}%
\bibitem [{\citenamefont {Usman}\ \emph {et~al.}(2016)\citenamefont {Usman},
  \citenamefont {Nitz}, \citenamefont {Harry}, \citenamefont {Biwer},
  \citenamefont {Brown}, \citenamefont {Cabero}, \citenamefont {Capano},
  \citenamefont {Dal~Canton}, \citenamefont {Dent}, \citenamefont {Fairhurst}
  \emph {et~al.}}]{usman2016pycbc}%
  \BibitemOpen
  \bibfield  {author} {\bibinfo {author} {\bibfnamefont {S.~A.}\ \bibnamefont
  {Usman}}, \bibinfo {author} {\bibfnamefont {A.~H.}\ \bibnamefont {Nitz}},
  \bibinfo {author} {\bibfnamefont {I.~W.}\ \bibnamefont {Harry}}, \bibinfo
  {author} {\bibfnamefont {C.~M.}\ \bibnamefont {Biwer}}, \bibinfo {author}
  {\bibfnamefont {D.~A.}\ \bibnamefont {Brown}}, \bibinfo {author}
  {\bibfnamefont {M.}~\bibnamefont {Cabero}}, \bibinfo {author} {\bibfnamefont
  {C.~D.}\ \bibnamefont {Capano}}, \bibinfo {author} {\bibfnamefont
  {T.}~\bibnamefont {Dal~Canton}}, \bibinfo {author} {\bibfnamefont
  {T.}~\bibnamefont {Dent}}, \bibinfo {author} {\bibfnamefont {S.}~\bibnamefont
  {Fairhurst}}, \emph {et~al.},\ }\bibfield  {title} {\bibinfo {title} {The
  {PyCBC} search for gravitational waves from compact binary coalescence},\
  }\href {https://doi.org/10.1088/0264-9381/33/21/215004} {\bibfield  {journal}
  {\bibinfo  {journal} {Class. Quantum Gravity}\ }\textbf {\bibinfo {volume}
  {33}},\ \bibinfo {pages} {215004} (\bibinfo {year} {2016})}\BibitemShut
  {NoStop}%
\bibitem [{\citenamefont {Roy}\ \emph {et~al.}(2019)\citenamefont {Roy},
  \citenamefont {Sengupta},\ and\ \citenamefont {Ajith}}]{temp_bank_soumen}%
  \BibitemOpen
  \bibfield  {author} {\bibinfo {author} {\bibfnamefont {S.}~\bibnamefont
  {Roy}}, \bibinfo {author} {\bibfnamefont {A.~S.}\ \bibnamefont {Sengupta}},\
  and\ \bibinfo {author} {\bibfnamefont {P.}~\bibnamefont {Ajith}},\ }\bibfield
   {title} {\bibinfo {title} {Effectual template banks for upcoming compact
  binary searches in advanced-ligo and virgo data},\ }\href
  {https://doi.org/10.1103/PhysRevD.99.024048} {\bibfield  {journal} {\bibinfo
  {journal} {Phys. Rev. D}\ }\textbf {\bibinfo {volume} {99}},\ \bibinfo
  {pages} {024048} (\bibinfo {year} {2019})}\BibitemShut {NoStop}%
\bibitem [{\citenamefont {Mukherjee}\ \emph {et~al.}(2021)\citenamefont
  {Mukherjee}, \citenamefont {Caudill}, \citenamefont {Magee}, \citenamefont
  {Messick}, \citenamefont {Privitera}, \citenamefont {Sachdev}, \citenamefont
  {Blackburn}, \citenamefont {Brady}, \citenamefont {Brockill}, \citenamefont
  {Cannon} \emph {et~al.}}]{PhysRevD.103.084047}%
  \BibitemOpen
  \bibfield  {author} {\bibinfo {author} {\bibfnamefont {D.}~\bibnamefont
  {Mukherjee}}, \bibinfo {author} {\bibfnamefont {S.}~\bibnamefont {Caudill}},
  \bibinfo {author} {\bibfnamefont {R.}~\bibnamefont {Magee}}, \bibinfo
  {author} {\bibfnamefont {C.}~\bibnamefont {Messick}}, \bibinfo {author}
  {\bibfnamefont {S.}~\bibnamefont {Privitera}}, \bibinfo {author}
  {\bibfnamefont {S.}~\bibnamefont {Sachdev}}, \bibinfo {author} {\bibfnamefont
  {K.}~\bibnamefont {Blackburn}}, \bibinfo {author} {\bibfnamefont
  {P.}~\bibnamefont {Brady}}, \bibinfo {author} {\bibfnamefont
  {P.}~\bibnamefont {Brockill}}, \bibinfo {author} {\bibfnamefont
  {K.}~\bibnamefont {Cannon}}, \emph {et~al.},\ }\bibfield  {title} {\bibinfo
  {title} {Template bank for spinning compact binary mergers in the second
  observation run of advanced ligo and the first observation run of advanced
  virgo},\ }\href {https://doi.org/10.1103/PhysRevD.103.084047} {\bibfield
  {journal} {\bibinfo  {journal} {Phys. Rev. D}\ }\textbf {\bibinfo {volume}
  {103}},\ \bibinfo {pages} {084047} (\bibinfo {year} {2021})}\BibitemShut
  {NoStop}%
\bibitem [{\citenamefont {Virtanen}\ \emph {et~al.}(2020)\citenamefont
  {Virtanen}, \citenamefont {Gommers}, \citenamefont {Oliphant}, \citenamefont
  {Haberland}, \citenamefont {Reddy}, \citenamefont {Cournapeau}, \citenamefont
  {Burovski}, \citenamefont {Peterson}, \citenamefont {Weckesser},
  \citenamefont {Bright} \emph {et~al.}}]{2020SciPy-NMeth}%
  \BibitemOpen
  \bibfield  {author} {\bibinfo {author} {\bibfnamefont {P.}~\bibnamefont
  {Virtanen}}, \bibinfo {author} {\bibfnamefont {R.}~\bibnamefont {Gommers}},
  \bibinfo {author} {\bibfnamefont {T.~E.}\ \bibnamefont {Oliphant}}, \bibinfo
  {author} {\bibfnamefont {M.}~\bibnamefont {Haberland}}, \bibinfo {author}
  {\bibfnamefont {T.}~\bibnamefont {Reddy}}, \bibinfo {author} {\bibfnamefont
  {D.}~\bibnamefont {Cournapeau}}, \bibinfo {author} {\bibfnamefont
  {E.}~\bibnamefont {Burovski}}, \bibinfo {author} {\bibfnamefont
  {P.}~\bibnamefont {Peterson}}, \bibinfo {author} {\bibfnamefont
  {W.}~\bibnamefont {Weckesser}}, \bibinfo {author} {\bibfnamefont
  {J.}~\bibnamefont {Bright}}, \emph {et~al.},\ }\bibfield  {title} {\bibinfo
  {title} {{{SciPy} 1.0: Fundamental Algorithms for Scientific Computing in
  Python}},\ }\href {https://doi.org/10.1038/s41592-019-0686-2} {\bibfield
  {journal} {\bibinfo  {journal} {Nature Methods}\ }\textbf {\bibinfo {volume}
  {17}},\ \bibinfo {pages} {261} (\bibinfo {year} {2020})}\BibitemShut
  {NoStop}%
\bibitem [{\citenamefont {Fasshauer}(2007)}]{doi:10.1142/6437}%
  \BibitemOpen
  \bibfield  {author} {\bibinfo {author} {\bibfnamefont {G.~E.}\ \bibnamefont
  {Fasshauer}},\ }\href {https://doi.org/https://doi.org/10.1142/6437} {\emph
  {\bibinfo {title} {{Meshfree Approximation Methods with Matlab}}}}\ (\bibinfo
   {publisher} {World Scientific},\ \bibinfo {year} {2007})\BibitemShut
  {NoStop}%
\bibitem [{\citenamefont {{Flyer}}\ \emph {et~al.}(2016)\citenamefont
  {{Flyer}}, \citenamefont {{Fornberg}}, \citenamefont {{Bayona}},\ and\
  \citenamefont {{Barnett}}}]{2016JCoPh}%
  \BibitemOpen
  \bibfield  {author} {\bibinfo {author} {\bibfnamefont {N.}~\bibnamefont
  {{Flyer}}}, \bibinfo {author} {\bibfnamefont {B.}~\bibnamefont {{Fornberg}}},
  \bibinfo {author} {\bibfnamefont {V.}~\bibnamefont {{Bayona}}},\ and\
  \bibinfo {author} {\bibfnamefont {G.~A.}\ \bibnamefont {{Barnett}}},\
  }\bibfield  {title} {\bibinfo {title} {{On the role of polynomials in RBF-FD
  approximations: I. Interpolation and accuracy}},\ }\href
  {https://doi.org/10.1016/j.jcp.2016.05.026} {\bibfield  {journal} {\bibinfo
  {journal} {Journal of Computational Physics}\ }\textbf {\bibinfo {volume}
  {321}},\ \bibinfo {pages} {21} (\bibinfo {year} {2016})}\BibitemShut
  {NoStop}%
\bibitem [{\citenamefont {{LIGO Scientific Collaboration and Virgo
  Collaboration}}(2017)}]{GW170817data}%
  \BibitemOpen
  \bibfield  {author} {\bibinfo {author} {\bibnamefont {{LIGO Scientific
  Collaboration and Virgo Collaboration}}},\ }\href
  {https://doi.org/https://doi.org/10.7935/K5B8566F} {\bibinfo {title} {{Data
  release for event GW170817}}} (\bibinfo {year} {2017}),\ \bibinfo {note}
  {accessed on 2023-04-20}\BibitemShut {NoStop}%
\bibitem [{\citenamefont {Abbott}\ \emph
  {et~al.}(2021{\natexlab{a}})\citenamefont {Abbott} \emph
  {et~al.}}]{RICHABBOTT2021100658}%
  \BibitemOpen
  \bibfield  {author} {\bibinfo {author} {\bibfnamefont {R.}~\bibnamefont
  {Abbott}} \emph {et~al.} (\bibinfo {collaboration} {{LIGO Scientific and
  Virgo Collaboration}}),\ }\bibfield  {title} {\bibinfo {title} {{Open data
  from the first and second observing runs of Advanced LIGO and Advanced
  Virgo}},\ }\href
  {https://doi.org/https://doi.org/10.1016/j.softx.2021.100658} {\bibfield
  {journal} {\bibinfo  {journal} {SoftwareX}\ }\textbf {\bibinfo {volume}
  {13}},\ \bibinfo {pages} {100658} (\bibinfo {year}
  {2021}{\natexlab{a}})}\BibitemShut {NoStop}%
\bibitem [{\citenamefont {Dhurandhar}\ and\ \citenamefont
  {Sathyaprakash}(1994)}]{PhysRevD.49.1707}%
  \BibitemOpen
  \bibfield  {author} {\bibinfo {author} {\bibfnamefont {S.~V.}\ \bibnamefont
  {Dhurandhar}}\ and\ \bibinfo {author} {\bibfnamefont {B.~S.}\ \bibnamefont
  {Sathyaprakash}},\ }\bibfield  {title} {\bibinfo {title} {Choice of filters
  for the detection of gravitational waves from coalescing binaries. ii.
  detection in colored noise},\ }\href
  {https://doi.org/10.1103/PhysRevD.49.1707} {\bibfield  {journal} {\bibinfo
  {journal} {Phys. Rev. D}\ }\textbf {\bibinfo {volume} {49}},\ \bibinfo
  {pages} {1707} (\bibinfo {year} {1994})}\BibitemShut {NoStop}%
\bibitem [{\citenamefont {Droz}\ \emph {et~al.}(1999)\citenamefont {Droz},
  \citenamefont {Knapp}, \citenamefont {Poisson},\ and\ \citenamefont
  {Owen}}]{PhysRevD.59.124016}%
  \BibitemOpen
  \bibfield  {author} {\bibinfo {author} {\bibfnamefont {S.}~\bibnamefont
  {Droz}}, \bibinfo {author} {\bibfnamefont {D.~J.}\ \bibnamefont {Knapp}},
  \bibinfo {author} {\bibfnamefont {E.}~\bibnamefont {Poisson}},\ and\ \bibinfo
  {author} {\bibfnamefont {B.~J.}\ \bibnamefont {Owen}},\ }\bibfield  {title}
  {\bibinfo {title} {{Gravitational waves from inspiraling compact binaries:
  Validity of the stationary-phase approximation to the Fourier transform}},\
  }\href {https://doi.org/10.1103/PhysRevD.59.124016} {\bibfield  {journal}
  {\bibinfo  {journal} {Phys. Rev. D}\ }\textbf {\bibinfo {volume} {59}},\
  \bibinfo {pages} {124016} (\bibinfo {year} {1999})}\BibitemShut {NoStop}%
\bibitem [{\citenamefont {Faye}\ \emph {et~al.}(2012)\citenamefont {Faye},
  \citenamefont {Marsat}, \citenamefont {Blanchet},\ and\ \citenamefont
  {Iyer}}]{Faye_2012}%
  \BibitemOpen
  \bibfield  {author} {\bibinfo {author} {\bibfnamefont {G.}~\bibnamefont
  {Faye}}, \bibinfo {author} {\bibfnamefont {S.}~\bibnamefont {Marsat}},
  \bibinfo {author} {\bibfnamefont {L.}~\bibnamefont {Blanchet}},\ and\
  \bibinfo {author} {\bibfnamefont {B.~R.}\ \bibnamefont {Iyer}},\ }\bibfield
  {title} {\bibinfo {title} {The third and a half-post-newtonian gravitational
  wave quadrupole mode for quasi-circular inspiralling compact binaries},\
  }\href {https://doi.org/10.1088/0264-9381/29/17/175004} {\bibfield  {journal}
  {\bibinfo  {journal} {Classical and Quantum Gravity}\ }\textbf {\bibinfo
  {volume} {29}},\ \bibinfo {pages} {175004} (\bibinfo {year}
  {2012})}\BibitemShut {NoStop}%
\bibitem [{\citenamefont {Blanchet}\ \emph {et~al.}(1995)\citenamefont
  {Blanchet}, \citenamefont {Damour}, \citenamefont {Iyer}, \citenamefont
  {Will},\ and\ \citenamefont {Wiseman}}]{PhysRevLett.74.3515}%
  \BibitemOpen
  \bibfield  {author} {\bibinfo {author} {\bibfnamefont {L.}~\bibnamefont
  {Blanchet}}, \bibinfo {author} {\bibfnamefont {T.}~\bibnamefont {Damour}},
  \bibinfo {author} {\bibfnamefont {B.~R.}\ \bibnamefont {Iyer}}, \bibinfo
  {author} {\bibfnamefont {C.~M.}\ \bibnamefont {Will}},\ and\ \bibinfo
  {author} {\bibfnamefont {A.~G.}\ \bibnamefont {Wiseman}},\ }\bibfield
  {title} {\bibinfo {title} {Gravitational-radiation damping of compact binary
  systems to second post-newtonian order},\ }\href
  {https://doi.org/10.1103/PhysRevLett.74.3515} {\bibfield  {journal} {\bibinfo
   {journal} {Phys. Rev. Lett.}\ }\textbf {\bibinfo {volume} {74}},\ \bibinfo
  {pages} {3515} (\bibinfo {year} {1995})}\BibitemShut {NoStop}%
\bibitem [{\citenamefont {Vallisneri}(2008)}]{vallisneri2008use}%
  \BibitemOpen
  \bibfield  {author} {\bibinfo {author} {\bibfnamefont {M.}~\bibnamefont
  {Vallisneri}},\ }\bibfield  {title} {\bibinfo {title} {Use and abuse of the
  {Fisher} information matrix in the assessment of gravitational-wave
  parameter-estimation prospects},\ }\href
  {https://doi.org/10.1103/PhysRevD.77.042001} {\bibfield  {journal} {\bibinfo
  {journal} {Phys. Rev. D}\ }\textbf {\bibinfo {volume} {77}},\ \bibinfo
  {pages} {042001} (\bibinfo {year} {2008})}\BibitemShut {NoStop}%
\bibitem [{\citenamefont {Iacovelli}\ \emph {et~al.}(2022)\citenamefont
  {Iacovelli}, \citenamefont {Mancarella}, \citenamefont {Foffa},\ and\
  \citenamefont {Maggiore}}]{Iacovelli_2022}%
  \BibitemOpen
  \bibfield  {author} {\bibinfo {author} {\bibfnamefont {F.}~\bibnamefont
  {Iacovelli}}, \bibinfo {author} {\bibfnamefont {M.}~\bibnamefont
  {Mancarella}}, \bibinfo {author} {\bibfnamefont {S.}~\bibnamefont {Foffa}},\
  and\ \bibinfo {author} {\bibfnamefont {M.}~\bibnamefont {Maggiore}},\
  }\bibfield  {title} {\bibinfo {title} {{GWFAST}: A fisher information matrix
  python code for third-generation gravitational-wave detectors},\ }\href
  {https://doi.org/10.3847/1538-4365/ac9129} {\bibfield  {journal} {\bibinfo
  {journal} {The Astrophysical Journal Supplement Series}\ }\textbf {\bibinfo
  {volume} {263}},\ \bibinfo {pages} {2} (\bibinfo {year} {2022})}\BibitemShut
  {NoStop}%
\bibitem [{\citenamefont {Hines}(2015)}]{RBF_github}%
  \BibitemOpen
  \bibfield  {author} {\bibinfo {author} {\bibfnamefont {T.}~\bibnamefont
  {Hines}},\ }\href {https://github.com/treverhines/RBF.git} {\bibinfo {title}
  {Python package containing the tools necessary for radial basis function
  {(RBF)} applications}} (\bibinfo {year} {2015})\BibitemShut {NoStop}%
\bibitem [{\citenamefont {Rocha}(2009)}]{ROCHA20091573}%
  \BibitemOpen
  \bibfield  {author} {\bibinfo {author} {\bibfnamefont {H.}~\bibnamefont
  {Rocha}},\ }\bibfield  {title} {\bibinfo {title} {On the selection of the
  most adequate radial basis function},\ }\href
  {https://doi.org/https://doi.org/10.1016/j.apm.2008.02.008} {\bibfield
  {journal} {\bibinfo  {journal} {Applied Mathematical Modelling}\ }\textbf
  {\bibinfo {volume} {33}},\ \bibinfo {pages} {1573} (\bibinfo {year}
  {2009})}\BibitemShut {NoStop}%
\bibitem [{\citenamefont {Khan}\ \emph {et~al.}(2016)\citenamefont {Khan},
  \citenamefont {Husa}, \citenamefont {Hannam}, \citenamefont {Ohme},
  \citenamefont {P{\"u}rrer}, \citenamefont {Forteza},\ and\ \citenamefont
  {Boh{\'e}}}]{khan2016frequency}%
  \BibitemOpen
  \bibfield  {author} {\bibinfo {author} {\bibfnamefont {S.}~\bibnamefont
  {Khan}}, \bibinfo {author} {\bibfnamefont {S.}~\bibnamefont {Husa}}, \bibinfo
  {author} {\bibfnamefont {M.}~\bibnamefont {Hannam}}, \bibinfo {author}
  {\bibfnamefont {F.}~\bibnamefont {Ohme}}, \bibinfo {author} {\bibfnamefont
  {M.}~\bibnamefont {P{\"u}rrer}}, \bibinfo {author} {\bibfnamefont {X.~J.}\
  \bibnamefont {Forteza}},\ and\ \bibinfo {author} {\bibfnamefont
  {A.}~\bibnamefont {Boh{\'e}}},\ }\bibfield  {title} {\bibinfo {title}
  {{Frequency-domain gravitational waves from nonprecessing black-hole
  binaries. {II.} A phenomenological model for the advanced detector era}},\
  }\href {https://doi.org/10.1103/PhysRevD.93.044007} {\bibfield  {journal}
  {\bibinfo  {journal} {Phys. Rev. D}\ }\textbf {\bibinfo {volume} {93}},\
  \bibinfo {pages} {044007} (\bibinfo {year} {2016})}\BibitemShut {NoStop}%
\bibitem [{\citenamefont {Abbott}\ \emph {et~al.}(2019)\citenamefont {Abbott},
  \citenamefont {Abbott}, \citenamefont {Abbott}, \citenamefont {Acernese},
  \citenamefont {Ackley}, \citenamefont {Adams}, \citenamefont {Adams},
  \citenamefont {Addesso}, \citenamefont {Adhikari}, \citenamefont {Adya} \emph
  {et~al.}}]{PhysRevX.9.011001}%
  \BibitemOpen
  \bibfield  {author} {\bibinfo {author} {\bibfnamefont {B.~P.}\ \bibnamefont
  {Abbott}}, \bibinfo {author} {\bibfnamefont {R.}~\bibnamefont {Abbott}},
  \bibinfo {author} {\bibfnamefont {T.~D.}\ \bibnamefont {Abbott}}, \bibinfo
  {author} {\bibfnamefont {F.}~\bibnamefont {Acernese}}, \bibinfo {author}
  {\bibfnamefont {K.}~\bibnamefont {Ackley}}, \bibinfo {author} {\bibfnamefont
  {C.}~\bibnamefont {Adams}}, \bibinfo {author} {\bibfnamefont
  {T.}~\bibnamefont {Adams}}, \bibinfo {author} {\bibfnamefont
  {P.}~\bibnamefont {Addesso}}, \bibinfo {author} {\bibfnamefont {R.~X.}\
  \bibnamefont {Adhikari}}, \bibinfo {author} {\bibfnamefont {V.~B.}\
  \bibnamefont {Adya}}, \emph {et~al.} (\bibinfo {collaboration} {LIGO
  Scientific Collaboration and Virgo Collaboration}),\ }\bibfield  {title}
  {\bibinfo {title} {Properties of the binary neutron star merger gw170817},\
  }\href {https://doi.org/10.1103/PhysRevX.9.011001} {\bibfield  {journal}
  {\bibinfo  {journal} {Phys. Rev. X}\ }\textbf {\bibinfo {volume} {9}},\
  \bibinfo {pages} {011001} (\bibinfo {year} {2019})}\BibitemShut {NoStop}%
\bibitem [{\citenamefont {Singer}\ and\ \citenamefont
  {Price}(2016{\natexlab{b}})}]{ligo_sky_map}%
  \BibitemOpen
  \bibfield  {author} {\bibinfo {author} {\bibfnamefont {L.~P.}\ \bibnamefont
  {Singer}}\ and\ \bibinfo {author} {\bibfnamefont {L.~R.}\ \bibnamefont
  {Price}},\ }\bibfield  {title} {\bibinfo {title} {Rapid bayesian position
  reconstruction for gravitational-wave transients},\ }\href
  {https://doi.org/10.1103/PhysRevD.93.024013} {\bibfield  {journal} {\bibinfo
  {journal} {Phys. Rev. D}\ }\textbf {\bibinfo {volume} {93}},\ \bibinfo
  {pages} {024013} (\bibinfo {year} {2016}{\natexlab{b}})}\BibitemShut
  {NoStop}%
\bibitem [{\citenamefont {Barsotti}\ \emph {et~al.}(2018)\citenamefont
  {Barsotti}, \citenamefont {Gras}, \citenamefont {Evans},\ and\ \citenamefont
  {Fritschel}}]{aLIGO_ZDHP}%
  \BibitemOpen
  \bibfield  {author} {\bibinfo {author} {\bibfnamefont {L.}~\bibnamefont
  {Barsotti}}, \bibinfo {author} {\bibfnamefont {S.}~\bibnamefont {Gras}},
  \bibinfo {author} {\bibfnamefont {M.}~\bibnamefont {Evans}},\ and\ \bibinfo
  {author} {\bibfnamefont {P.}~\bibnamefont {Fritschel}},\ }\href
  {https://dcc.ligo.org/LIGO-T1800044/public} {\emph {\bibinfo {title} {{The
  updated Advanced LIGO design curve}}}},\ \bibinfo {type} {Tech. Rep.}\
  \bibinfo {number} {LIGO-T1800044-v5}\ (\bibinfo  {institution} {LIGO
  Scientific Collaboration},\ \bibinfo {year} {2018})\BibitemShut {NoStop}%
\bibitem [{\citenamefont {Manzotti}\ and\ \citenamefont
  {Dietz}(2012)}]{AdVirgo}%
  \BibitemOpen
  \bibfield  {author} {\bibinfo {author} {\bibfnamefont {A.}~\bibnamefont
  {Manzotti}}\ and\ \bibinfo {author} {\bibfnamefont {A.}~\bibnamefont
  {Dietz}},\ }\href@noop {} {\bibinfo {title} {Prospects for early localization
  of gravitational-wave signals from compact binary coalescences with advanced
  detectors}} (\bibinfo {year} {2012}),\ \Eprint
  {https://arxiv.org/abs/1202.4031} {arXiv:1202.4031 [gr-qc]} \BibitemShut
  {NoStop}%
\bibitem [{\citenamefont {Gerber}\ and\ \citenamefont
  {Furrer}(2019)}]{RJ-2019-030}%
  \BibitemOpen
  \bibfield  {author} {\bibinfo {author} {\bibfnamefont {F.}~\bibnamefont
  {Gerber}}\ and\ \bibinfo {author} {\bibfnamefont {R.}~\bibnamefont
  {Furrer}},\ }\bibfield  {title} {\bibinfo {title} {{optimParallel: An R
  Package Providing a Parallel Version of the L-BFGS-B Optimization Method}},\
  }\href {https://doi.org/10.32614/RJ-2019-030} {\bibfield  {journal} {\bibinfo
   {journal} {{The R Journal}}\ }\textbf {\bibinfo {volume} {11}},\ \bibinfo
  {pages} {352} (\bibinfo {year} {2019})}\BibitemShut {NoStop}%
\bibitem [{\citenamefont {Gerber}(2020)}]{florian_gerber_2020_3888570}%
  \BibitemOpen
  \bibfield  {author} {\bibinfo {author} {\bibfnamefont {F.}~\bibnamefont
  {Gerber}},\ }\href {https://doi.org/10.5281/zenodo.3888570} {\bibinfo {title}
  {florafauna/optimparallel-python: test zenodo}} (\bibinfo {year}
  {2020})\BibitemShut {NoStop}%
\bibitem [{\citenamefont {Abbott}\ \emph
  {et~al.}(2021{\natexlab{b}})\citenamefont {Abbott}, \citenamefont {Abbott},
  \citenamefont {Abraham}, \citenamefont {Acernese}, \citenamefont {Ackley},
  \citenamefont {Adams}, \citenamefont {Adams}, \citenamefont {Adhikari},
  \citenamefont {Adya}, \citenamefont {Affeldt} \emph {et~al.}}]{NSBH_GW20015}%
  \BibitemOpen
  \bibfield  {author} {\bibinfo {author} {\bibfnamefont {R.}~\bibnamefont
  {Abbott}}, \bibinfo {author} {\bibfnamefont {T.~D.}\ \bibnamefont {Abbott}},
  \bibinfo {author} {\bibfnamefont {S.}~\bibnamefont {Abraham}}, \bibinfo
  {author} {\bibfnamefont {F.}~\bibnamefont {Acernese}}, \bibinfo {author}
  {\bibfnamefont {K.}~\bibnamefont {Ackley}}, \bibinfo {author} {\bibfnamefont
  {A.}~\bibnamefont {Adams}}, \bibinfo {author} {\bibfnamefont
  {C.}~\bibnamefont {Adams}}, \bibinfo {author} {\bibfnamefont {R.~X.}\
  \bibnamefont {Adhikari}}, \bibinfo {author} {\bibfnamefont {V.~B.}\
  \bibnamefont {Adya}}, \bibinfo {author} {\bibfnamefont {C.}~\bibnamefont
  {Affeldt}}, \emph {et~al.},\ }\bibfield  {title} {\bibinfo {title}
  {Observation of gravitational waves from two neutron star{\textendash}black
  hole coalescences},\ }\href {https://doi.org/10.3847/2041-8213/ac082e}
  {\bibfield  {journal} {\bibinfo  {journal} {The Astrophysical Journal
  Letters}\ }\textbf {\bibinfo {volume} {915}},\ \bibinfo {pages} {L5}
  (\bibinfo {year} {2021}{\natexlab{b}})}\BibitemShut {NoStop}%
\bibitem [{\citenamefont {Borne}\ and\ \citenamefont
  {Wende}(2019)}]{Borne2019DomainDM}%
  \BibitemOpen
  \bibfield  {author} {\bibinfo {author} {\bibfnamefont {S.~L.}\ \bibnamefont
  {Borne}}\ and\ \bibinfo {author} {\bibfnamefont {M.}~\bibnamefont {Wende}},\
  }\bibfield  {title} {\bibinfo {title} {Domain decomposition methods in
  scattered data interpolation with conditionally positive definite radial
  basis functions},\ }\href@noop {} {\bibfield  {journal} {\bibinfo  {journal}
  {Comput. Math. Appl.}\ }\textbf {\bibinfo {volume} {77}},\ \bibinfo {pages}
  {1178} (\bibinfo {year} {2019})}\BibitemShut {NoStop}%
\bibitem [{\citenamefont {Pankow}\ \emph
  {et~al.}(2015{\natexlab{b}})\citenamefont {Pankow}, \citenamefont {Brady},
  \citenamefont {Ochsner},\ and\ \citenamefont
  {O’Shaughnessy}}]{Pankow_2015}%
  \BibitemOpen
  \bibfield  {author} {\bibinfo {author} {\bibfnamefont {C.}~\bibnamefont
  {Pankow}}, \bibinfo {author} {\bibfnamefont {P.}~\bibnamefont {Brady}},
  \bibinfo {author} {\bibfnamefont {E.}~\bibnamefont {Ochsner}},\ and\ \bibinfo
  {author} {\bibfnamefont {R.}~\bibnamefont {O’Shaughnessy}},\ }\bibfield
  {title} {\bibinfo {title} {Novel scheme for rapid parallel parameter
  estimation of gravitational waves from compact binary coalescences},\
  }\bibfield  {journal} {\bibinfo  {journal} {Physical Review D}\ }\textbf
  {\bibinfo {volume} {92}},\ \href {https://doi.org/10.1103/physrevd.92.023002}
  {10.1103/physrevd.92.023002} (\bibinfo {year}
  {2015}{\natexlab{b}})\BibitemShut {NoStop}%
\bibitem [{\citenamefont {Zhang}\ \emph {et~al.}(2017)\citenamefont {Zhang},
  \citenamefont {Zhao},\ and\ \citenamefont {Levesley}}]{Zhang:2017aa}%
  \BibitemOpen
  \bibfield  {author} {\bibinfo {author} {\bibfnamefont {Q.}~\bibnamefont
  {Zhang}}, \bibinfo {author} {\bibfnamefont {Y.}~\bibnamefont {Zhao}},\ and\
  \bibinfo {author} {\bibfnamefont {J.}~\bibnamefont {Levesley}},\ }\bibfield
  {title} {\bibinfo {title} {Adaptive radial basis function interpolation using
  an error indicator},\ }\href {https://doi.org/10.1007/s11075-017-0265-5}
  {\bibfield  {journal} {\bibinfo  {journal} {Numerical Algorithms}\ ,\
  \bibinfo {pages} {441}} (\bibinfo {year} {2017})}\BibitemShut {NoStop}%
\bibitem [{\citenamefont {Del~Pozzo}\ \emph {et~al.}(2013)\citenamefont
  {Del~Pozzo}, \citenamefont {Li}, \citenamefont {Agathos}, \citenamefont {Van
  Den~Broeck},\ and\ \citenamefont {Vitale}}]{DelPozzo:2013ala}%
  \BibitemOpen
  \bibfield  {author} {\bibinfo {author} {\bibfnamefont {W.}~\bibnamefont
  {Del~Pozzo}}, \bibinfo {author} {\bibfnamefont {T.~G.~F.}\ \bibnamefont
  {Li}}, \bibinfo {author} {\bibfnamefont {M.}~\bibnamefont {Agathos}},
  \bibinfo {author} {\bibfnamefont {C.}~\bibnamefont {Van Den~Broeck}},\ and\
  \bibinfo {author} {\bibfnamefont {S.}~\bibnamefont {Vitale}},\ }\bibfield
  {title} {\bibinfo {title} {{Demonstrating the feasibility of probing the
  neutron star equation of state with second-generation gravitational wave
  detectors}},\ }\href {https://doi.org/10.1103/PhysRevLett.111.071101}
  {\bibfield  {journal} {\bibinfo  {journal} {Phys. Rev. Lett.}\ }\textbf
  {\bibinfo {volume} {111}},\ \bibinfo {pages} {071101} (\bibinfo {year}
  {2013})},\ \Eprint {https://arxiv.org/abs/1307.8338} {arXiv:1307.8338
  [gr-qc]} \BibitemShut {NoStop}%
\bibitem [{\citenamefont {Dietrich}\ \emph {et~al.}(2019)\citenamefont
  {Dietrich}, \citenamefont {Samajdar}, \citenamefont {Khan}, \citenamefont
  {Johnson-McDaniel}, \citenamefont {Dudi},\ and\ \citenamefont
  {Tichy}}]{Dietrich_2019}%
  \BibitemOpen
  \bibfield  {author} {\bibinfo {author} {\bibfnamefont {T.}~\bibnamefont
  {Dietrich}}, \bibinfo {author} {\bibfnamefont {A.}~\bibnamefont {Samajdar}},
  \bibinfo {author} {\bibfnamefont {S.}~\bibnamefont {Khan}}, \bibinfo {author}
  {\bibfnamefont {N.~K.}\ \bibnamefont {Johnson-McDaniel}}, \bibinfo {author}
  {\bibfnamefont {R.}~\bibnamefont {Dudi}},\ and\ \bibinfo {author}
  {\bibfnamefont {W.}~\bibnamefont {Tichy}},\ }\bibfield  {title} {\bibinfo
  {title} {Improving the nrtidal model for binary neutron star systems},\
  }\bibfield  {journal} {\bibinfo  {journal} {Physical Review D}\ }\textbf
  {\bibinfo {volume} {100}},\ \href
  {https://doi.org/10.1103/physrevd.100.044003} {10.1103/physrevd.100.044003}
  (\bibinfo {year} {2019})\BibitemShut {NoStop}%
\bibitem [{\citenamefont {Pratten}\ \emph
  {et~al.}(2021{\natexlab{b}})\citenamefont {Pratten}, \citenamefont
  {Garc{\'{\i} }a-Quir{\'{o}}s}, \citenamefont {Colleoni}, \citenamefont
  {Ramos-Buades}, \citenamefont {Estell{\'{e}}s}, \citenamefont {Mateu-Lucena},
  \citenamefont {Jaume}, \citenamefont {Haney}, \citenamefont {Keitel},
  \citenamefont {Thompson} \emph {et~al.}}]{IMRPhenomXPHM}%
  \BibitemOpen
  \bibfield  {author} {\bibinfo {author} {\bibfnamefont {G.}~\bibnamefont
  {Pratten}}, \bibinfo {author} {\bibfnamefont {C.}~\bibnamefont {Garc{\'{\i}
  }a-Quir{\'{o}}s}}, \bibinfo {author} {\bibfnamefont {M.}~\bibnamefont
  {Colleoni}}, \bibinfo {author} {\bibfnamefont {A.}~\bibnamefont
  {Ramos-Buades}}, \bibinfo {author} {\bibfnamefont {H.}~\bibnamefont
  {Estell{\'{e}}s}}, \bibinfo {author} {\bibfnamefont {M.}~\bibnamefont
  {Mateu-Lucena}}, \bibinfo {author} {\bibfnamefont {R.}~\bibnamefont {Jaume}},
  \bibinfo {author} {\bibfnamefont {M.}~\bibnamefont {Haney}}, \bibinfo
  {author} {\bibfnamefont {D.}~\bibnamefont {Keitel}}, \bibinfo {author}
  {\bibfnamefont {J.~E.}\ \bibnamefont {Thompson}}, \emph {et~al.},\ }\bibfield
   {title} {\bibinfo {title} {Computationally efficient models for the dominant
  and subdominant harmonic modes of precessing binary black holes},\ }\bibfield
   {journal} {\bibinfo  {journal} {Physical Review D}\ }\textbf {\bibinfo
  {volume} {103}},\ \href {https://doi.org/10.1103/physrevd.103.104056}
  {10.1103/physrevd.103.104056} (\bibinfo {year}
  {2021}{\natexlab{b}})\BibitemShut {NoStop}%
\bibitem [{\citenamefont {García-Quirós}\ \emph {et~al.}(2020)\citenamefont
  {García-Quirós}, \citenamefont {Colleoni}, \citenamefont {Husa},
  \citenamefont {Estellés}, \citenamefont {Pratten}, \citenamefont
  {Ramos-Buades}, \citenamefont {Mateu-Lucena},\ and\ \citenamefont
  {Jaume}}]{IMRPhenomXHM}%
  \BibitemOpen
  \bibfield  {author} {\bibinfo {author} {\bibfnamefont {C.}~\bibnamefont
  {García-Quirós}}, \bibinfo {author} {\bibfnamefont {M.}~\bibnamefont
  {Colleoni}}, \bibinfo {author} {\bibfnamefont {S.}~\bibnamefont {Husa}},
  \bibinfo {author} {\bibfnamefont {H.}~\bibnamefont {Estellés}}, \bibinfo
  {author} {\bibfnamefont {G.}~\bibnamefont {Pratten}}, \bibinfo {author}
  {\bibfnamefont {A.}~\bibnamefont {Ramos-Buades}}, \bibinfo {author}
  {\bibfnamefont {M.}~\bibnamefont {Mateu-Lucena}},\ and\ \bibinfo {author}
  {\bibfnamefont {R.}~\bibnamefont {Jaume}},\ }\bibfield  {title} {\bibinfo
  {title} {Multimode frequency-domain model for the gravitational wave signal
  from nonprecessing black-hole binaries},\ }\bibfield  {journal} {\bibinfo
  {journal} {Physical Review D}\ }\textbf {\bibinfo {volume} {102}},\ \href
  {https://doi.org/10.1103/physrevd.102.064002} {10.1103/physrevd.102.064002}
  (\bibinfo {year} {2020})\BibitemShut {NoStop}%
\bibitem [{\citenamefont {Narola}\ \emph {et~al.}(2023)\citenamefont {Narola},
  \citenamefont {Janquart}, \citenamefont {Meijer}, \citenamefont {Haris},\
  and\ \citenamefont {Broeck}}]{narola2023relative}%
  \BibitemOpen
  \bibfield  {author} {\bibinfo {author} {\bibfnamefont {H.}~\bibnamefont
  {Narola}}, \bibinfo {author} {\bibfnamefont {J.}~\bibnamefont {Janquart}},
  \bibinfo {author} {\bibfnamefont {Q.}~\bibnamefont {Meijer}}, \bibinfo
  {author} {\bibfnamefont {K.}~\bibnamefont {Haris}},\ and\ \bibinfo {author}
  {\bibfnamefont {C.~V.~D.}\ \bibnamefont {Broeck}},\ }\href@noop {} {\bibinfo
  {title} {Relative binning for complete gravitational-wave parameter
  estimation with higher-order modes and precession, and applications to
  lensing and third-generation detectors}} (\bibinfo {year} {2023}),\ \Eprint
  {https://arxiv.org/abs/2308.12140} {arXiv:2308.12140 [gr-qc]} \BibitemShut
  {NoStop}%
\bibitem [{gwo()}]{gwosc_web}%
  \BibitemOpen
  \href {https://gwosc.org/} {\bibinfo {title} {gwosc.org}}\BibitemShut
  {NoStop}%
\bibitem [{\citenamefont {{L. Pathak}}(2023)}]{meshfree_github}%
  \BibitemOpen
  \bibfield  {author} {\bibinfo {author} {\bibnamefont {{L. Pathak}}},\ }\href
  {https://github.com/lalit-pathak/prompt_sky_localization_meshfree_pe}
  {\bibinfo {title} {github repository of the relevant codes used in this
  paper}} (\bibinfo {year} {2023})\BibitemShut {NoStop}%
\bibitem [{\citenamefont {Abbott}\ \emph
  {et~al.}(2021{\natexlab{c}})\citenamefont {Abbott}, \citenamefont {Abbott},
  \citenamefont {Acernese}, \citenamefont {Ackley}, \citenamefont {Adams},
  \citenamefont {Adhikari}, \citenamefont {Adhikari}, \citenamefont {Adya},
  \citenamefont {Affeldt}, \citenamefont {Agarwal} \emph
  {et~al.}}]{abbott2021gwtc-3}%
  \BibitemOpen
  \bibfield  {author} {\bibinfo {author} {\bibfnamefont {R.}~\bibnamefont
  {Abbott}}, \bibinfo {author} {\bibfnamefont {T.}~\bibnamefont {Abbott}},
  \bibinfo {author} {\bibfnamefont {F.}~\bibnamefont {Acernese}}, \bibinfo
  {author} {\bibfnamefont {K.}~\bibnamefont {Ackley}}, \bibinfo {author}
  {\bibfnamefont {C.}~\bibnamefont {Adams}}, \bibinfo {author} {\bibfnamefont
  {N.}~\bibnamefont {Adhikari}}, \bibinfo {author} {\bibfnamefont
  {R.}~\bibnamefont {Adhikari}}, \bibinfo {author} {\bibfnamefont
  {V.}~\bibnamefont {Adya}}, \bibinfo {author} {\bibfnamefont {C.}~\bibnamefont
  {Affeldt}}, \bibinfo {author} {\bibfnamefont {D.}~\bibnamefont {Agarwal}},
  \emph {et~al.},\ }\bibfield  {title} {\bibinfo {title} {Gwtc-3: compact
  binary coalescences observed by ligo and virgo during the second part of the
  third observing run},\ }\bibfield  {journal} {\bibinfo  {journal} {arXiv
  preprint arXiv:2111.03606}\ }\href
  {https://doi.org/https://doi.org/10.48550/arXiv.2111.03606}
  {https://doi.org/10.48550/arXiv.2111.03606} (\bibinfo {year}
  {2021}{\natexlab{c}})\BibitemShut {NoStop}%
\bibitem [{\citenamefont {Abbott}\ \emph
  {et~al.}(2021{\natexlab{d}})\citenamefont {Abbott}, \citenamefont {Abbott},
  \citenamefont {Abraham}, \citenamefont {Acernese}, \citenamefont {Ackley},
  \citenamefont {Adams}, \citenamefont {Adams}, \citenamefont {Adhikari},
  \citenamefont {Adya}, \citenamefont {Affeldt}, \citenamefont {Agarwal},
  \citenamefont {Agathos} \emph {et~al.}}]{Abbott_2021}%
  \BibitemOpen
  \bibfield  {author} {\bibinfo {author} {\bibfnamefont {R.}~\bibnamefont
  {Abbott}}, \bibinfo {author} {\bibfnamefont {T.~D.}\ \bibnamefont {Abbott}},
  \bibinfo {author} {\bibfnamefont {S.}~\bibnamefont {Abraham}}, \bibinfo
  {author} {\bibfnamefont {F.}~\bibnamefont {Acernese}}, \bibinfo {author}
  {\bibfnamefont {K.}~\bibnamefont {Ackley}}, \bibinfo {author} {\bibfnamefont
  {A.}~\bibnamefont {Adams}}, \bibinfo {author} {\bibfnamefont
  {C.}~\bibnamefont {Adams}}, \bibinfo {author} {\bibfnamefont {R.~X.}\
  \bibnamefont {Adhikari}}, \bibinfo {author} {\bibfnamefont {V.~B.}\
  \bibnamefont {Adya}}, \bibinfo {author} {\bibfnamefont {C.}~\bibnamefont
  {Affeldt}}, \bibinfo {author} {\bibfnamefont {D.}~\bibnamefont {Agarwal}},
  \bibinfo {author} {\bibfnamefont {M.}~\bibnamefont {Agathos}}, \emph
  {et~al.},\ }\bibfield  {title} {\bibinfo {title} {Observation of
  gravitational waves from two neutron star{\textendash}black hole
  coalescences},\ }\href {https://doi.org/10.3847/2041-8213/ac082e} {\bibfield
  {journal} {\bibinfo  {journal} {The Astrophysical Journal Letters}\ }\textbf
  {\bibinfo {volume} {915}},\ \bibinfo {pages} {L5} (\bibinfo {year}
  {2021}{\natexlab{d}})}\BibitemShut {NoStop}%
\end{thebibliography}%

\clearpage

\appendix
\onecolumngrid

\section{Effect of different choices of Gaussian distributed nodes on the posteriors}
\label{appendix:A}
The distribution of input nodes plays an important role in the meshfree PE algorithm as described in Section~\ref{subsec:nodesplacement}. One would like to place nodes intelligently so as to provide good coverage near the support of the log-likelihood function, which falls very steeply around its mode. Here, we show the effect of node placement on the posteriors obtained. We choose a certain fraction of RBF nodes ($\ngauss$) from a normal (Gaussian) distribution described by the covariance matrix. The remaining nodes are distributed uniformly over the sample space. We find that the overall accuracy increases as we choose a higher fraction of Gaussian distributed nodes (particularly on the $\iota$ and $d_L$ parameters). Although, the estimations of $\alpha$, $\delta$, and $\mathcal{M}$ are quite robust against different choices of $\ngauss$.

\begin{figure*}[bh]
\centering
\includegraphics[width=0.85\linewidth]{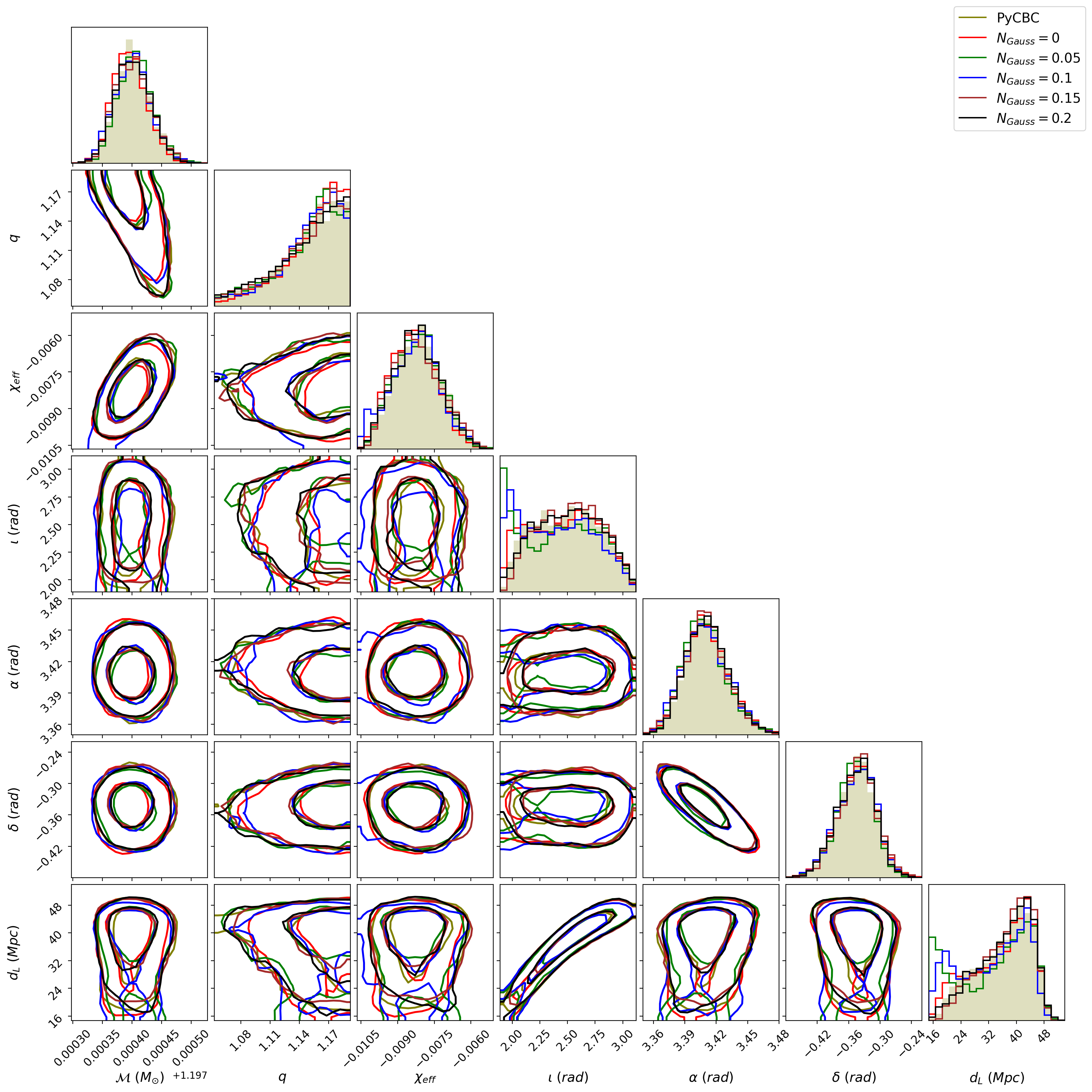}
\caption{The effect of choosing different values of $\ngauss$ (fraction of Gaussian nodes) on the reconstructed parameters.}
\label{fig:corner_plot_Ngauss}
\end{figure*}

\section{GW200115: An NSBH event}
\label{appendix:B}
We present the results of reconstructing the GW200115 NSBH event using the meshfree and the standard \pycbc ~ likelihood (for reference). 
The seismic cutoff frequency is set to $20$ Hz, assuming the \texttt{IMRPhenomD} waveform model. We used a $64$ seconds data segment around the GW200115 NSBH event, and PSD was generated using the same data with \texttt{median-mean} estimation. The center of the sample space was chosen as the MAP values of the PE samples obtained by previous LIGO analysis~\cite{abbott2021gwtc-3}. 

The parameters for RBF interpolation are as follows: $N = 1100$, $N_{\text{Gauss}} = 110$, $\nu = 10$, $\phi = \exp(-\epsilon \, r^2)$, and $\epsilon = 20$. The sampling parameters for \texttt{dynesty} nested sampler were the following: (i) \texttt{nlive} $= 2500$, (ii) \texttt{walks} $= 350$, (iii) \texttt{sample} = ``rwalk'', and (iv) \texttt{dlogz} $ = 0.1$. 

The meshfree likelihood-based PE process was completed in $\sim 16$ minutes when performed on $64$ cores. The skymaps are broadly in agreement with results published by the LVK collaboration~\cite{Abbott_2021}.

\begin{figure*}[tbh]
\centering
\includegraphics[width=0.75\linewidth]{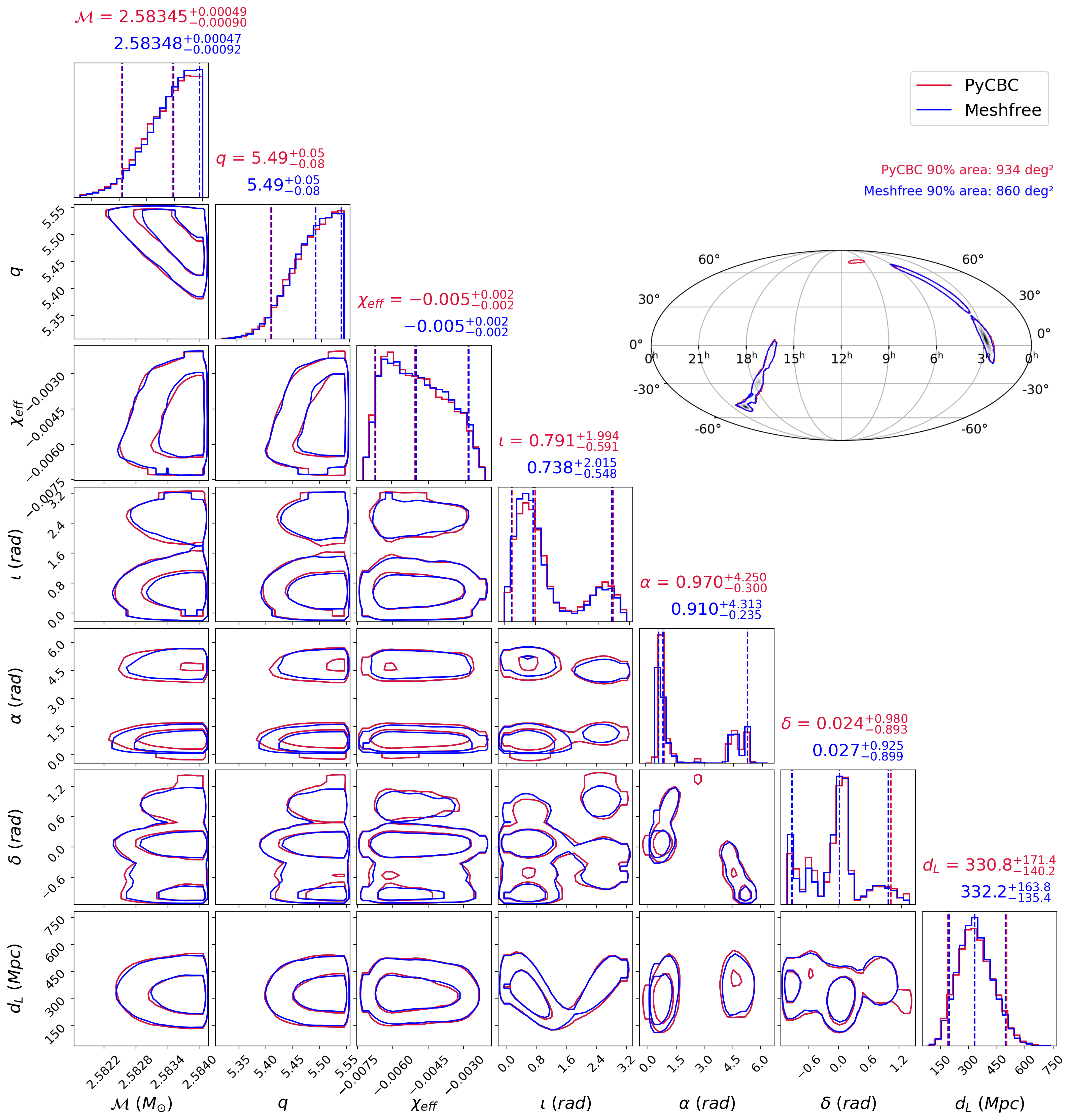}
\caption{The corner plot shows the posterior distributions of ten-dimensional parameters using two different methods: the meshfree method and the standard \pycbc ~ likelihood. The estimated distributions from both methods exhibit good agreement, as indicated by the median values in the marginalized posterior titles. In the inset, the skymaps are also shown, which are in agreement with results published by the LVK collaboration~\cite{Abbott_2021}.}
\label{fig:corner_plot_GW200115}
\end{figure*}

\section{Likelihood errors}
\label{appendix:C}
To assess the accuracy of the meshfree method to approximate the likelihood, we generated $10^4$ random query points from the prior distribution specified in Table~\ref{tab:priordistr}. Then, we calculated the likelihood using both the meshfree and PyCBC likelihood methods. The median absolute error for both \TaylorF~and \texttt{IMRPhenomD} waveform models were found to be $\mathcal{O}(10^{-2})$.

\begin{figure*}[hbt!]
\begin{subfigure}{0.49\linewidth}
    \includegraphics[width=\linewidth]{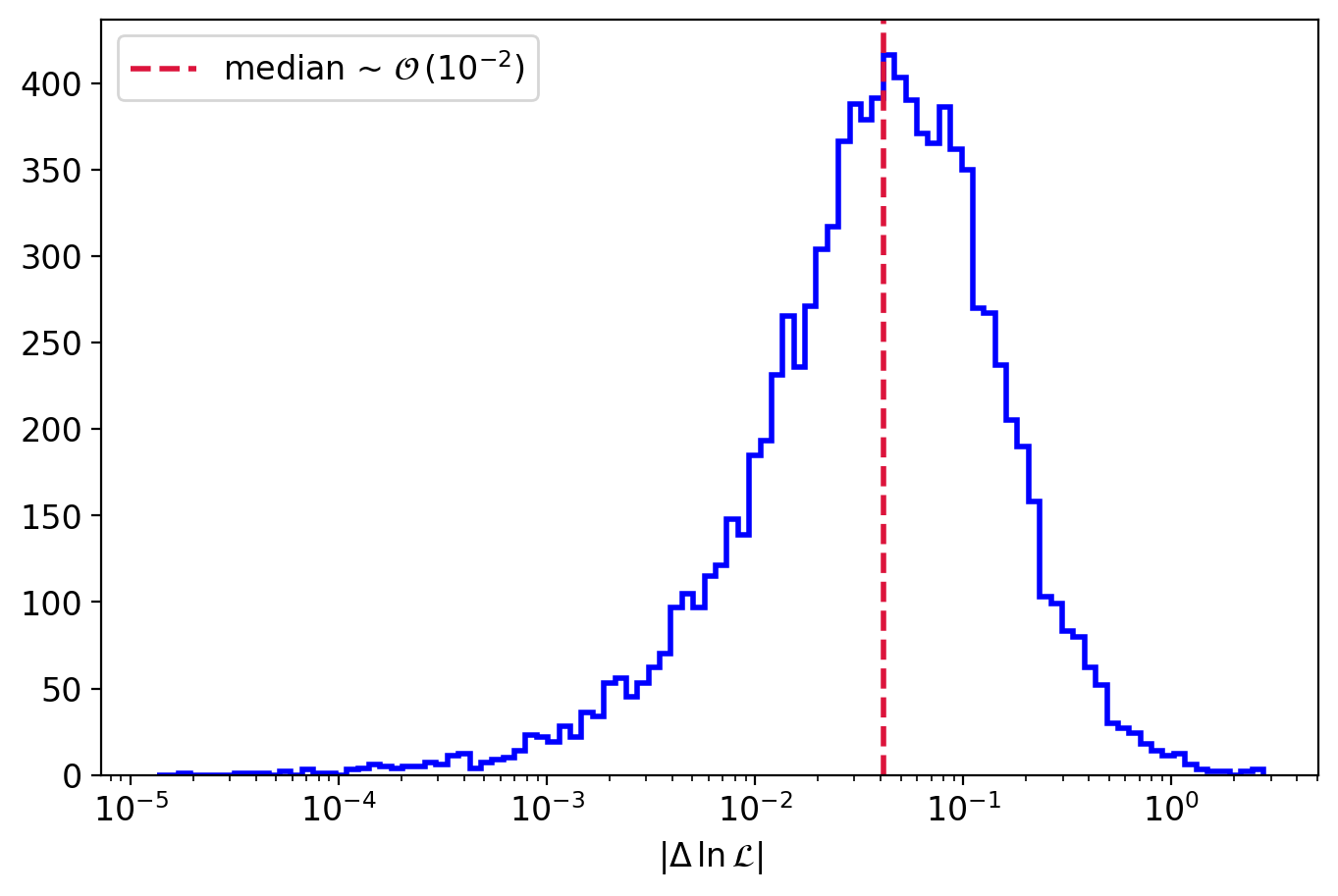}
    \caption{\TaylorF~signal model}
    \label{figerror:TF2}
\end{subfigure}\hfill
\begin{subfigure}{0.49\linewidth}
    \includegraphics[width=\linewidth]{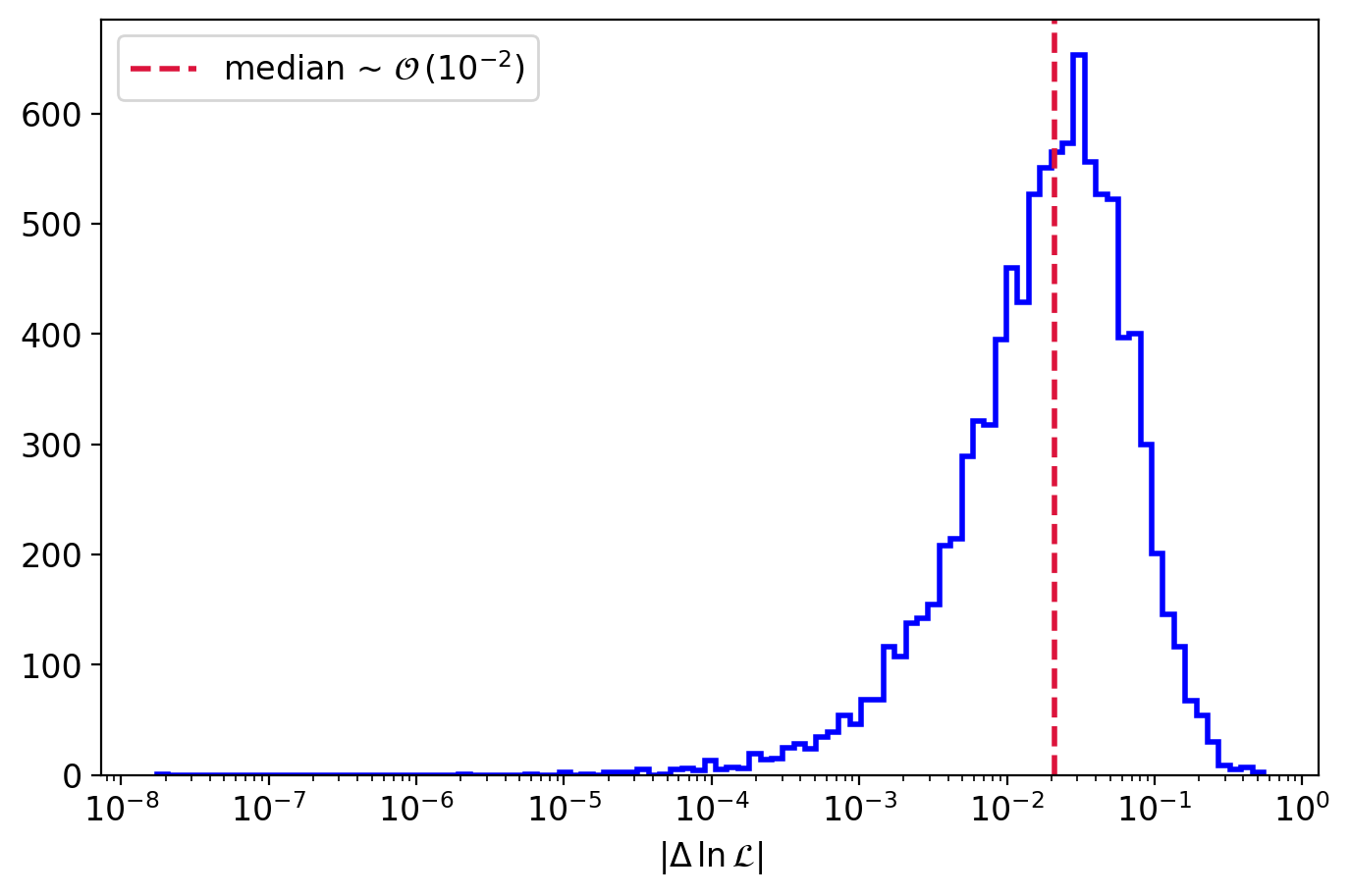}
    \caption{\texttt{IMRPhenomD} signal model}
    \label{figerror:IMR}
\end{subfigure}
\caption{The likelihood errors ($\Delta \ln \mathcal{L} \equiv \ln \mathcal{L}_{\text{PyCBC}} - \ln \mathcal{L}_{\text{RBF}}$) are shown for both \TaylorF~and \texttt{IMRPhenomD} waveform models (on GW170817 event). The red-dashed line represents the median absolute error ($|\Delta \ln \mathcal{L}|$) that are $\sim \mathcal{O} \, (10^{-2})$. The are very few points with high errors ($|\cdot| \sim 4$($2$) for \TaylorF(\texttt{IMRPhenomD})) which are not shown here.}
\label{fig:error}
\end{figure*}

\end{document}